\documentclass[twocolumn,twocolappendix,tighten,dvipsnames]{aastex63}

\usepackage{comment}
\usepackage[encapsulated]{CJK}
\usepackage{ucs}
\usepackage[utf8x]{inputenc}

\usepackage{eht}

\usepackage{savesym}
\savesymbol{tablenum}
\usepackage[group-separator={,}]{siunitx}
\restoresymbol{SIX}{tablenum}

\newcommand\subsubsubsection[1]{\paragraph{#1}}

\newcommand{\aprilvii}{April~7\xspace}

\newcommand{\vam}{VA morphology\xspace}
\newcommand{\mring}{m-ring\xspace}
\newcommand{\mrings}{m-rings\xspace}
\newcommand{\Mring}{M-ring\xspace}

\begin{document}

%
\newcounter{iPap}\setcounter{iPap}{5}
\newcommand{\ehtsubtitle}{This is just the GAL for now}

\ifnum\value{iPap}=1 \renewcommand{\ehtsubtitle}{The Shadow of the Supermassive Black Hole in the Center of the Milky Way}\fi
\ifnum\value{iPap}=2 \renewcommand{\ehtsubtitle}{EHT and Multi-wavelength Observations, Data Processing, and Calibration}\fi
\ifnum\value{iPap}=3 \renewcommand{\ehtsubtitle}{Imaging of the Galactic Centre Supermassive Black Hole}\fi
\ifnum\value{iPap}=4 \renewcommand{\ehtsubtitle}{Variability, morphology, and black hole mass}\fi
\ifnum\value{iPap}=5 \renewcommand{\ehtsubtitle}{Testing Astrophysical Models of the Galactic Center Black Hole}\fi
\ifnum\value{iPap}=6 \renewcommand{\ehtsubtitle}{Testing the Black Hole Metric}\fi

\title{%
First Sagittarius A* Event Horizon Telescope Results.
\Roman{iPap}. \ehtsubtitle}

\shorttitle{\ehtsubtitle}

\author[0000-0002-9475-4254]{Kazunori Akiyama}
\affiliation{Massachusetts Institute of Technology Haystack Observatory, 99 Millstone Road, Westford, MA 01886, USA}
\affiliation{National Astronomical Observatory of Japan, 2-21-1 Osawa, Mitaka, Tokyo 181-8588, Japan}
\affiliation{Black Hole Initiative at Harvard University, 20 Garden Street, Cambridge, MA 02138, USA}

\author[0000-0002-9371-1033]{Antxon Alberdi}
\affiliation{Instituto de Astrof\'{\i}sica de Andaluc\'{\i}a-CSIC, Glorieta de la Astronom\'{\i}a s/n, E-18008 Granada, Spain}

\author{Walter Alef}
\affiliation{Max-Planck-Institut f\"ur Radioastronomie, Auf dem H\"ugel 69, D-53121 Bonn, Germany}

\author[0000-0001-6993-1696]{Juan Carlos Algaba}
\affiliation{Department of Physics, Faculty of Science, Universiti Malaya, 50603 Kuala Lumpur, Malaysia}

\author[0000-0003-3457-7660]{Richard Anantua}
\affiliation{Black Hole Initiative at Harvard University, 20 Garden Street, Cambridge, MA 02138, USA}
\affiliation{Center for Astrophysics $|$ Harvard \& Smithsonian, 60 Garden Street, Cambridge, MA 02138, USA}
\affiliation{Department of Physics \& Astronomy, The University of Texas at San Antonio, One UTSA Circle, San Antonio, TX 78249, USA}

\author[0000-0001-6988-8763]{Keiichi Asada}
\affiliation{Institute of Astronomy and Astrophysics, Academia Sinica, 11F of Astronomy-Mathematics Building, AS/NTU No. 1, Sec. 4, Roosevelt Rd, Taipei 10617, Taiwan, R.O.C.}

\author[0000-0002-2200-5393]{Rebecca Azulay}
\affiliation{Departament d'Astronomia i Astrof\'{\i}sica, Universitat de Val\`encia, C. Dr. Moliner 50, E-46100 Burjassot, Val\`encia, Spain}
\affiliation{Observatori Astronòmic, Universitat de Val\`encia, C. Catedr\'atico Jos\'e Beltr\'an 2, E-46980 Paterna, Val\`encia, Spain}
\affiliation{Max-Planck-Institut f\"ur Radioastronomie, Auf dem H\"ugel 69, D-53121 Bonn, Germany}

\author[0000-0002-7722-8412]{Uwe Bach}
\affiliation{Max-Planck-Institut f\"ur Radioastronomie, Auf dem H\"ugel 69, D-53121 Bonn, Germany}

\author[0000-0003-3090-3975]{Anne-Kathrin Baczko}
\affiliation{Max-Planck-Institut f\"ur Radioastronomie, Auf dem H\"ugel 69, D-53121 Bonn, Germany}

\author{David Ball}
\affiliation{Steward Observatory and Department of Astronomy, University of Arizona, 933 N. Cherry Ave., Tucson, AZ 85721, USA}

% ...

\collaboration{0}{The Event Horizon Telescope Collaboration}

\ifnum\value{iPap}=1 \include{./SAL1}\fi
\ifnum\value{iPap}=2 \include{./SAL2}\fi
\ifnum\value{iPap}=3 \include{./SAL3}\fi
\ifnum\value{iPap}=4 \include{./SAL4}\fi
\ifnum\value{iPap}=5 \author[0000-0001-9197-932X]{Tin Lok Chan}
\affiliation{Department of Physics, The Chinese University of Hong Kong, Shatin, N. T., Hong Kong}

\author[0000-0003-3462-0817]{Richard Qiu}
\affiliation{Department of Physics, Harvard University, 17 Oxford Street, Cambridge, MA 02138, USA}
\affiliation{John A. Paulson School of Engineering and Applied Sciences, Harvard University, 29 Oxford Street, Cambridge, MA 02138, USA,}

\author[0000-0003-0220-5723]{Sean Ressler}
\affiliation{Kavli Institute For Theoretical Physics, 2411 Kohn Hall, Santa Barbara, CA 93111 USA}

\author[0000-0001-7448-4253]{Chris White}
\affiliation{Department of Astrophysical Sciences, Peyton Hall, Princeton University, Princeton, NJ 08544, USA}

\fi
\ifnum\value{iPap}=6 \include{./SAL6}\fi

\shortauthors{The EHT Collaboration et al.}


\begin{abstract}
  In this paper, we provide a first physical interpretation for the Event Horizon Telescope (EHT)'s 2017 observations of \sgra.
  Our main approach is to compare resolved EHT data at $230\GHz$ and unresolved non-EHT observations from radio to X-ray wavelengths to predictions from a library of models based on time-dependent general relativistic magnetohydrodynamics (GRMHD) simulations, including aligned, tilted, and stellar wind-fed simulations; radiative transfer is performed assuming both thermal and non-thermal electron distribution functions. We test the models against 11 constraints drawn from EHT 230\GHz data and observations at 86\GHz, 2.2\um, and in the X-ray.
  All models fail at least one constraint.
  Light curve variability provides a particularly severe constraint, failing nearly all strongly magnetized (MAD) models and a large fraction of weakly magnetized (SANE) models.
  A number of models fail only the variability constraints.  We identify a promising cluster of these models, which are MAD and have inclination $i \le 30\degree$.
  They  have accretion rate $(5.2$--$9.5)\times10^{-9}\msun\yr^{-1}$, bolometric luminosity $(6.8$--$9.2)\times10^{35}\ergps$, and outflow power $(1.3$--$4.8)\times10^{38}\ergps$.
  We also find that: all models with $i \ge 70\degree$ fail at least two constraints, as do all models with equal ion and electron temperature;  exploratory, non-thermal model sets tend to have higher 2.2\um flux density; the population of cold electrons is limited by X-ray constraints due to the risk of bremsstrahlung overproduction.
  Finally we discuss physical and numerical limitations of the models, highlighting the possible importance of kinetic effects and duration of the simulations.
\end{abstract}




\section{Introduction}
\label{sec:intro}

The center of the Milky Way contains a massive compact object that is likely a supermassive black hole \citep{2019Sci...365..664D, 2019A&A...625L..10G}.
The putative black hole is surrounded by hot plasma that is visible across 17 decades in electromagnetic frequency.
Hereafter, we will use \sgra to refer to the supermassive black hole candidate and the hot plasma.

\sgra is one of the most studied objects on the sky, both observationally and theoretically.
A key  characteristic of the \sgra system is its extremely low overall luminosity with respect to the Eddington limit.
The low luminosity suggests that matter falls onto \sgra's central object in the form of a radiatively inefficient/advection dominated accretion flow (RIAF/ADAF, as proposed by \citealt{1977ApJ...214..840I,1982Natur.295...17R,1994ApJ...428L..13N, 1995ApJ...444..231N, 1995ApJ...452..710N, 1996A&AS..120C.287N, 1998ApJ...492..554N,2014ARA&A..52..529Y}) rather than in the form of a radiatively efficient thin disk \citep{1973A&A....24..337S}.
Since the nearly flat radio spectrum of \sgra is similar to radio spectra observed in jets from Active Galactic Nuclei, it has also been suggested that the majority of the \sgra emission could be produced by a jet launched by an accreting black hole rather than matter falling through the black hole event horizon \citep{1993A&A...278L...1F, 2000A&A...362..113F}.

Models of magnetized RIAFs/ADAFs have been constructed using semi-analytic prescriptions \citep[e.g.,][]{1995Natur.374..623N,2000ApJ...541..234O, 2009ApJ...697...45B,2011ApJ...735..110B} and using time-dependent General Relativistic Magnetohydrodynamics (GRMHD) simulations \citep[e.g.,][]{2000ApJ...528..462H, 2003ApJ...589..458D, 2003ApJ...589..444G, 2007CQGra..24S.235G, 2012ApJS..201....9F, 2014ApJ...796...22F, 2016ApJS..225...22W, 2017ApJS..231...17A, 2018JPhCS1031a2008O, Olivares2019, 2019ApJS..243...26P, Liska2019}.
Semi-analytic RIAF/ADAF models typically do not include relativistic jets or outflows, but those are naturally produced in GRMHD simulation and contribute to the observed emission.
GRMHD simulations also naturally produce variability, which is observed in \sgra at multiple wavelengths.

GRMHD simulations of ADAFs show that ADAF-like inflows are not unique.
In particular two dramatically different modes are observed, depending on the magnetic flux interior to the black hole equator: the standard and normal evolution (SANE) mode, in which the midplane magnetic field pressure is less than the gas pressure and magnetic fields are turbulent; and the magnetically arrested disk (MAD) mode, in which magnetic fields are strong and organized and can even disrupt accretion.
An outstanding question about \sgra is whether the flow is in MAD or SANE mode, or possibly in a third mode that results from wind-fed accretion \citep{2020ApJ...896L...6R}.

The energy distribution of electrons in the emitting plasma is also not known.
Because emission is driven by the synchrotron process, this is critical in determining the observational appearance of the source.
In particular the energy per electron may increase with latitude in the flow, leading to a jet or outflow that outshines an equatorial inflow.

The question of whether emission is dominated by an inflow or outflow is intimately tied to the problem of what drives an outflow, if there is one.
In GRMHD simulations of black hole accretion the strength of the outflow depends sensitively on the black hole spin (e.g., \citealt{M87PaperV}, hereafter \citetalias{M87PaperV} or \citealt{2022MNRAS.511.3795N}).
At large spin GRMHD simulations produce powerful jets driven by extraction of black hole spin energy via the \citet{1977MNRAS.179..433B} process.
A spatially resolved study of \sgra may thus also constrain the black hole spin and provide direct evidence for black hole energy extraction.

Previously published GRMHD models of \sgra generically predict source sizes at millimeter wavelengths consistent with observational data \citep[e.g.,][]{2008Natur.455...78D, 2009ApJ...706..497M, 2009ApJ...703L.142D,2010ApJ...717.1092D};
the radio spectral shape is similar to jet emission \citep[e.g.,][]{2013A&A...559L...3M, 2017MNRAS.467.3604R}, and the source linear polarization requires strongly magnetized flow or non-thermal electrons \citep{2015Sci...350.1242J, 2017ApJ...837..180G, 2020MNRAS.494.4168D}.

A major difficulty in determining the nature of \sgra radio emission is caused by the interstellar scattering screen that distorts our view of the Galactic Center up to $\lambda \sim 1\mm$ wavelengths
\citep[see][and references therein]{2018arXiv180501242P, 2018ApJ...865..104J,2019ApJ...871...30I}.
The Event Horizon Telescope (EHT) is a very-long-baseline interferometric (VLBI) experiment operating at $230\GHz$ or wavelength $\lambda = 1.3\mm$ (see \citealt{M87PaperII}, hereafter \citetalias{M87PaperII}, for an introduction to the instrument).
EHT operates at high enough frequency to penetrate the scattering screen, with angular resolution sufficient to directly image structures in the immediate vicinity of the black hole event horizon.

In April~2017 the EHT observed \sgra (among other sources, including the core of the M87 galaxy, see \citealt{M87PaperI}, hereafter \citetalias{M87PaperI}) and produced the first ever horizon scale images of the source.
We report the results of these observations in \citet{PaperII}, hereafter \citetalias{PaperII} and \citet{PaperIII}, hereafter \citetalias{PaperIII}, characterize the basic properties of the emission visible in the EHT images in \citealt{PaperIV}, hereafter \citetalias{PaperIV}, and discuss implications for tests of general relativity in \citealt{PaperVI}, hereafter \citetalias{PaperVI}.
The main goal of this paper \citepalias{PaperV} is to provide the first comprehensive physical interpretation of the EHT~2017 \sgra datasets.

This paper is structured as follows.
Section~\ref{sec:models} describes our main assumptions, a one-zone source models, and a standard simulation and synthetic image library used to model near-horizon emission from \sgra.
Our model library assumes that general relativity is valid and the spacetime around \sgra is described by the Kerr metric  \citep{1963PhRvL..11..237K}.
A discussion of \sgra observations in the context of alternative theories of gravity can be found in  \citet{PaperVI}, hereafter \citetalias{PaperVI}.
Our model library is based on time-dependent GRMHD simulations that, combined with general relativistic radiative transfer models, result in images and broadband spectra of the models.
The library of simulated images was used in \citetalias{PaperIII} and \citetalias{PaperIV}, to validate the \sgra EHT imaging and parameter estimation algorithms.
In Section~\ref{sec:observations}, we describe the  observational constraints that are used in the present work to test theoretical models of \sgra.
These data comprise a subset of EHT~2017 observations and other non-EHT historical or other data.
In Section~\ref{sec:comparisons}, we describe model scoring procedures and use our model library to infer physical properties of \sgra system.
We discuss model limitations, results in the context of previous studies and outlook for future \sgra theoretical research directions in Section~\ref{sec:discussions}.
Finally, we conclude in Section~\ref{sec:conclusions}.

This paper is supplemented with several appendices.
In Appendix~\ref{app:numerical},  discusses numerical details of our simulations.
In Appendix~\ref{app:variability}, discusses the impact of physical  and numerical effects on the model variability.
In Appendix~\ref{app:tables}, summarizes the results of applying constraints to our fiducial models in an extended set of figures.

\section{Astrophysical Models}
\label{sec:models}

\subsection{Basic Assumptions}
\label{sec:basic}

We assume the mass of and distance to \sgra are
\begin{align}
  \mbh &= 4.14  \times 10^6 \msun, \label{eq:mass} \\
  D    &= 8.127 \kpc,              \label{eq:dist}
\end{align}
which are approximately the mean of the values reported by \citet{2019Sci...365..664D} and \citet{2019A&A...625L..10G}, which differ from each other by about 4\%.
The distance is consistent with maser parallax measurements \citep{2019ApJ...885..131R}.

We also assume that \sgra is a black hole described by the Kerr metric.
The dimensionless spin, $\abh \equiv Jc/G\mbh^2$, is a free parameter with $-1 < \abh < 1$, where $J$, $G$, and $c$ are the black hole spin angular momentum, gravitational constant, and speed of light, respectively.
Following \citetalias{M87PaperV}, we use
$\abh > 0$ to indicate that the angular momentum of the accretion flow and black hole are parallel (the accretion flow is ``prograde'') and
$\abh < 0$ to indicate that the angular momentum of the accretion flow and black hole are antiparallel (``retrograde'')\footnote{For tilted disks the sign of $\abh$ is the sign of ${\bf J}\cdot{\bf L}$ where ${\bf J}$ is black hole spin angular momentum and ${\bf L}$ is accretion flow orbital angular momentum.}.

The implied characteristic length
\begin{equation}
  r_\mathrm{g}         \equiv G\mbh/ c^2    \simeq 6.1\times10^{11}\cm,
\end{equation}
the characteristic time
\begin{equation}
  t_\mathrm{g}         \equiv G\mbh/ c^3    \simeq 20.4\sec,
\end{equation}
and the angular scale
\begin{equation}
  \vartheta_\mathrm{g} \equiv G\mbh/(c^2 D) \simeq 5.03\uas.
\end{equation}
The expected diameter of the black hole shadow is $2\sqrt{27} G\mbh/(c^2 D)$ for $\abh = 0$.
For $|\abh| > 0$ the shadow is noncircular and its size and shape depend on $\abh$ and inclination $i$ (the angle between the line of sight and the spin axis); its width can be as small as $9 G\mbh/(c^2 D)$ for $\abh = 1$ and $i = 90\degree$ \citep{1973blho.conf..215B}.

If the emitting plasma is ionized hydrogen then the Eddington luminosity
\begin{align}
  L_\mathrm{Edd}
  \equiv 4\pi G\mbh c m_p/\sigma_\mathrm{T}
  = 5.2 \times 10^{44}\ergps,
\end{align}
where symbols have their usual meaning.
The corresponding Eddington accretion rate
\begin{align}
  \dot\mbh_\mathrm{Edd}
  \equiv L_\mathrm{Edd}/(0.1 c^2)
  &= 5.8 \times 10^{24} \gm \sec^{-1} \nonumber\\
  &= 0.09 \msun \yr^{-1},
\end{align}
where the nominal efficiency is 10\%.
The bolometric luminosity of \sgra is $L_\mathrm{bol} \sim 10^{35}\ergps$ in a quiescent, non-flaring state, so that
\begin{align}
 \frac{L_\mathrm{bol}}{L_\mathrm{Edd}} = 1.9 \times 10^{-10} \left(\frac{L_\mathrm{bol}}{10^{35} \ergps}\right),
\end{align}
an extremely small Eddington ratio.

\subsection{One-Zone Model}

Here we motivate the more complicated models that follow using  a simple one-zone model, following \citetalias{M87PaperV}
and one-zone models developed in the literature over many decades \citep[e.g.][]{1996IAUS..169..169F}.

Consider a uniform sphere of plasma with radius $r = 5\rg$, comparable to the observed size of \sgra at $230\GHz$ (\citetalias{PaperIII}, \citetalias{PaperIV}), with uniform magnetic field oriented at $\pi/3$ to the line-of-sight.
In turbulent astrophysical plasmas, it is common for the gas pressure to be comparable to the magnetic pressure, so we set $n_i \kB T_i + n_e \kB T_e = B^2/(8\pi)$, where $T_i \equiv$ ion temperature, $T_e \equiv$ electron temperature, $\kB \equiv$ Boltzmann constant, and $B \equiv$ magnetic field strength.
The plasma is collisionless (as shown below), and it is plausible that the ions are preferentially heated, so we assume $T_i = 3 T_e$.
If the ions are sub-virial by a factor of $3$, as commonly seen in relativistic MHD simulations, i.e., $(3/2) k T_i \sim (1/3) (1/2) (G M m_p/r)$, then the ions are nonrelativistic and the electrons are relativistic, with $\Theta_e \equiv  \kB T_e / (m_e c^2) \sim 10$.

Assuming a thermal distribution of electron energies (eDF) and therefore a thermal synchrotron emissivity $j_\nu$ \citep[e.g.,][]{2011ApJ...737...21L} and assuming optically thin emission, the flux density from a uniform sphere, $F_\nu = (4/3)\pi r^3 j_\nu D^{-2} 10^{23}\,\mathrm{Jy}$.
Requiring $F_\nu = 2.4\,\mathrm{Jy}$, the average measured by ALMA during the 2017 campaign \citep{Wielgus2022}, yields a nonlinear equation for the electron density $n_e$ with solution
\begin{align}
  n_e &\simeq 1.0\times10^6\cm^{-3},\\
  B   &\simeq 29\,\mathrm{Gauss}.
  \label{eq:onezone}
\end{align}
This is consistent with a similar one-zone model fit to archival \sgra millimeter spectra \citep{2019ApJ...881L...2B}.
The synchrotron optical depth $\tau_\mathrm{sync} = r j_\nu/B_\nu \simeq 0.4$, where $B_\nu$ is the Planck function, so the optically thin approximation is marginal.

The one-zone model has electron scattering optical depth  $\tau_e = \sigma_\mathrm{T} n_e r \simeq 2\times10^{-6}$ and thus the Compton parameter $y = 16 \Theta_e^2 \max(\tau_e,\tau_e^2) \simeq 0.003$ is small.
Synchrotron cooling therefore dominates Compton cooling.

The synchrotron cooling timescale for electrons $t_\mathrm{cool} \equiv u_e/\Lambda$ where $u_e = 3 n_e k T_e$ is the electron internal energy and $\Lambda \simeq 5.4 B^2 e^4 n_e \Theta_e^2 /(c^3 m_e^2)$ is the synchrotron cooling rate for a thermal population of electrons with $\Theta_e \gtrsim 1$ (see Appendix~A in \citealt{2011ApJ...735....9M}; finite optical depth reduces $\Lambda$).
Therefore $t_\mathrm{cool}=2.3 \times 10^4\sec \simeq 1.1 \times 10^3 \tg$, which is longer than the inflow timescale $r/v^r \sim r^{3/2}$.
This suggests that radiative cooling can be neglected in the plasma models.%
\footnote{The cooling time for 2.2$\um$ emitting electrons is $\sim 60 (B/(30 {\rm G}))^{-3/2} \tg$, so cooling is a more significant source of uncertainty for 2.2$\um$ emission.}
More detailed calculations confirm this estimate  \citep{2018MNRAS.478.5209C,  2020MNRAS.499.3178Y}.\footnote{If \sgra is fed by stellar winds then the inflowing plasma may be mainly helium \citep{2019MNRAS.482L.123R}; this changes the one-zone model slightly.
Helium accretion is discussed in \cite{Wong_2022}.}

The one-zone model solution implies that the mean free path to Coulomb scattering is large compared to $\rg$, i.e. the source plasma is collisionless.
At $\Theta_e \sim 1$, for example, the electron-electron Coulomb scattering cross section is comparable to the Thomson cross section, and the mean free path is therefore $\sim \tau_e^{-1} \rg$.
The electron-ion Coulomb scattering mean free path is even longer, and the electrons and ions are therefore poorly coupled.
This is consistent with our assumption that the ions and electrons can have different temperatures  \citep{1976ApJ...204..187S,1977ApJ...214..840I, 1982Natur.295...17R} and motivates consideration of non-thermal (unrelaxed) electron distribution functions \citep[see][]{2000ApJ...541..234O, 2009ApJ...701..521C, 2014A&A...570A...7M, 2018A&A...612A..34D, 2021NatAs.tmp..218C, Chatterjee2021, 2021arXiv211203933E, Scepi2021, 2022A&A...660A.107F}.

\subsection{Numerical Models}

\begin{deluxetable*}{cccccccccc}
  \label{tab:GRMHDmodels}
  \tablecaption{EHT GRMHD Simulation Library}
  \tablehead{
    \colhead{Setup}                &
    \colhead{Code}                 &
    \colhead{$\abh$}               &
    \colhead{Mode}                 &
    \colhead{$\Gamma_\mathrm{ad}$} &
    \colhead{$t_\mathrm{final}$ }  &
    \colhead{$r_{\rm out}$}        &
    \colhead{Resolution}%
  }
  \startdata
  torus    & {\kharma}$^a$   & 0, $\pm 0.5$, $\pm 0.94$   & MAD/SANE & $\frac{4}{3}/\frac{4}{3}$  & \no{30000}  & \no{1000}     & $288\times128\times128$     \\
  torus    & {\bhac}$^b$     & 0, $\pm 0.5$, $\pm 0.94$   & MAD/SANE & $\frac{4}{3}/\frac{4}{3}$  & \no{30000}  & \no{3333}     & $512\times192\times192$     \\
  torus    & {\hamr}$^c$     & 0, $\pm 0.5$, $\pm 0.94$   & MAD/SANE & $\frac{13}{9}/\frac{5}{3}$ & \no{35000}  & \no{1000}/200 & $348/240\times192\times192$ \\
  torus    & {\koral}$^d$    & \!\!\!\!\!\!0, $\pm 0.3$,  %
  $\pm 0.5$, $\pm 0.7$, $\pm 0.9$\!\!\!\!\!\!\!\!\!\!\!\! & MAD      & $\frac{13}{9}$             & \no{101000} & \no{100000}  & $288\times192\times144$     \\
  tilted   & {\hamr}$^e$     & $0.94$                     & SANE$^f$  & $\frac{5}{3}$              & \no{105000} & \no{100000}  & $448\times144\times240$     \\
  wind-fed & {\athenapp}$^g$ & 0                          & MAD     & $\frac{5}{3}$              & \no{20000}  & \no{2400}    & $356\times128\times128$
  \enddata
  \tablecomments{Summary of the EHT \sgra GRMHD simulation library.
    The last column is $N_1 \times N_2 \times N_3$, with coordinate
    $x_1$ monotonic in radius, $x_2$ monotonic in colatitude $\theta$,
    and $x_3$ proportional to longitude $\phi$.
    The first four entries use aligned torus initial conditions.
    The last two entries are tilted accretion models and two
    realizations of the wind-fed accretion model which differ in
    stellar wind magnetization.
    Times are given in units of $G M/c^3 = 20.4\sec$ and radii in units
    of $G M/c^2$.%
  }
  \tablerefs{%
    $^a$see \citet{2021JOSS....6.3336P}; \kharma is a GPU-enabled version of the {\tt iharm3d} code.
    $^b$\citet{2017ComAC...4....1P, Olivares2019, 2021MNRAS.506..741M, 2021NatAs.tmp..218C}.
    $^c$\citet{Liska2019}.
    $^d$\citet{2022MNRAS.511.3795N}.
    $^e$\citet{Chatterjee2020}.
    $^f$ $\phi/\phi_{crit} \simeq 0.8$.
    $^g$\citet{2016ApJS..225...22W, 2020ApJ...896L...6R}.%
  }
\end{deluxetable*}

The one-zone model is too simple for comparison with EHT data.  In particular it does not predict EHT image morphology, and it fails to model emission that arises outside the near-horizon region, including 86GHz emission and X-ray emission.
Steady spherical accretion models \citep[e.g.,][]{2000ApJ...528L..13F} go one  step beyond the one zone model, incorporating relativistic gravity and a radially extended flow.
Steady, disk-like (RIAF) accretion models in the Kerr metric go still further and include rotation and departures from spherical symmetry \citep[e.g.,][]{2009ApJ...697...45B, 2009ApJ...706..960H,2018ApJ...863..148P}.
Steady phenomenological models do not, however, self-consistently capture fluctuations in the flow.
That requires either a statistical model \citep{2021ApJ...906...39L} or a time-dependent numerical simulation.
Here we
use numerical simulations,
adopt an ideal GRMHD model for the flow,
employ simple parameterized models to assign an electron distribution function, and
solve the radiative transfer equation along geodesics to produce simulated images.

\subsubsection{Plasma Flow Model}

\begin{figure*}
  \centering
  \includegraphics[width=0.425\textwidth]{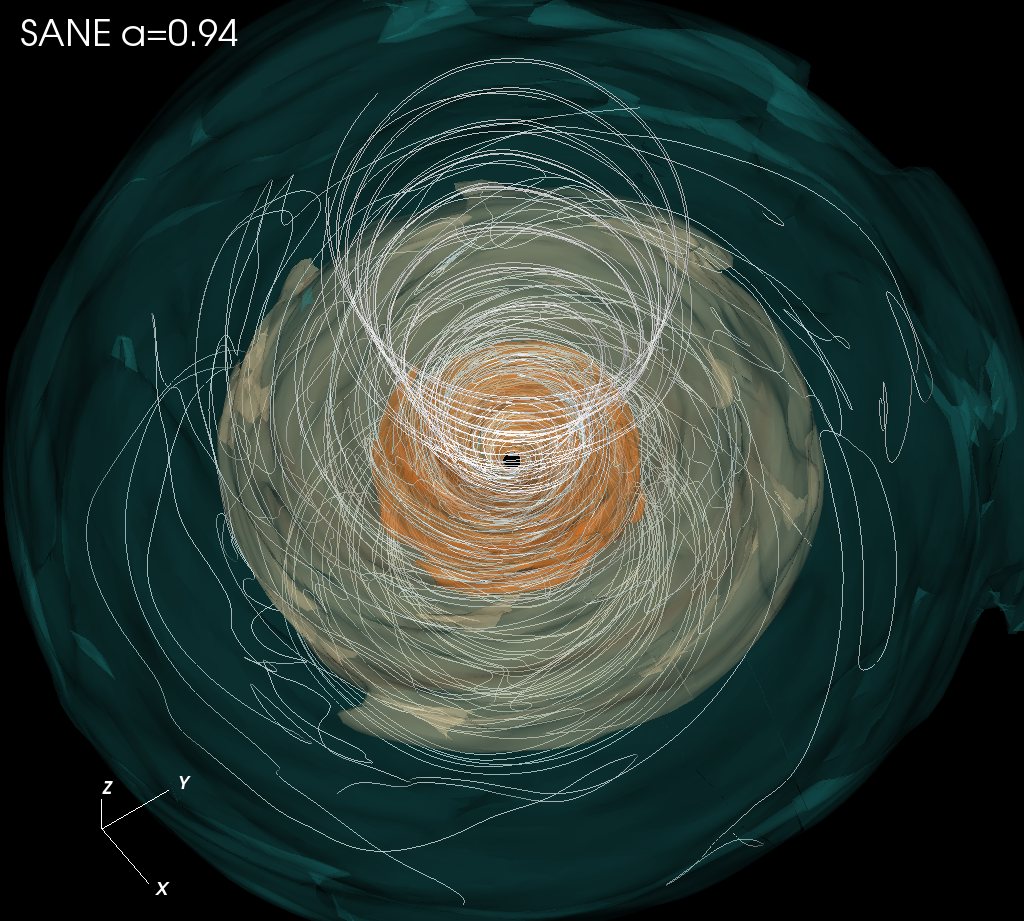}\hspace{1.5pt}%
  \includegraphics[width=0.425\textwidth]{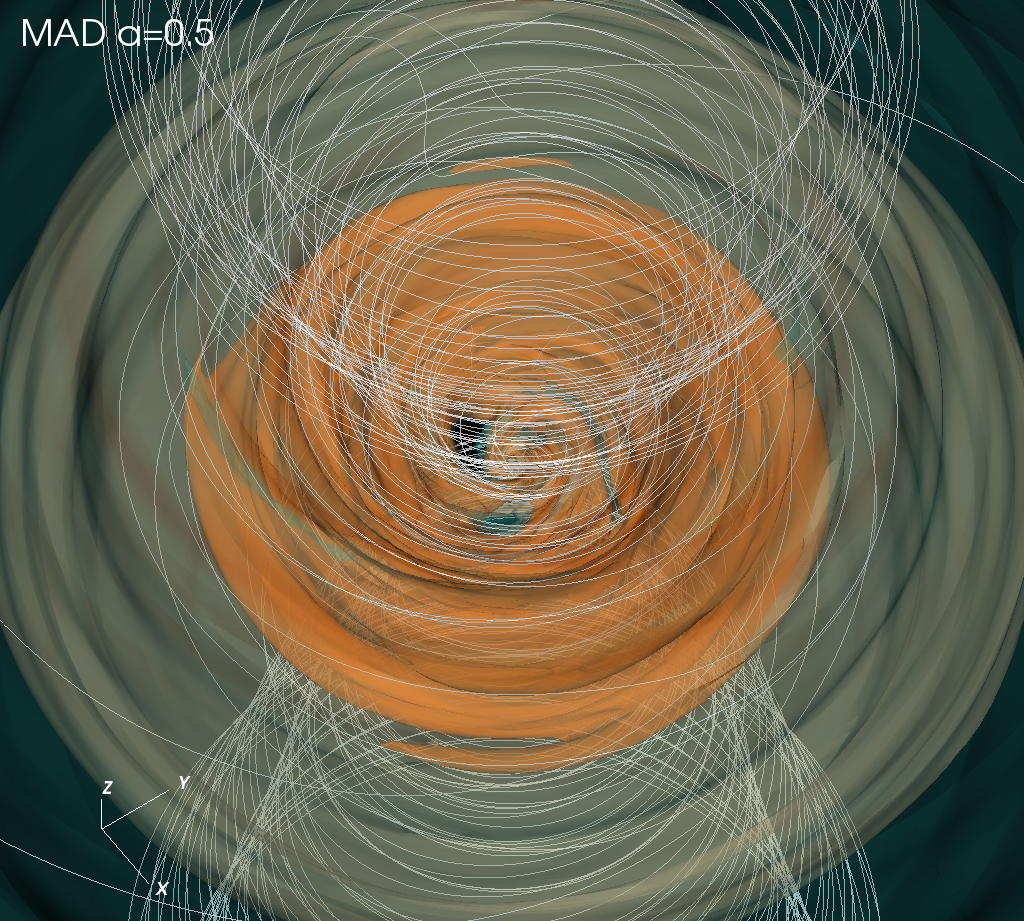}\\
  \includegraphics[width=0.425\textwidth]{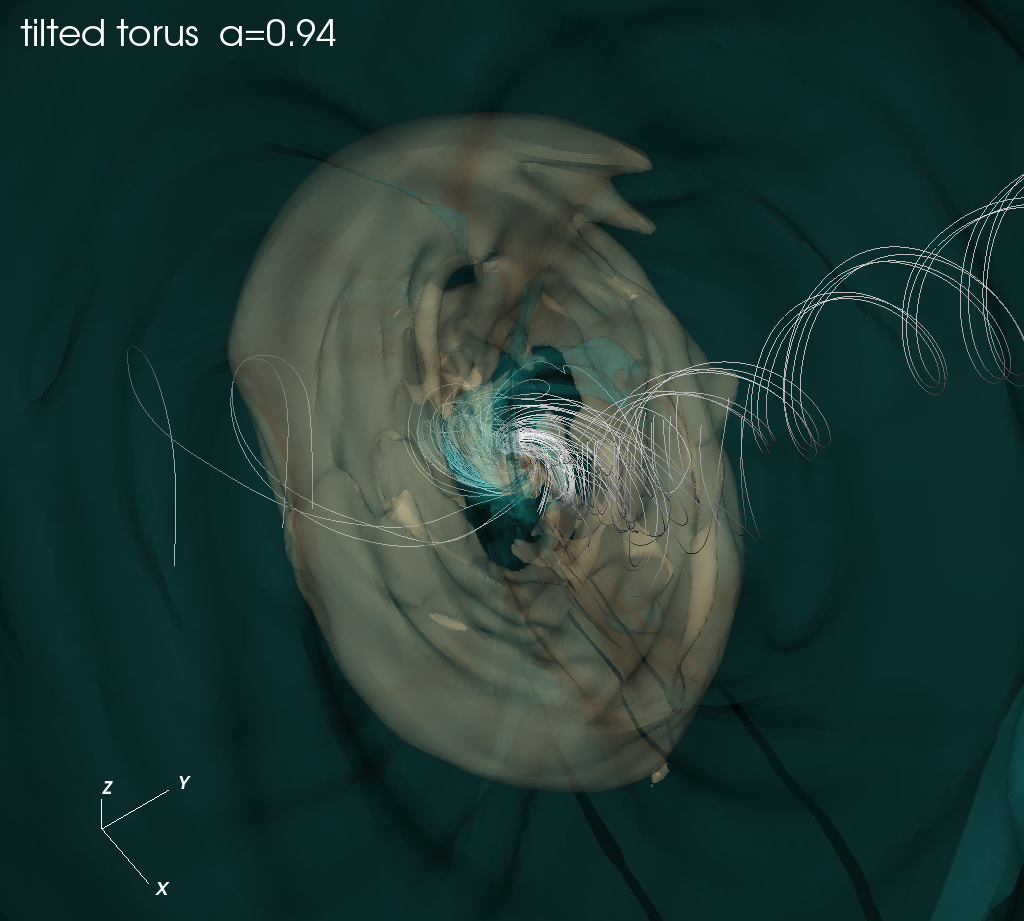}\hspace{1.5pt}%
  \includegraphics[width=0.425\textwidth]{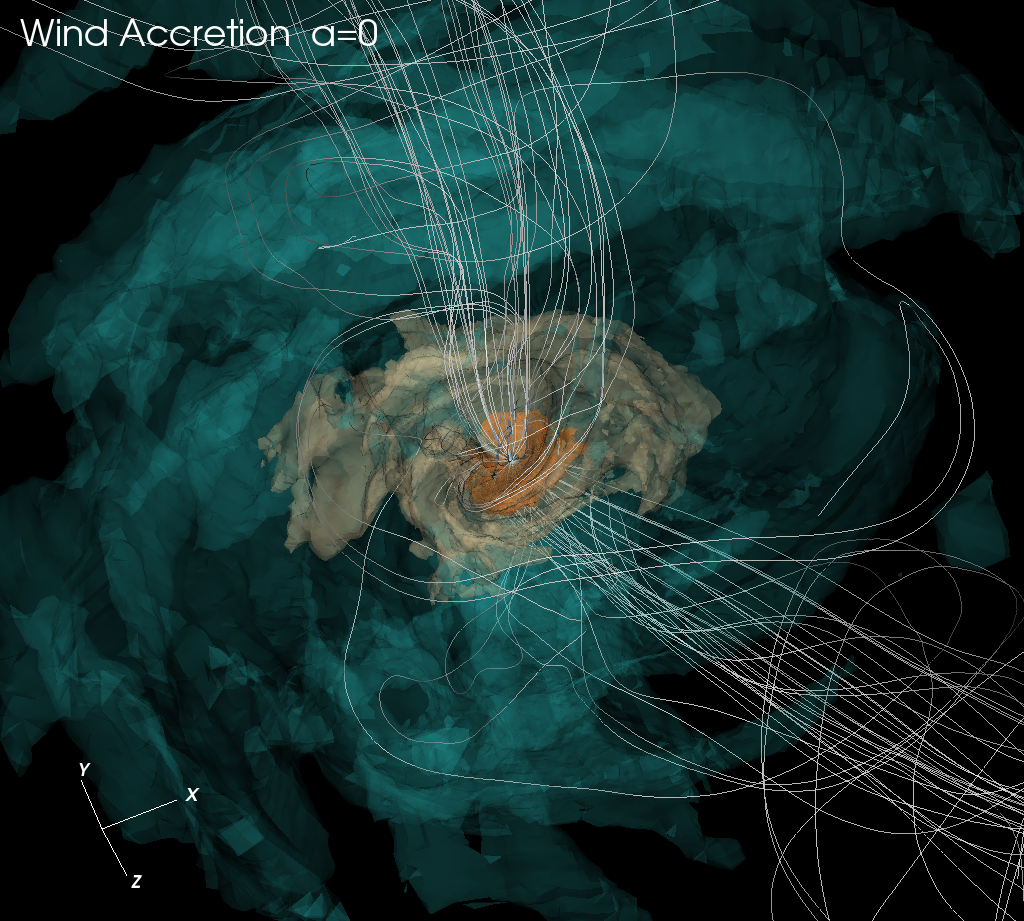}
  \caption{3-D overview of selected GRMHD simulations of \sgra in our library.
    The color marks constant dimensionless density surfaces and lines follow magnetic field lines.  The magnetic field lines shown are only those which are attached to the inner part of the accretion flow, at $r\approx 5~\rg$.
    Two top panels show accretion simulations with default torus initial condition and
    two bottom panels show non-standard accretion models.
    The spin is aligned with z-axis.}
  \label{fig:GRMHD}
\end{figure*}

\begin{figure*}
  \centering
  \includegraphics[width=0.9\textwidth]{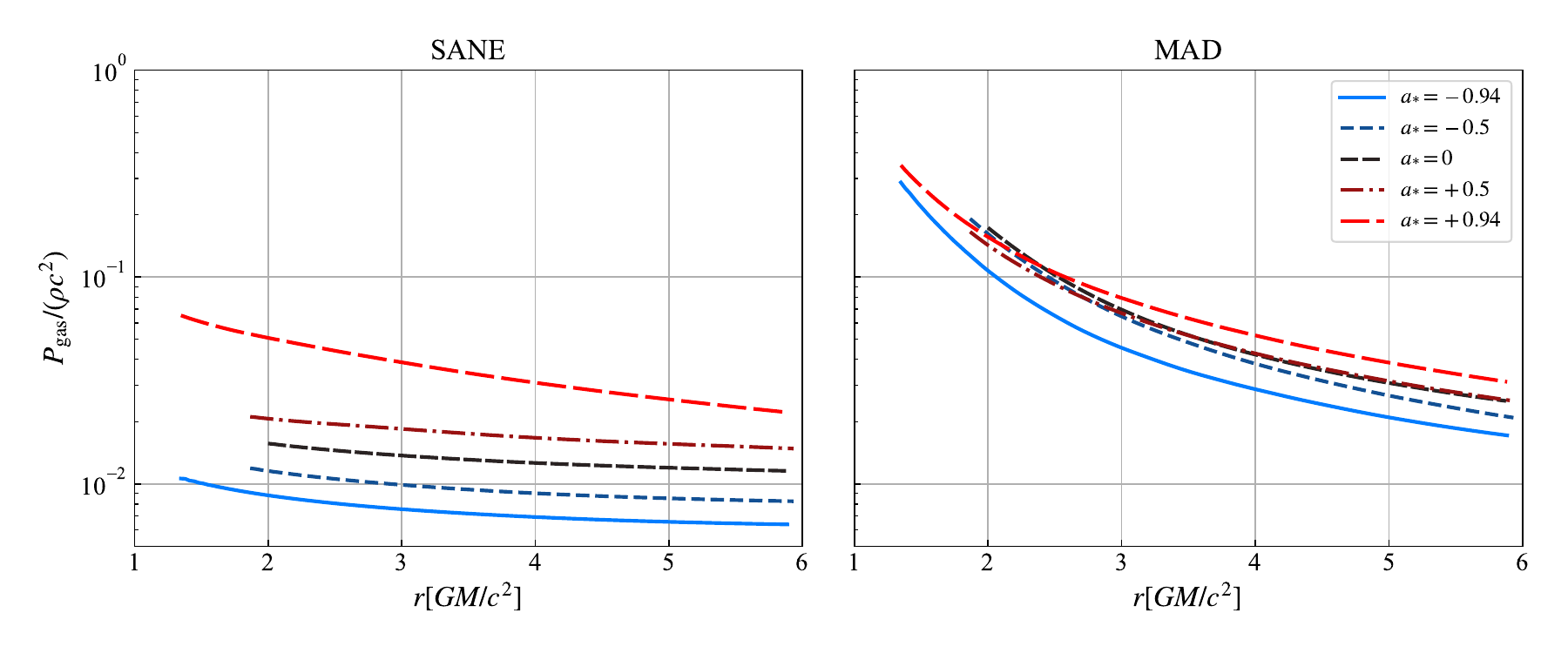}
  \caption{Time- and azimuth-averaged profiles of midplane dimensionless gas temperature $P/(\rho c^2)$ in \kharma fiducial GRMHD simulations.
    Evidently MAD models are hotter than SANE, and both MADs and SANEs grow hotter as the black hole spin $\abh$ increases.
    The hottest models are $\abh = 0.94$ MAD models.}
  \label{fig:grmhd_temp}
\end{figure*}

We model the plasma flow around \sgra using ideal, non-radiative GRMHD in the Kerr metric,
with $\abh$ a free parameter \citep[see e.g.,][]{1999ApJ...522..727K,2001MNRAS.326L..41K,2003ApJ...589..444G, 2003ApJ...589..458D, 2005ApJ...635..723A, 2007A&A...473...11D}.

We integrate the GRMHD equations in three spatial dimensions using multiple algorithms:
\kharma   \citep{2021JOSS....6.3336P},
\bhac     \citep{2017ComAC...4....1P,Olivares2019},
\hamr     \citep{Liska2019},
\koral    \citep{2013MNRAS.429.3533S}, and
\athenapp \citep{2016ApJS..225...22W};
see \citet{2019ApJS..243...26P} and \citet{Olivares_et_al} for comparisons of GRMHD codes.
All simulations assume constant adiabatic index  $\Gamma_\mathrm{ad}$.

Unless stated otherwise the initial conditions for the GRMHD simulations are constant-angular-momentum hydrodynamic equilibrium tori \citep{1976ApJ...207..962F}, with orbital angular momentum that is parallel or antiparallel to the black hole spin.
The tori are seeded with a weak, poloidal magnetic field.  The simulations use varying torus pressure maximum radius
(from $\sim 15\,\rg$ to $40\,\rg$), peak temperature, adiabatic index, and initial field configurations.  These variations permit us to test the robustness of our results (see Appendix \ref{app:numerical}).

The torus initial conditions are motivated by the notion that the near-horizon flow, where most of the emission is generated (\citetalias{M87PaperV}), relaxes to a statistically steady state that is nearly independent of the flow at larger radius.
This notion is challenged in the stellar wind-fed models of \cite{2020ApJ...896L...6R}, which are included in our study.

All simulations are run in Kerr-Schild-like coordinates, which are regular on the horizon.
Unless stated otherwise, boundary conditions are outflow at the inner boundary, which is located inside the event horizon, and outflow at the outer boundary, which is located at $r_{\rm max} \ge \no{1000}\,\rg$.
Most simulations are evolved to $t_\mathrm{final} \ge  \no{30000}\,\tg$.

Once the evolution has started, the magnetorotational instability \citep[MRI,][]{1992ApJ...400..610B}, and possibly other instabilities such as, for MAD models,  magnetic Rayleigh-Taylor instabilities \citep{2018MNRAS.478.1837M}, drive the torus to a turbulent, fluctuating state.
Defining $P_\mathrm{gas} \equiv$ gas pressure and $P_\mathrm{mag} \equiv B^2 / (8\pi) \equiv$ magnetic pressure, the standard accretion flow models can be divided by latitude into three zones:
\emph{i})~an equatorial inflow,
\emph{ii})~a mid-latitude disk wind/corona with  $\beta  \equiv P_\mathrm{gas} / P_\mathrm{mag} \sim 1$, and
\emph{iii})~a polar ``funnel'' with $\sigma \equiv B^2/(4\pi \rho c^2) \gg 1$.

The magnetic flux through the horizon, characterized by $\phi \equiv \Phi_{\mathrm{BH}} (\dot{M} r_\mathrm{g}^2 c)^{-1/2}$ ($\Phi_{\rm BH} \equiv$ magnetic flux interior to the black hole equator, $\dot{M} \equiv$ mass accretion rate) divides the outcome into two states:
the magnetically arrested disk (MAD) state \citep[e.g.,][]{1974Ap&SS..28...45B, 2003ApJ...592.1042I, 2003PASJ...55L..69N, 2011MNRAS.418L..79T},
and the Standard and Normal Evolution (SANE) state \citep[e.g.,][]{2003ApJ...589..444G, 2003ApJ...599.1238D, 2012MNRAS.426.3241N}.
MAD models have $\phi \sim \phi_{\rm crit} \sim 60$.\footnote{In the Lorentz-Heaviside units commonly used in GRMHD simulations $\phi_\mathrm{crit}$ is smaller by a factor of $(4\pi)^{1/2} \simeq 3.545$.}
In MAD models, magnetic flux accretes onto the hole until $\phi \gtrsim \phi_\mathrm{crit}$.  Accretion of additional flux leads to flux expulsion events so that the flow maintains $\phi \sim \phi_\mathrm{crit}$.  Our SANE models, in contrast, typically have $\phi \sim 1$.

We consider two GRMHD simulations with initial conditions that differ from the fiducial aligned torus: strongly magnetized non-MAD tilted torus simulations \citep{Liska2018, Chatterjee2020} and a simulation in which \sgra is fed directly by winds from stars in its vicinity \citep{2020ApJ...896L...6R}.
The wind-fed simulations result in a mode of accretion that is similar to MAD but typically has lower mean angular momentum and is less well organized.
The wind-fed models have $\abh = 0$.

The GRMHD simulation library is summarized in Table~\ref{tab:GRMHDmodels}.
Figure~\ref{fig:GRMHD} shows a few examples of GRMHD simulations for an aligned SANE, an aligned MAD, a tilted torus, and a wind-fed simulation.
These simulations vary in numerical method and in numerical resolution.
We present more information on numerical methods in Appendices~\ref{app:numerical} and \ref{app:variability}.

The gas temperature profile is a critical feature of the GRMHD simulations.
Figure \ref{fig:grmhd_temp} shows the time- and azimuth-averaged profiles of the midplane dimensionless gas temperature $P/(\rho c^2)$ in a set of aligned GRMHD simulations.
The temperature profiles exhibit trends with spin and magnetic state (MAD or SANE) that drive many of the trends seen in the radiative models: MAD models are a factor of several hotter than SANE models and both MAD and SANE become hotter as $\abh$ increases.

\subsubsection{Radiative Transfer Model}

Synthetic images are generated from the GRMHD simulations in a radiative transfer step.  The transfer step requires
\emph{i})~a model for the electron distribution function (hereafter eDF);
\emph{ii})~assignment of a density scale to the GRMHD simulation;
\emph{iii})~the inclination $i$ (angle between the torus angular momentum and the line of sight)
\emph{iv})~a numerical integration performed as a post-processing step that assumes that the plasma evolution is unaffected by radiation.

\subsubsubsection{Electron Distribution Function}
\label{sec:eDF}

Thermal models have electron energies distributed according to the Maxwell-J{\"u}ttner distribution function:
\begin{align}\label{eq:thermaleDF}
  \frac{1}{n_e}\frac{dn_e}{d\gamma} = \frac{\gamma^2 \sqrt{1-1/\gamma^2}} {\Theta_e K_2(1/\Theta_e)} \exp\left(-\frac{\gamma}{\Theta_e}\right);
\end{align}
where $K_2$ is a modified Bessel function of the second kind and $\gamma$ is the electron Lorentz factor.
Recall $\Theta_e = \kB T_e/(m_e c^2)$, which is determined by the ion-electron temperature ratio $R \equiv T_i/T_e$:
\begin{align}\label{eq:te_vs_R}
  T_e=\frac{2 m_p u}{3 \kB \rho (2+R)}.
\end{align}
Here $u$ and $\rho$ are the internal energy density and rest-mass density from the GRMHD simulation, and we have assumed that the ions are nonrelativistic with adiabatic index $5/3$ and the electrons are relativistic with  adiabatic index $4/3$.
Thermal models are motivated by the idea that wave-particle scattering drives partial relaxation of the eDF, even though Coulomb scattering is ineffective.

The temperature ratio depends on a balance between microphysical dissipation, radiative cooling, and fluid transport.
Models for collisionless dissipation vary widely in their predictions for the ratio of heat deposited in ions and electrons, but depend most strongly on the local magnetic field strength.
This motivates a prescription in which the temperature ratio depends solely on the plasma $\beta \equiv P_\mathrm{gas}/P_\mathrm{mag}$ \citep{2015ApJ...799....1C}.
We adopt the same model as \citetalias{M87PaperV} and \citetalias{M87PaperVIII}, where
\begin{equation}\label{eq:rhigh_prescription}
  R = \frac{T_i}{T_e} = \Rh \frac{b^2}{b^2+1} + \Rl \frac{1}{b^2+1},
\end{equation}
\citep{2016A&A...586A..38M}
and $b \equiv \beta/\beta_\mathrm{crit}$.
This model has three free parameters: $\beta_\mathrm{crit}$, $\Rl$, and $\Rh$.
We fix $\Rl = 1$ (consistent with the long cooling time in \sgra; see discussion in \citealt{M87PaperVIII}) and $\beta_\mathrm{crit} = 1$, but allow $\Rh$ to vary from 1 to 160.  When $\Rh \gg 1$ emission is shifted away from the midplane and toward the poles.

In \emph{non-thermal} models, the eDF has a power-law tail extending to high energy.
We explore two implementations:
\emph{i}) a power-law distribution function
\begin{align}
  \frac{1}{n_e} \frac{d n_e}{d\gamma} &=
  \frac{p-1}{\gamma_{\min}^{1-p} - \gamma_{\vphantom{i}\max}^{1-p}} \gamma^{-p},
  \label{eq:non-thermaleDF}
\end{align}
with power-law index $p$ and upper and lower limits $\gamma_{\min}$ and $\gamma_{\vphantom{i}\max}$; and
\emph{ii}) a so-called $\kappa$ distribution function, inspired by observations of the solar wind and by results of collisionless plasma simulations \citep[e.g.,][and references therein]{2015JPlPh..81e3201K}
\begin{align}
  \frac{1}{n_e} \frac{d n_e}{d\gamma} =
  \gamma \sqrt{\gamma^2-1} \left(1+\frac{\gamma-1}{\kappa w}\right)^{-(\kappa+1)},
  \label{eq:kappaeDF}
\end{align}
which has width parameter $w$ and power-law index parameter $\kappa$.

Evidently, any eDF assignment scheme is an approximation since the eDF depends in general on both local conditions and particle histories.
Notice that we also assume the eDF is isotropic and neglect electron-positron pairs.

Once the eDF is specified, the radiative transfer coefficients (emissivities, absorptivities, and rotativities) can be readily calculated; see \cite{2021ApJ...921...17M} for a recent summary.

\subsubsubsection{Model Scaling}

With the exception of the stellar wind-fed simulations, the GRMHD simulations considered in this work contain a characteristic speed, $c$, but are otherwise scale-free; they set $GM = c = 1$.
Physical scales are assigned during the radiative transfer step.
The black hole mass $\mbh$ fixes the length unit $\rg$ and time unit $\tg$.
Because the GRMHD simulations are non-selfgravitating, one is free to set a density scale, or equivalently the accretion rate $\dot{M}$ or plasma mass scale $\Munit$.

The plasma mass scale parameter $\Munit$ controls the plasma emissivity and the plasma optical depth and thus the source brightness.
We adjust $\Munit$ iteratively until the time-averaged 230\GHz flux densities of the models are within a few percent of the $2.4\,\mathrm{Jy}$ mean observed during the 2017 campaign.
Notice that, in this work, model parameters are always varied at constant time-averaged millimeter flux density.

\subsubsubsection{Radiative Transfer Calculation}

\begin{figure*}
  \centering
  \includegraphics[width=\textwidth]{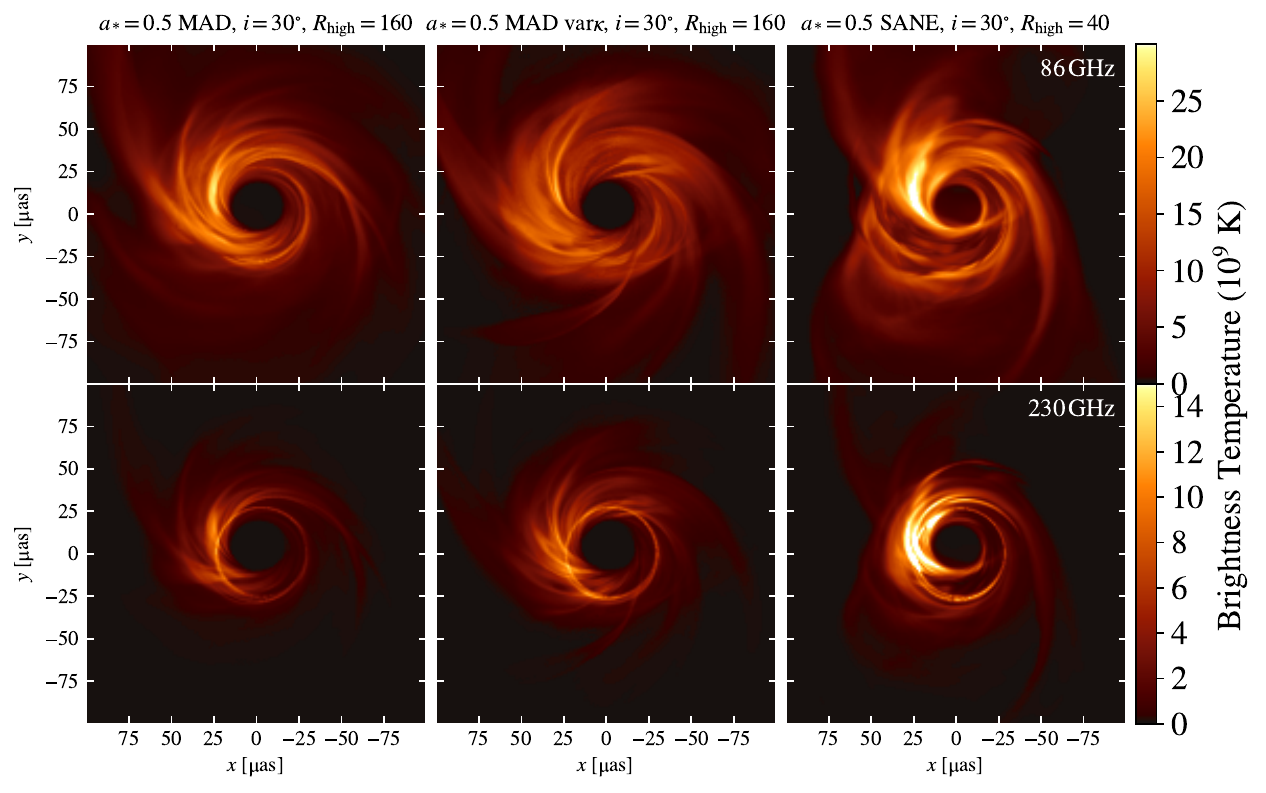}
  \caption{Example images from the model library.
    Left column: thermal MAD from the best-bet region of parameter space; middle column: non-thermal variable $\kappa$ MAD; right column: thermal SANE model.
    Top row: 86\GHz images; bottom row: 230\GHz images.
    Color represents intensity, or equivalently brightness temperature.
    Angular momentum of the accretion flow projected onto the image points up.
    These are relatively successful models satisfying most of the observational  constraints.}
  \label{fig:example_imgs}
\end{figure*}

\begin{figure*}
  \centering
  \includegraphics[width=\textwidth]{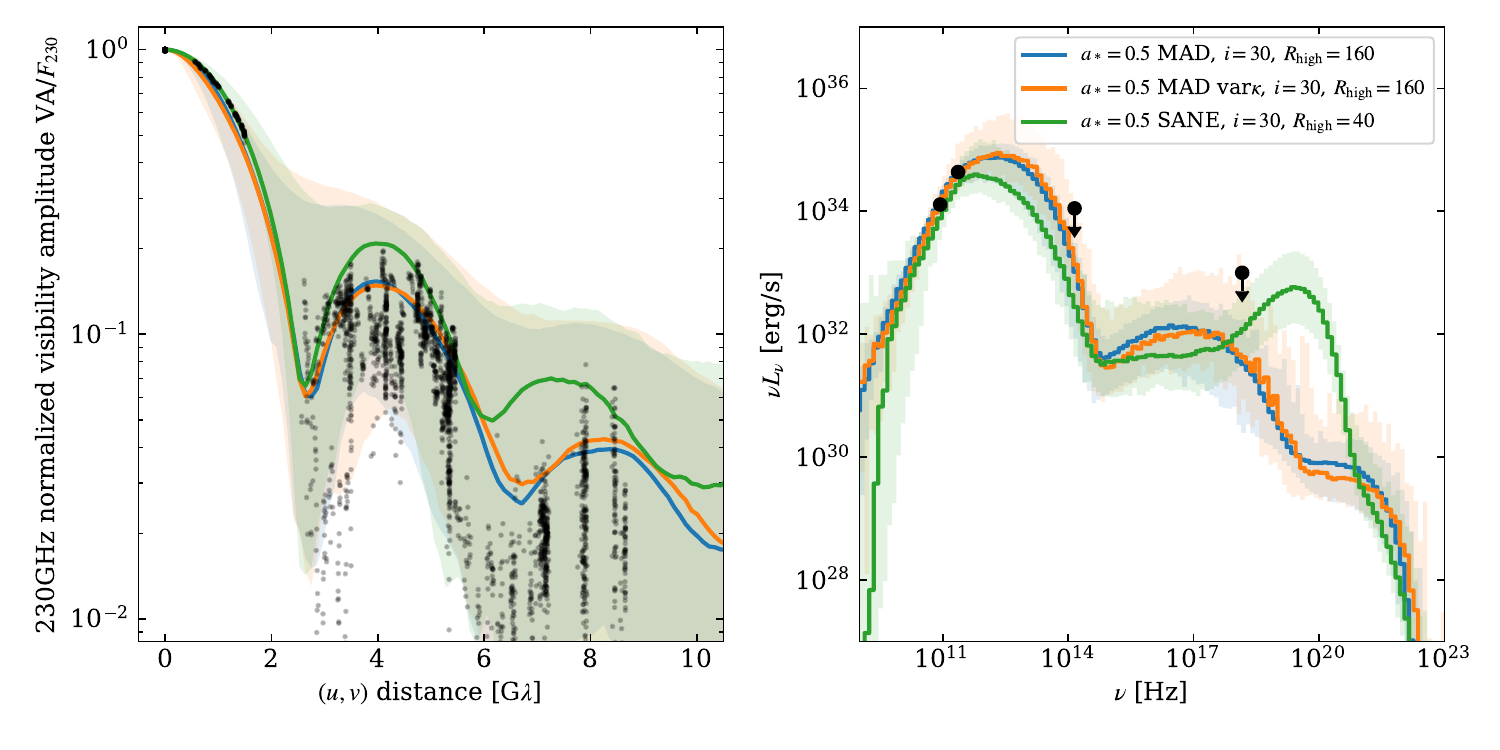}
  \caption{Visibility amplitudes (VAs; left) and SEDs (right) of the three examples models compared with the calibrated EHT~2017 data.
    Black symbols show observations.
    Blue, orange, and green are the models shown in Figure~\ref{fig:example_imgs}.
    Observed VAs are 1\,minute incoherently averaged data from the HOPS pipeline on \aprilvii.
    Model VAs for a single snapshot are shown as a solid line for a section in the $(u,v)$ plane at position angle $0\degree$. The band shows the 1st through 99th percentile over all position angles and all times.  No noise is included in the model VAs in this figure.
    Model SEDs (right) show a solid line for the mean SED and a band for the range across snapshots.}
  \label{fig:example_vas_seds}
\end{figure*}

Given an eDF, density scale $\Munit$, inclination $i$, and radiative transfer coefficients based on local properties of the plasma, the emergent radiation is obtained by integrating the radiative transfer equation.
We use two classes of numerical methods: observer-to-emitter ray tracing to generate synthetic images ({\tt ipole}, \citealt{2018MNRAS.475...43M}, {\tt BHOSS}, \citealt{2012A&A...545A..13Y}), and emitter-to-observer Monte Carlo to generate spectral energy distributions (SEDs, using {\tt grmonty}, \citealt{2009ApJS..184..387D}).

All radiative transport calculations are carried out using the fast light approximation, in which plasma variables are read from a GRMHD output file at constant Kerr-Schild time and are assumed not to change during ray tracing.
Including light travel time effects in the model introduces minor changes to light curves and images \citep{2010ApJ...717.1092D, 2021MNRAS.508.4282M}.
Further detail on numerical methods is given Appendix~\ref{app:radtrans}.
Comparisons of radiative transfer codes \citep{2020ApJ...897..148G, Prather_et_al_2022} show that differences between codes do not contribute substantially to the error budget.

Images are produced at 86\GHz, 230\GHz and 2.2\um.
Direct imaging includes synchrotron and bremsstrahlung \citep[both ion-electron and electron-electron; see][for a recent review]{2020ApJ...898...50Y}.
Unless stated otherwise the image library has a field of view (full width), resolution (pixel count), and half-width angular size of: $800\uas$, $200 \times 200$, $80 \vartheta_\mathrm{g}$ at 86\GHz; $200\uas$, $400 \times 400$, $20 \vartheta_\mathrm{g}$ at 230\GHz; and $100\uas$, $200\times 200$, $10 \vartheta_\mathrm{g}$ at $2.2\um$.

SEDs are produced for narrow bins in inclination angle.
At each inclination, the SED is averaged over azimuth.
The SED includes synchrotron, bremsstrahlung, and Compton scattering.

We find that $2.2\um$ emission is usually dominated by synchrotron, but occasionally $2.2\um$ synchrotron is so weak that Compton scattering dominates.
We also find that the X-ray can be dominated by either Compton scattering or bremsstrahlung, with the latter dominating in models with a large population of cold electrons at large radius.
Figures~\ref{fig:example_imgs} and \ref{fig:example_vas_seds} show examples of model images and multiwavelength SEDs from our library.

The GRMHD simulation-derived temperatures are unreliable in regions where $\sigma \equiv B^2/(8\pi\rho c^2)$ is large, because truncation error in integration of the total energy equation produces large fractional errors in temperature.
All radiative transfer models therefore set the emissivity, absorptivity, and inverse-Compton scattering cross-sections to $0$ for the regions with $\sigma > \sigma_\mathrm{cut} = 1$.

\subsection{Summary of \texorpdfstring{\sgra}{Sgr A*} Model Library}

A summary of radiative transfer calculations is given in Table~\ref{tab:radiativemodels}.
The entire image library contains $6$ simulation sets,  $\sim 1.8$\,million images at each of 86\GHz, 230\GHz, and $2.2\um$, and $\sim 1.3$\,million SEDs.
The images and SEDs together occupy about $50$\,terabytes.

We refer to the thermal, $\Rh$ models as ``fiducial'' models, and the remainder as ``exploratory'' models that test the effect of incorporating changes in the eDF or initial conditions.
Nearly all exploratory models (exceptions are described in the discussion) are imaged over $5 \times 10^3 G M/c^3$, in comparison to $\ge 10^4 G M/c^3$ for the fiducial models.
The sampling noise in the exploratory models is therefore larger than in the fiducial models and thus they cannot be tested as rigorously.

The library contains multiple, redundant models for the fiducial models and variable $\kappa$ models.
This provides some control over the systematic uncertainties associated with variations in GRMHD simulation setup and algorithms.

\begin{deluxetable*}{ccccccc}\label{tab:radiativemodels}
\tablecaption{EHT Model Library}
\tablehead{
  \colhead{Simulation}           &%
  \colhead{Transfer Code}        &%
  \colhead{$\Rh$}                &%
  \colhead{Inclination}          &%
  \colhead{SED}                  &%
  \colhead{$\Delta t/(10^3\tg)$} &%
  \colhead{Notes}%
  }
\startdata
\multicolumn{7}{c}{\bf Fiducial models}\\
\hline
\multicolumn{4}{l}{\it Thermal $\Rh$ models} & & &\\
\kharma& \ipole & 1, 10, 40, 160 &  10, 30, ..., 170 &  Yes & 15--30 & \\
\bhac  & \bhoss & 1, 10, 40, 160 &  10, 30, ..., 90  &  Yes & 20--30 & \\
\hamr  & \bhoss & 1, 40, 160     &  10, 50, 90       &  Yes & 20--35 & \\
\koral & \ipole & 20             &  10, 30, ..., 170 &  No  & 5--100 & \\
\hline
\multicolumn{7}{c}{\bf Exploratory models}\\
\hline
\multicolumn{4}{l}{\it Thermal $\Rh$ models} & &  &\\
\hamr Tilted   & \bhoss & 1, 40, 160 & 10, 50, 90 &  Yes & 100--103 & \\
Wind Accretion & \ipole & 13, 28     & N/A        &  No  & 10       &  \\
\hline
\multicolumn{4}{l}{\it Thermal critical $\beta$ model} & & & \\
\kharma & \ipole & N/A &  10, 50, 90 &  No & 30--35 &  \\
\hline
\multicolumn{4}{l}{\it Thermal + power-law models} & & & \\
\hamr &  \bhoss & 1, 40, 160 &  10, 50, 90 &  No & 30--35 & $p = 4$ \\
\hline
\multicolumn{4}{l}{\it Thermal + $\kappa$ models} & & & \\
\bhac & \bhoss & 1, 10, 40, 160 &  10, 30, ..., 90     &  No  & 25--30 & $\kappa = 5$ \\
\bhac & \bhoss & 1, 10, 40, 160 &  10, 30, ..., 90     &  No  & 25--30 & $\kappa = 3.5 (\epsilon_0 = 0.05)$\\
\bhac & \bhoss & 1, 10, 40, 160 &  10, 30, ..., 90     &  No  & 25--30 & $\kappa = 3.5 (\epsilon_0=0.10)$ \\
\bhac & \bhoss & 1, 10, 40, 160 &  10, 30, ..., 90     &  No  & 25--30 & $\kappa = 3.5 (\epsilon_0=0.20)$ \\
\bhac & \bhoss & 1, 10, 40, 80, 160 &  10, 30, ..., 90 &  No  & 25--30 & variable $\kappa=\kappa(\beta, \sigma)$ \\
\hamr & \ipole & 1, 10, 40, 160 &  10, 30, ..., 90     &  Yes & 30--35 & variable $\kappa=\kappa(\beta, \sigma)$ \\
\enddata
\tablecomments{%
  Summary of the EHT \sgra model library.
  All models are imaged at $86\GHz$, $230\GHz$, and $2.2\um$ and some (column 5) also have spectral energy distributions.
  For the wind-fed accretion model the viewing angle is set by the stellar orbits and $\Rh$ is set so the model matches the observed 230\GHz flux; $\Rh = 13, 28$ for models with weak and strong stellar wind magnetizations, respectively \citep{2020ApJ...896L...6R}.
}

\end{deluxetable*}

\section{Observational Constraints}\label{sec:observations}

\sgra is one of the most frequently observed objects on the sky: it has been observed with a slew of telescopes over 5 decades in time and 17 decades in electromagnetic frequency.
We must select a manageable subset of this data to constrain our models.
In doing so we have attempted to identify constraints
\emph{i})~that are believed to be uncorrelated, so that each tests a distinct aspect of the model;
\emph{ii})~that use data that can be simulated with the models; and
\emph{iii})~that are based on EHT 2017 $230\GHz$ VLBI data or that are based on emission produced within or close to the $230\GHz$ emitting region and \emph{iv})
that are observed contemporaneously or near-contemporaneously with the EHT 2017 campaign.

The selected constraints are described in detail below.
In brief, the 11 constraints can be divided into 3 classes.
The first class uses EHT data and compares estimates of source size, morphology of the visibility amplitude distribution, and three parameters of the best-fit \mring image model (5 constraints).
The second class uses non-EHT data, including 86\GHz flux density and source size, the median $2.2\um$ flux density, and the X-ray luminosity (4 constraints).
The third class considers variability, including the 230\GHz source-integrated variability and the visibility amplitude (VA) variability based on EHT data (2 constraints).

The selected constraints are heterogeneous and it is not yet possible to combine them in a consistent, fully satisfactory way.
Indeed, uncertainties in the data and the models are not well enough understood to make that possible.
In this first analysis we set a pass/fail criterion for each constraint and consider the implications of various combinations of constraints.

As the number of constraints increases so does the probability of wrongly rejecting a model.
Consider a set of $N$ constraints, and for each assigns a probability $p$ that the model is consistent with the data.
The model is rejected if $p < p_c$.
Then the probability that the model is wrongly rejected by a single constraint is $p_c$.
Applying all $N$ constraints, the probability that the model is wrongly rejected is $1 - (1 - p_c)^N$; for $N = 11$ and $p_c = 0.01$, this is $\approx N p_c \simeq 10\%$.
Each of $N$ constraints must therefore be able to reject a model with probability $\ll 1/N$ or the model scoring is meaningless.

The confidence with which a model can be evaluated is limited by sampling noise.
Many constraints (e.g. 86\GHz flux density) compare an observation to a distribution of synthetic observations from a model.
Time series of synthetic observations are not yet well characterized, but most have a correlation time $\tau\sim$ few $\times 100\,\tg$.
If the model decorrelates on timescales longer than $\tau$ then a model of duration $T$ yields $\sim T/\tau$ independent samples,\footnote{In what follows we must sometimes estimate how many independent samples are available in a time series.
Rather than estimating $\tau$  model-by-model we uniformly assume $\tau \simeq 500 \tg$.
The analysis is insensitive to this choice.} and thus a fractional error in moments of the distribution $\sim (T/\tau)^{-1/2} = 0.18 (T/\no{15000})^{-1/2}(\tau/500)^{1/2}$.
Increasing the number of constraints, then, requires increasing the duration of the GRMHD simulations.

Evidently the models have significant sampling noise, which we control for in part by using three redundant fiducial models.
Nevertheless one should not attach too much significance to the success or failure of individual models.

\subsection{EHT Observational Constraints}

We test the models against EHT interferometric data
in three ways.
First, we compare an estimate of the source size (``second moment'')
against an estimate based on short baseline visibility amplitudes
(VAs).
Second, we check the location of the first minimum and the long
baselines values of the VAs (``\vam'').
Finally, using a variant of a procedure from \citetalias{PaperIV}, we
compare fits for the diameter, width, and asymmetry of an \mring (a
parameterized image-plane model, ``\mring constraints'') to
distributions based on synthetic data generated from the model library.

\subsubsection{230\GHz VLBI Pre-Image Size}
\label{sec:sz}

The source size can be characterized using the second moments of the
source image on the sky.
The second moments in the image domain map to second derivatives of
the visibilities near zero baseline in the \uv domain, so short
baseline VAs can be used to directly estimate the source size.

This procedure is used in \citetalias{PaperII} to set an upper limit
of 95\uas full width at half maximum (FWHM) and lower limit of 38\uas FWHM for the second moment along a direction
through the source corresponding to the orientation of the short
baselines (SMT-LMT and ALMA-LMT). This is done without any assumption about the structure of the source and is therefore quite permissive.

These limits do not include scattering.  The scattering kernel is estimated to have 16.2\uas FWHM along the relevant EHT baselines. To descatter the sky image size, we subtract this value in quadrature, which produces a scattering-corrected 93.6\uas FWHM upper limit and 34.4\uas FWHM lower limit.

To score a model we evaluate the second moment tensor for each
simulated 230\GHz image and find its eigenvalues
$\lambda_\mathrm{maj}^2/(8\ln 2)$ and $\lambda_\mathrm{min}^2/(8\ln
2)$, where $\lambda_\mathrm{maj}$ and $\lambda_\mathrm{min}$ are the
major and minor axis FWHM.
The image is deemed compliant if there exists any position angle for which the second moment would satisfy the size constraints, i.e. it is compliant if for any $\lambda$ such that $\lambda_\mathrm{min} \le \lambda \le \lambda_\mathrm{maj}$, $\lambda$ lies between the scattering-corrected upper and lower limits.
We reject models with compliance fraction $< 0.01$.

\subsubsection{230\GHz VLBI Visibility Amplitude Morphology}

The second constraint provides a morphological check on the VAs.
We ask two questions of each model snapshot:
\emph{i})~is the first minimum in the visibilities---``the null''---at about the right place, and
\emph{ii})~are the long-baseline VAs comparable to the data?
The null locations and long-baseline amplitudes are sensitive to
the source structure.
For example, if the source is a simple, circularly symmetric ring of
finite width then the location of the first minimum depends only on
the ring diameter, while the VAs on long baselines depend mainly on
ring width.
GRMHD models are more complicated, with fluctuations in the null locations and long-baseline amplitudes (e.g., \citealt{2018ApJ...856..163M}, \citetalias{M87PaperV}).

We compare with data from \aprilvii, which has the best \uv
coverage near the minima in the VAs.
The first visibility minimum in both the N-S and E-W directions in the data always occurs between 2.5--3.5\,$\mathrm{G}\lambda$ \citepalias[see][for details]{PaperII}.
For the long-baseline interval between 6--8\,$\mathrm{G}\lambda$ in the data,
the VAs have $\lesssim 4\%$ of the zero-baseline flux.
One complication when comparing models to data on long baselines is
the effect of interstellar scattering.
Diffractive scattering effectively convolves the image with a smooth
kernel and can reduce the amplitudes to about $\sim 50\%$ of their
descattered values in the 6--8\,$\mathrm{G}\lambda$ range; refractive
scattering, on the other hand, introduces noise at all baselines of
order 0.5--3\%, depending on the characteristics of the
scattering screen \citep{2018arXiv180501242P, 2018ApJ...865..104J}.

To apply this constraint, we compute the VA of each model snapshot
along position angles (PAs) $0\degree$, $45\degree$, $90\degree$,
$135\degree$ (because of Hermitian symmetry we need only consider PAs in the 0--$180\degree$ range).
We find the first minimum numerically and compute the median VAs
between 6 and 8\,$\mathrm{G}\lambda$.
We classify a snapshot as compliant if
\emph{i})~for at least one position angle the first minimum falls
between 2.5 and 3.5\,$\mathrm{G}\lambda$ and
\emph{ii})~at no position angle do the median VAs exceed $4\% / 50\% = 0.08$ of the zero-baseline flux.
We reject models with compliance fraction $< 1\%$.

\subsubsection{230\GHz \Mring Fitting}

Following \citetalias{PaperIV}, we fit an \mring image plane model to snapshots from EHT data and from simulated data and then compare the distributions of fit parameters.

The \mring is a $\delta$ function in radius with diameter $d$ multiplied by a truncated (up to $m = 3$; notice that Paper IV truncates at $m = 4$) Fourier series, convolved with a Gaussian of width $w$.
The model also contains a centered Gaussian component, with amplitude and width as free parameters, to absorb large scale emission and emission interior to the ring.\footnote{In \citetalias{PaperIV} this is called an mG-ring.}
The \mring model has 10 parameters.\footnote{The 10 parameters are: ring diameter, ring width, fraction of the flux in the Gaussian component, width of the Gaussian, and six parameters describing the amplitude and phase of the three Fourier components} We use 3 of the parameters that are well constrained and physically interpretable: the \mring diameter $d$, the \mring width $w$ (FWHM of the convolving Gaussian), and the $m=1$ relative amplitude $\beta_1$ (the ``asymmetry'').  For more details about the \mring model see Section 4.3 of \citetalias{PaperIV}.

We fit the \mring independently to snapshots consisting of 2-minute intervals of EHT data (this averaging interval is consistent with that used in \citetalias{PaperIV}).
Over these short intervals, we approximate the source as static.
Uncertainties in the fitted \mring parameters are dominated by the limited baseline coverage during these snapshots rather than by calibration uncertainties or thermal noise.
Because snapshots that are close in time sample nearly identical baselines, they do not provide additional model constraints.

To compare fitted \mring parameters from EHT data to those from  synthetic data, we select a subset of ten 120-second scans  that have detections on more than 10 baselines and integration times at all stations $> 40\sec$.
The selected scans are as widely separated in time as possible so that they sample distinct baseline coverage, with an average separation of $\simeq \no{1240}\sec \simeq 60 \,\tg$, which is small compared to the VA correlation time in the models \citep{Georgiev_2022}.
Note the selected scans overlap with those found in \cite{Farah_2022}.
Only small changes in model selection were observed if any one scan was removed from the comparison.
The data were descattered before fitting, that is, the VA were divided by the scattering kernel.

The maximum likelihood \mring parameters for the ten selected EHT scans are listed in Table~\ref{tab:mringfits}.
Evidently the fit parameters are noisy.
The fits for $d$ range from 39\uas to 84\uas, for $w$ from 9\uas to 21\uas, and for $\beta_1$ from $0.04$ to $0.48$.
The variation in fit parameters could be caused by source variability, thermal noise, or gain variations.
In the models the main driver of fit variations is source variability.

\begin{deluxetable}{ccccc}  \label{tab:mringfits}
  \tablecaption{\Mring Fits to EHT Observations}
  \tablehead{ %
    \colhead{Scan \#} & %
    \colhead{$t$ [UTC hrs]} & %
    \colhead{$d$ [$\mu$as]} & %
    \colhead{$w$ [$\mu$as] } & %
    \colhead{$\beta_1$} %
  }
  \startdata
  111 & 11.28 & 83.87 & 8.87  & 0.122 \\
  121 & 11.78 & 57.09 & 13.98 & 0.220 \\
  125 & 11.92 & 55.63 & 16.46 & 0.132 \\
  130 & 12.35 & 40.68 & 19.08 & 0.039 \\
  134 & 12.62 & 57.22 & 17.22 & 0.368 \\
  142 & 12.92 & 58.80 & 17.55 & 0.208 \\
  149 & 13.28 & 52.31 & 21.16 & 0.278 \\
  155 & 13.75 & 38.94 & 18.17 & 0.482 \\
  163 & 14.05 & 56.22 & 19.86 & 0.470 \\
  171 & 14.38 & 39.48 & 17.71 & 0.408 \\
  \enddata
  \tablecomments{\mring fits to selected 120s scans from \aprilvii.
    Column 2 gives UTC in hours for the observation.
    Columns 3-5 give best-fit parameters for the \mring diameter, width, and asymmetry parameter respectively.}
\end{deluxetable}

For the models, we read in a series of model images, generate synthetic data for each image for each scan at four position angles (0\degree, 45\degree, 90\degree, 135\degree), and fit \mrings to the synthetic data.
This produces a distribution of \mring parameters for each model.

The synthetic data is generated as follows.
A model image $I(x,y)$ is Fourier transformed to complex visibilities $V(u,v)$ with an assumed position angle then sampled on baselines $i$ drawn from the comparison scan, $V_i \equiv V(u_i,v_i)$.
Normally distributed thermal noise $\delta V_{\mathrm{th},i}$ with amplitude based on telescope performance during the scan is added, and multiplicative, normally distributed noise with unit variance $N$ is added to crudely model gain corrections: $\tilde{V}_i = V_i (1 + \epsilon N) + \delta V_{\mathrm{th},i}$.
We set $\epsilon = 0.05$, but no substantial changes in fit parameters were observed for $\epsilon = 0.02$.
We then fit to the VAs $|\tilde{V}_i|$ and closure phases.\footnote{Maximum likelihood \mring parameters were found for each scan using the Julia package \texttt{Comrade.jl} \citep{comrade} in combination with a differential evolution-based optimizer
found in the Julia package \texttt{Metaheuristics.jl}.  The set of scripts used for the fits can be found in the GitHub repository \url{https://github.com/ptiede/EHTGRMHDCal}.}

We sample the model images once per $500\,\tg$, which is comparable to a correlation time.
A model with a $\no{15000}\,\tg$ imaging window thus produces $30$ fits per scan per position angle.

In comparing the models to the data we
\emph{i}) generate the distribution of fit parameters at each position angle;
\emph{ii}) use a Kolmogorov-Smirnov (KS) test to compare the distribution of $\sim 300$ synthetic data fits with the distribution of $10$ observational fits, and obtain a p-value (what is the probability they are drawn from the same underlying distribution?);
\emph{iii}) average the p-values over the four sampled position angles (i.e., marginalize over position angle; the models do not show a significant position angle preference); and
\emph{iv}) reject the model if $p < 0.01$.

\subsection{Non-EHT Constraints}

In addition to the EHT data, the SED of \sgra is well constrained in \citetalias{PaperII} and thus potentially useful for model selection.
We limit comparison to three bands: 86\GHz, 2.2\um, and X-ray.

\subsubsection{86\GHz Flux}

The Global Millimeter VLBI Array (GMVA) observed \sgra on April 3, 2017, just 3 days ($\simeq \no{13000}\,\tg$) before the EHT campaign.
\citet{2019ApJ...871...30I} estimate that the compact flux during this observation was $F_{86} = 2.0 \pm 0.2\,\mathrm{Jy}$ ($2\sigma$ errors).

To test the models we compute a library of 86\GHz images for all GRMHD snapshots for all models, and integrate over them to obtain  $F_{86}$.
We assume normally distributed measurement errors with $\sigma = 0.1\,\mathrm{Jy}$ and convolve the $F_{86}$ distribution for each model with the resulting Gaussian.
We reject models where the value of the error-broadened cumulative distribution function (CDF) at 2.0\,Jy is <1\% or >99\%.

\subsubsection{86\GHz Image Size}\label{sec:86size}

The GMVA observations from April 3, 2017 constrain the FWHM of the source major axis.
Notice that two different values for the major axis FWHM have been published in the literature: $120 \pm 34\uas$ \citep{2019ApJ...871...30I}
${\rm FWHM}_{maj} = 146^{+11}_{-12}\uas$ \citep[95\% confidence][]{2021ApJ...915...99I}.
We adopt the later analysis.

We compute the major axis FWHM for each image in the 86\GHz image library.
We assume normally distributed errors with $\sigma = 6\uas$ and convolve the model major axis distribution with the normal distribution.
We reject models with error-broadened CDF $< 1\%$ or $> 99\%$ at $146\uas$.

Our synthetic 86\GHz images have a 800\uas field of view.
A 200\uas field of view cuts off enough emission that the major axis is biased downward in many models by $\sim 20\%$.
Increasing the field of view beyond 800\uas has negligible effect.

\subsubsection{\texorpdfstring{$2.2\um$}{2um} Median Flux Density Constraint}\label{subsec:nir}

\sgra has a quiescent and a flaring component in the near-infrared (NIR), with flares occurring a few times per day
(1 day $\simeq \no{4200}\,\tg$) \citep{2018ApJ...863...15W}.
Since there is as yet no generally accepted model for NIR flares, we accept models that do not produce flares (indeed none of our models reliably produce flares, even those with non-thermal eDFs).
Our working hypothesis is that the models can be made to produce flares by introducing a process that accelerates a small fraction of electrons into a  NIR-bright tail of the eDF.  If the model overproduces quiescent 2.2\um emission, however, then we reject it.

\sgra had a median 2.2\um flux $= 0.8 \pm 0.3\,\mathrm{mJy}$ in 2017  \citep[][see Table 1]{2020A&A...638A...2G}.
The median flux density likely overestimates the median quiescent flux density since it includes flares.

We compute the model median 2.2\um flux density using one of two procedures.
If a full SED\textemdash which includes Compton scattering\textemdash is available then we use it.
The SEDs are generated by the \grmonty Monte Carlo code \citep{2009ApJS..184..387D, 2022ApJS..259...64W}.
If a full SED is not available (see Table \ref{tab:radiativemodels}) then we compute a $2.2\um$ image that includes only synchrotron emission (synchrotron absorption is negligible at $2.2\um$ for \sgra).

A rigorous model evaluation procedure would correct for the upward bias in median quiescent flux density from flares and allow for errors in the model and observed median flux density, but these refinements are sufficiently uncertain that, instead, we set a conservative threshold of $1.0$\,mJy and reject the model if its
median $2.2\um$ flux density exceeds threshold.

\subsubsection{X-ray Luminosity Constraints}

\sgra flares in the X-ray less than about once per day \citep[see][and references therein]{2018MNRAS.473..306Y}.
\emph{Chandra} observations during the 2017 campaign suggest a conservative upper limit on the median (quiescent) $\nu L_\nu$ at $6\,\mathrm{keV}$ of $10^{33}\ergps$ (\citetalias{PaperII}).

As for the model $2.2\um$ flux density, we estimate $\left.\nu L_\nu\right|_{h\nu=6\,\mathrm{keV}}$ in two ways.
The SED, which incorporates Compton scattering and bremsstrahlung, is used if it is available.
If the SED is not available then we compute an X-ray image that includes only bremsstrahlung (which dominates the X-ray emission in thermal SANE models with $\Rh = 40$, $160$) enabling us to eliminate a few additional models.

We reject the model if its median $\left.\nu L_\nu\right|_{h\nu=6\,\mathrm{keV}} > 10^{33}\ergps$.

\subsection{Variability}

\sgra shows variability on a wide range of timescales.
This is expected: fluctuations in stellar wind feeding at the scale of the S-stars plausibly introduce long timescale variations, while turbulence at smaller radii, down to the scale of the event horizon, introduces a spectrum of shorter timescale variations.
Quantitative comparison of observed variability to the models is therefore a potentially powerful tool for model selection.

We consider two variability measures: one characterizes variability in the 230\GHz light curves \citep{Wielgus2022} and a second  characterizes variability of VAs in EHT data (\citetalias{PaperIV}; \citealt{NoiseModeling}).

\subsubsection{230 \GHz light curves}

We compare variability in the models to light curve observations of \sgra from 2005--2017 using the 3-hour {\em modulation index} $\mi{3}$, where $\mi{\Delta T} \equiv \sigma_{\Delta T}/\mu_{\Delta T}$, $\sigma_{\Delta T}$ is the standard deviation measured over an interval $\Delta T$ (in hours), and $\mu_{\Delta T}$ is the mean measured over the same interval.

Following \citet{2015ApJ...812..103C} we use $\mi{\Delta T}$ because it is easy to describe, easy to compute, commonly used in the literature (in the X-ray astronomy literature it is ``rms \%''), and closely related to the structure function, since the expectation value for $\sigma_T^2$ is given by an integral over the structure function \citep[see][]{Lee_2022}.

We use $\Delta T = 3$\,hours ($\sim 530\,\tg$) because it is long enough to be comparable to the characteristic timescale measured in damped random walk fits to the ALMA light curve \citep[see Table 10 of][]{Wielgus2022} but short enough that the model light curves provide a sample that is large enough to be constraining.
In extracting a sample of $\mi{3}$ from the light curves we use as many 3-hour segments as possible, equally spaced away from the light curve endpoints and each other, and calculate $\mi{3}$ on each segment.
We treat consecutive measurements of $\mi{3}$ as independent, consistent with the minimal correlation expected for a damped random walk \citep{Lee_2022}.

We must select an observed distribution of $\mi{3}$.
The April 7 data alone provide only a weak constraint because there are only 3 samples.
The $\mi{3}$ measured from EHT 2017 observations on April 5--11 provide 7 samples, while the $\mi{3}$ measured from all available light curves longer than 3\,hours, including earlier SMA and CARMA data (the ``historical distribution''; see \citealt{Wielgus2022}) yields 42 samples.
The 2017 distribution is consistent with being drawn from the historical distribution, although April 6 has one of the quietest segments on record, and April 11 one of the most variable.
We selected the historical distribution and note that the 2017 distribution rejects slightly {\em more} models but leads to identical conclusions.

For each model we use a two-sample KS test to estimate the probability $p$ that the model and observed $\mi{3}$s are drawn from the same underlying distribution.
We reject the model if $p < 0.01$.

Through the KS test, the strength of the $\mi{3}$ constraint depends on the number of data and model samples.
The fiducial models have duration $10^4$ or $1.5\times 10^4\,\tg$ (18 or 28 samples), whereas most exploratory models have duration $5 \times 10^3\,\tg$ (9 samples).
The $\mi{3}$ constraint is therefore weaker for the exploratory models: an exploratory model that passes the constraint may be more variable than a fiducial model that fails.

\subsubsection{EHT Structural Variability}

Fluctuations in the spatial structure of the source produce fluctuations in the VAs.
Here we compare  the power spectrum of structural variability from EHT observations with predictions from GRMHD models.

A nonparametric technique to measure the variance of the spatially-detrended VAs at a location in the $(u,v)$-plane is described in \citet{NoiseModeling} and briefly summarized here.
We use EHT observations of \sgra from April 5, 6, 7, and 10 (April 11 was excluded).
To remove correlations associated with variations in the total flux, we normalize the VA data with the contemporaneous intrasite light curve \citep{Georgiev_2022}.
The light curve-normalized visibility amplitudes are then linearly detrended, and variances are computed and azimuthally averaged \citep{NoiseModeling}.
The resulting $\sigma_\text{var}^2 (|u|)$ is a measure of the fractional structural variability as a function of baseline length $|u|$.
The $\sigma_\text{var}^2 (|u|)$ is included in an inflated error budget when making images of and fitting models to the 2017 EHT observations of \sgra \citepalias{PaperIII, PaperIV}.

We measured this quantity from the GRMHD simulations \citep[see][for details]{Georgiev_2022}.
For all simulations reported here, $\sigma_\text{var}^2$ is well-approximated by a broken power law with parameters that are nearly universal among simulations.
The $\sigma_\text{var}^2$ is measured over a four-day period, which is longer than the typical model duration.
We therefore expect that model values will be biased downward compared to the data. Furthermore, each GRMHD simulation can only give one draw from a distribution that is broader than if the simulation spanned 4 days. This secondary effect negates the downward bias, which is further unimportant as we do not exclude models for being not variable enough. To measure the larger broadness of the distribution, we use multiple simulations with the same parameters and subdivide the analysis of long simulations into windows.
The uncertainties in the measurement from the GRMHD simulations due to simulation resolution, the fast-light approximation, and code differences are small compared to the uncertainty due to the variability of $\sigma_\text{var}^2$ due to short simulations \citep{Georgiev_2022}.

The measured $\sigma_\text{var}^2$ is well characterized by a power law for $2~{\rm G}\lambda < |u| < 6~{\rm G}\lambda$ \citep{Georgiev_2022}.
For comparison with the models presented here, we distill the $\sigma_{\text{var}}^2$ to two numbers: the amplitude $\afour^2$ at $4~{\rm G}\lambda$ and a power law index $b$.
Because the normalization is done in the center of the fit range the estimated $\log_{10}(\afour^2) = -3.4 \pm 0.1$ and $b=2.4\pm0.8$ are essentially uncorrelated.

Model predictions for $\afour^2$ and $b$ are computed using the power spectral densities from \citet{Georgiev_2022}\footnote{\citet{Georgiev_2022} gives the power spectral density of the complex visibility, $\langle\hat{P}(|u|)\rangle$, rather than the VA, and thus $\sigma_\text{var}^2=\langle \hat{P}\rangle/2$.}.
The anisotropic diffractive scattering kernel from \citet{Johnson_2018} is applied to $\sigma_\text{var}^2(|u|)$ and averaged over relative orientations of the major axis of the scattering kernel and the black hole spin.
These estimates are then azimuthally averaged, and the parameters $\afour^2$ and $b$ are determined from a least-squares linear fit to $\sigma_\text{var}^2(|u|)$ in $2~{\rm G}\lambda < |u| < 6~{\rm G}\lambda$.

For each model the fits for $\afour^2$ and $b$ are done separately on each window of length $5 \times 10^3\,\tg$, giving at most three measurements for most models.
This makes a direct comparison with the measured value difficult, as the model distribution is poorly constrained.

\citet{Georgiev_2022} estimates the typical width of a model distribution is $\log_{10}(\afour^2) \pm 0.1$.
We can obtain a rough estimate for how the models fare compared to the measurement by taking the mean across windows, assuming the width of the distribution is $\sigma = 0.1$, and comparing this with the observed distribution under the assumption that both are distributed normally.
We reject models with error-broadened CDF <1\% or >99\% at $\log_{10}(\afour^2) = -3.4$.

\section{Model Comparison}\label{sec:comparisons}

\begin{figure*}
  \centering\hspace{18pt}
  \includegraphics[width=6.5in]{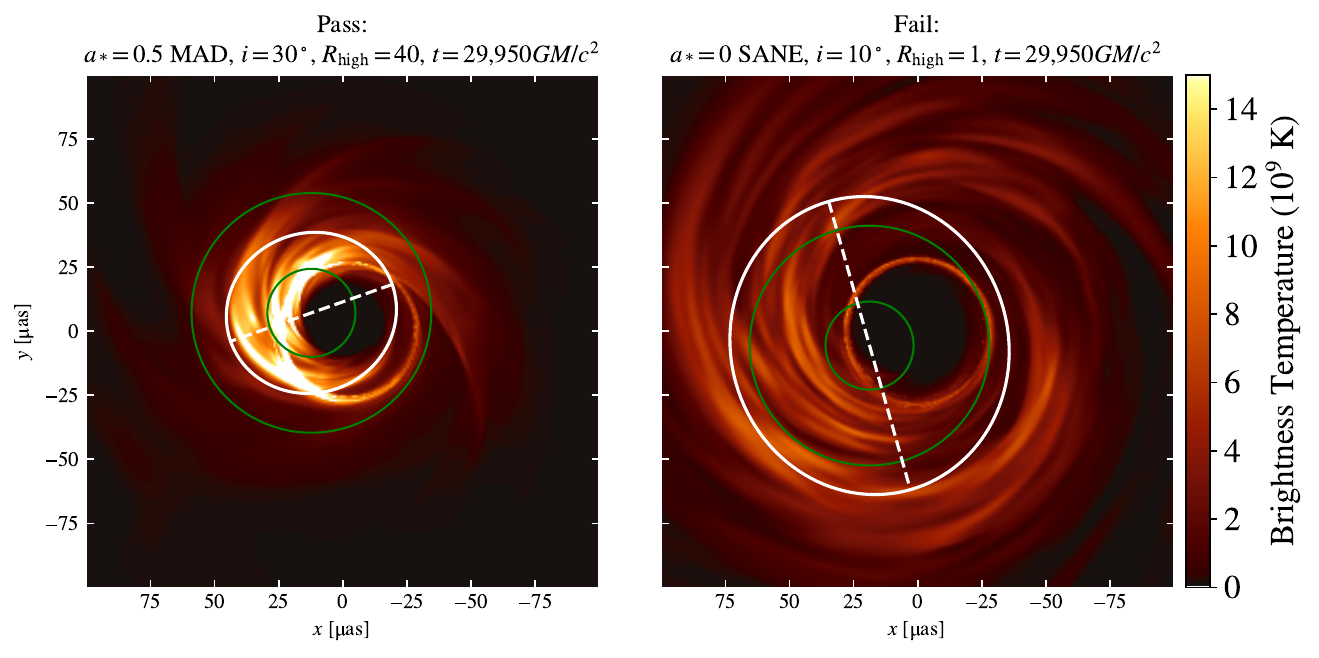}
  \caption{Pre-image size constraint example.
    Left: passing snapshot;
    right: failing snapshot.
    The model is rejected if $< 1\%$ of model snapshots pass.
    The solid white ellipse  represents the second moments of the image, and the dashed line shows the major axis.
    The two green circles show the observed lower and upper limits from \citetalias{PaperIII}.
    The snapshot is rejected if the minor axis is larger than the upper limit, or if the major axis is smaller than the lower limit.}
  \label{fig:passfail_sz}
\end{figure*}

\begin{figure*}
  \centering\hspace{18pt}
  \includegraphics[width=6.5in]{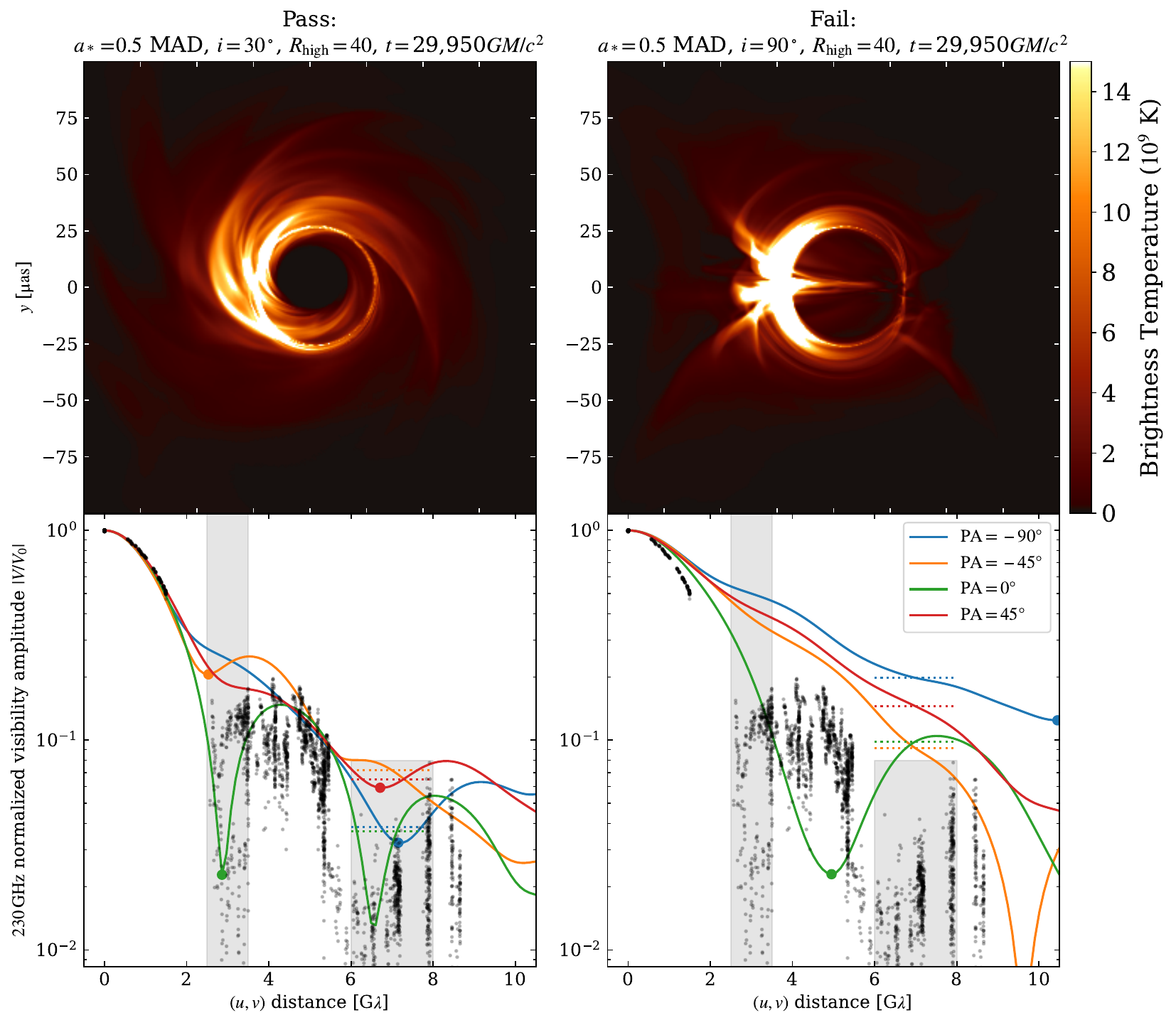}
  \caption{Visibility amplitude (VA) morphology constraint example.
    \emph{Top}: snapshot images;
    \emph{Bottom}: visibility amplitude.
    \emph{Left}: passing snapshot;
    \emph{right}: failing snapshot.
    In the bottom row,
    the solid lines in each plot show visibility amplitudes on a section through the origin in the \uv domain, at four position angles (PAs), where $0\degree$ is parallel to the projected angular momentum vector of the accretion flow.
    Solid black points show data from \aprilvii.
    The \vam constraint requires that for at least one PA the first minimum in VA fall within the left grey band, and for all PAs the median of the VAs lie inside the right grey band (see Section~\ref{sec:constraints}, \emph{Visibility Amplitude Morphology}, for details).
    Evidently the snapshot at right fails {\em both} conditions.
    The top row shows the corresponding snapshots in the image domain.
    The color bar on the right is for both images.}
  \label{fig:passfail_va}
\end{figure*}

\begin{figure*}
  \centering
  \includegraphics[width=6.5in]{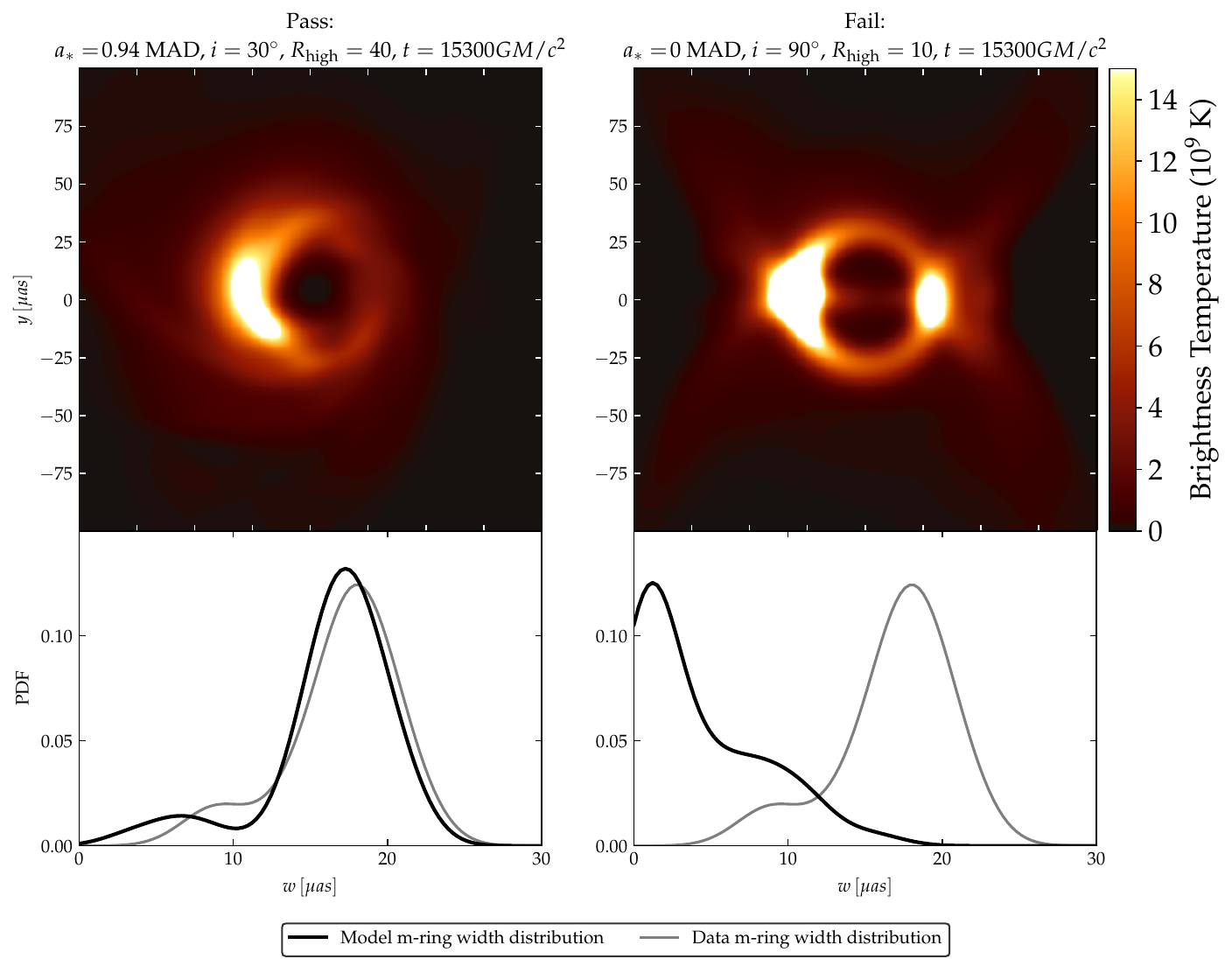}
  \caption{\Mring width constraint example.
    \emph{Top left}: snapshot from model that satisfies the constraint, Gaussian blurred to $20\uas$;
    \emph{top right}: snapshot from model that fails the constraint, Gaussian blurred to $20\uas$;
    \emph{bottom left}: observed distribution of \mring widths over our 10 selected scans (light grey) and passing model  distribution over 10 scans and 4 position angles (black);
    \emph{bottom right}: observed distribution of \mring widths over our 10 selected scans (light grey) and failing model  distribution (black) over 10 scans and 4 PAs.}
  \label{fig:mring_width_example}
\end{figure*}

\subsection{Fiducial Models}\label{subsec:thermal}

We start with the fiducial models.
Recall that these have aligned (prograde or retrograde) accretion flows, thermal eDFs, and electron temperature assigned according to the $\Rh$ model, as in \citetalias{M87PaperV}, and include the  \kharma, \bhac, and \hamr model sets listed at the top of Table~\ref{tab:radiativemodels}.

A set of plots showing how the three, redundant fiducial model sets fare for each constraint is provided in Appendix \ref{app:tables}.
Table~\ref{tab:passfraction_thermal} summarizes the fraction of fiducial \kharma, \bhac, and \hamr models that pass each constraint.

\subsubsection{EHT Constraints}\label{sec:constraints}

\subsubsubsection{Second Moment}

Without assuming a ring, the EHT data allow a wide range of second moments.
The second moment constraint passes $98\%$ of all models.  Here and in what follows, the quoted passing fraction for the model describes the fraction of points in parameter space for which the existing model sets (\kharma, \bhac, and when present \hamr) agree that the model passes the constraint.
In short, nearly all fiducial models are about the right size once we use the 230\GHz to fix the mass unit $\mathcal{M}$.
The few rejected models are $\abh \le 0$, face-on, SANE models with $\Rh = 1$.
These models have extended emission on scales large compared to the critical impact parameter $b_c = \sqrt{27}\,\rg$.
The right panel of Figure~\ref{fig:passfail_sz} shows an example of one of these failed models.  The left panel shows an example of a passing model.

\subsubsubsection{Visibility Amplitude Morphology}
\label{sec:vam}

The \vam constraint tests the null location and long-baseline VAs.  Figure~\ref{fig:passfail_va} shows an example of a passing and failing model.
The constraint disfavors edge-on models at positive spin and a few large $\Rh$ SANE models.
This is mainly because the edge-on models contain bright spots, corresponding to the approaching side of the rotating accretion flow, and faint rings, so the first nulls get washed out by the bright features.
The \vam constraint passes 79\% of all models.

\subsubsubsection{\Mring Fits}

\begin{figure*}
 \centering
 \includegraphics[width=\textwidth]{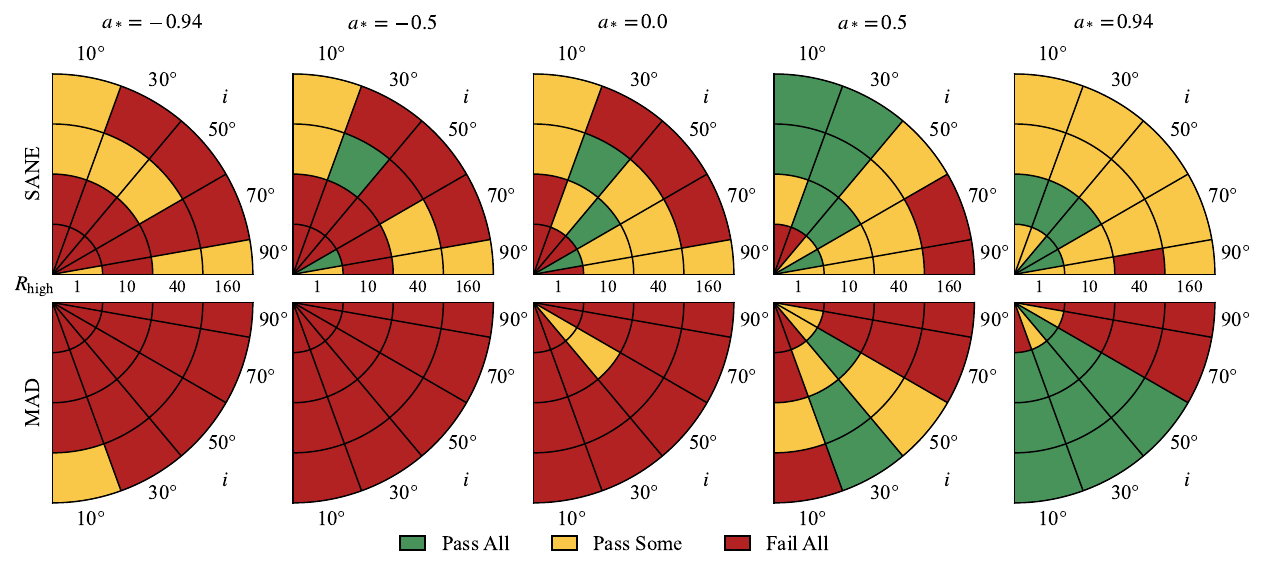}
  \caption{Pass/fail plot for the \mring widths.
    Green indicates that the \kharma, \bhac, and \hamr models pass, yellow that one or two of the fiducial models fail, and red that all three fail.
    The inclination coverage is not uniform: \bhac and \kharma models cover all 5 inclinations while \hamr models cover $i = 10\degree$, $50\degree$, $90\degree$ only.
    The $i = 30\degree$, $70\degree$ wedges therefore include only the \bhac and \kharma models.}
  \label{fig:mring_width_salsa}
\end{figure*}

\begin{figure*}
  \centering
  \includegraphics[width=\textwidth]{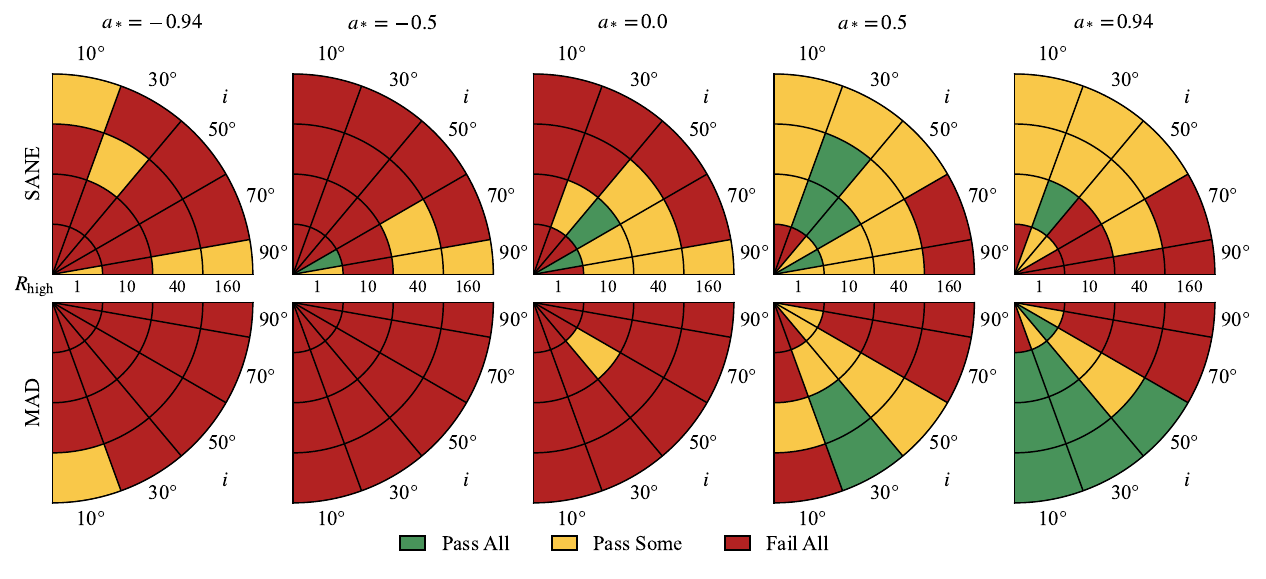}
  \caption{Combined EHT constraints (logical {\em and}) including the second moment, \vam, and \mring fit constraints.
    Green indicates that the \kharma, \bhac, and \hamr fiducial models pass, yellow that one or two of the fiducial models fail, and red that all three fail.
    The inclination coverage is not uniform: \bhac and \kharma models cover all 5 inclinations while \hamr models cover $i = 10\degree$, $50\degree$, $90\degree$ only.}
  \label{fig:all_EHT_constraints}
\end{figure*}

The \mring asymmetry, diameter, and width are treated as separate constraints.
Recall that we compare the distribution from the data to that from the model using a two-sample KS test.

The asymmetry parameter is typically not well constrained.
Many rejected models are at high inclination and have $\abh =0.94$.
These models have asymmetries that are large and detectable because Doppler boosting concentrates emission in an equatorial spot on the approaching side of the disk.
The asymmetry parameter constraint passes 91\% of all models.

The \mring diameter, which depends on the diameter of the shadow and the ring width, is better constrained than the asymmetry parameter and varies systematically from model to model.
The ring diameter constraint passes 54\% of all models.

Most of the models that fail are low inclination models with ring diameters that are too large. Only two \bhac models fail because the ring diameter is too small.  Most of the rejected models are low inclination models at $\abh < 0$.

The \mring width $w$ is the most tightly constrained of the three \mring parameters.
Although the closure phases constrain $w$ as well, it is easiest to see how $w$ affects visibility amplitudes at long baselines.
For example, for a circularly symmetric ring the VAs are a Bessel function multiplied by a Gaussian with width $\sim 1/w$.  Increasing $w$ therefore decreases the amplitude of the long baselines.
Figure~\ref{fig:mring_width_example} shows examples of models that pass and fail the \mring width constraint.

Figure~\ref{fig:mring_width_salsa} summarizes the pass/fail status of the fiducial models for the \mring width.
All rejected models have median $w$ that is below the median of the data, $ \simeq 17.5\uas$.
The rejected models include all MAD models at $\abh \le 0$ and all edge-on ($i = 90\degree$) models in the \kharma, \bhac, and \hamr fiducial models.
MAD models exhibit a strong trend toward smaller $w$ as $i$ increases.
SANE models exhibit a similar but weaker trend.
The SANE model images have  higher optical depth, broader rings, and more substructure than the MAD models.
Their $w$ distributions are typically broad, with mode well below $17.5\uas$.
Only for $\abh = 0.94$, where the optical depth is lower due to higher temperatures in the emitting region, do most of the models exhibit a sharply peaked $w$ distribution centered at $17.5\uas$.

\subsubsubsection{EHT Constraint Summary}

We can combine all EHT constraint cuts with a logical {\em and} operation.
The results are summarized in Figure~\ref{fig:all_EHT_constraints}.
Evidently EHT data alone are capable of discriminating between models.
The edge-on ($i = 90\degree$) models all fail, with some failing \mring width, diameter, asymmetry and the \vam  constraint.
The cuts clearly favor $\abh > 0$ models, with a few exceptions.
There are two clusters of models that do not fail any constraints in any models: positive spin MAD models at low inclination, and positive spin SANE models, also at low inclination.

\subsubsection{Non-EHT Constraints}

\begin{figure*}[h!]
  \centering
  \includegraphics[height=3.25in]{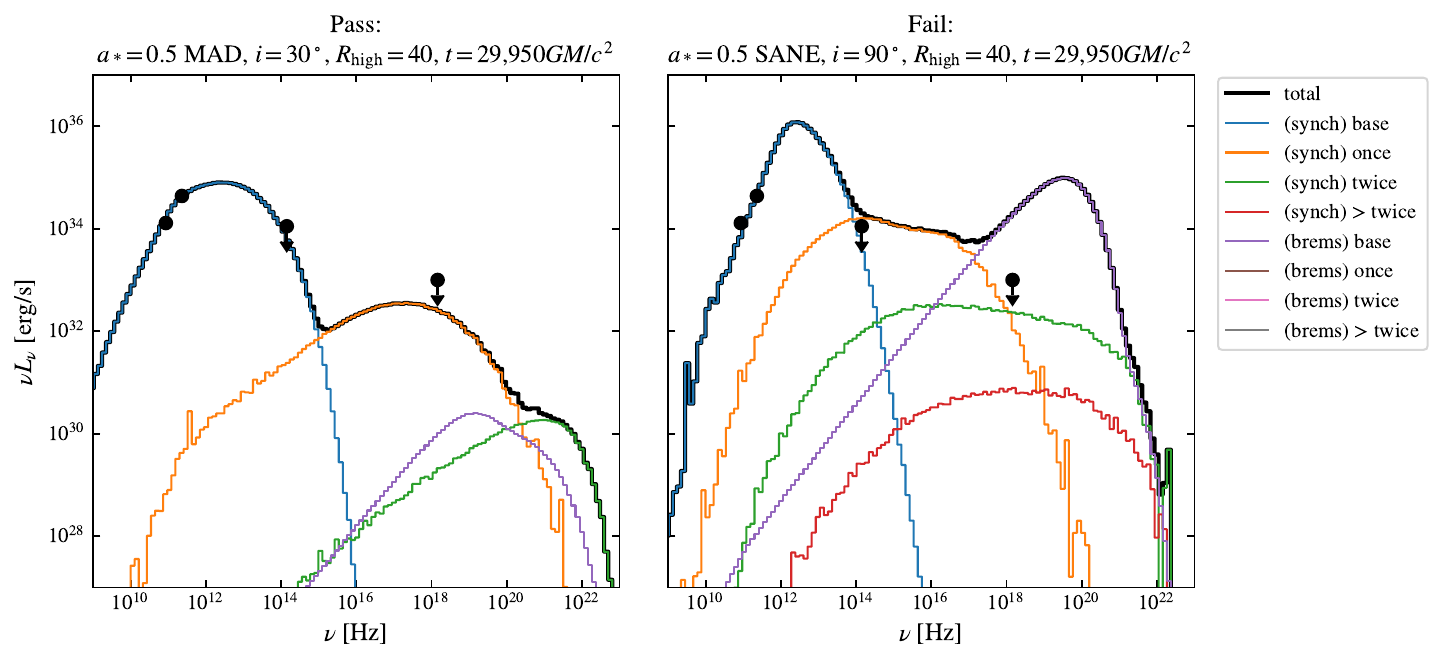}
  \caption{%
    Non-EHT flux density constraint example.
    \emph{Left}: passing model with SED close to the measured 86\GHz point and below the quiescent $2.2\um$ and X-ray points.
    \emph{Right}: failing model with inconsistent (strongly rising) millimeter wavelength spectral index, overproduction of $2.2\um$ due to strong Comptonization, and overproduction of X-rays by bremsstrahlung.}
  \label{fig:passfail_sed}
\end{figure*}

\begin{figure*}
  \centering
  \includegraphics[width=\textwidth]{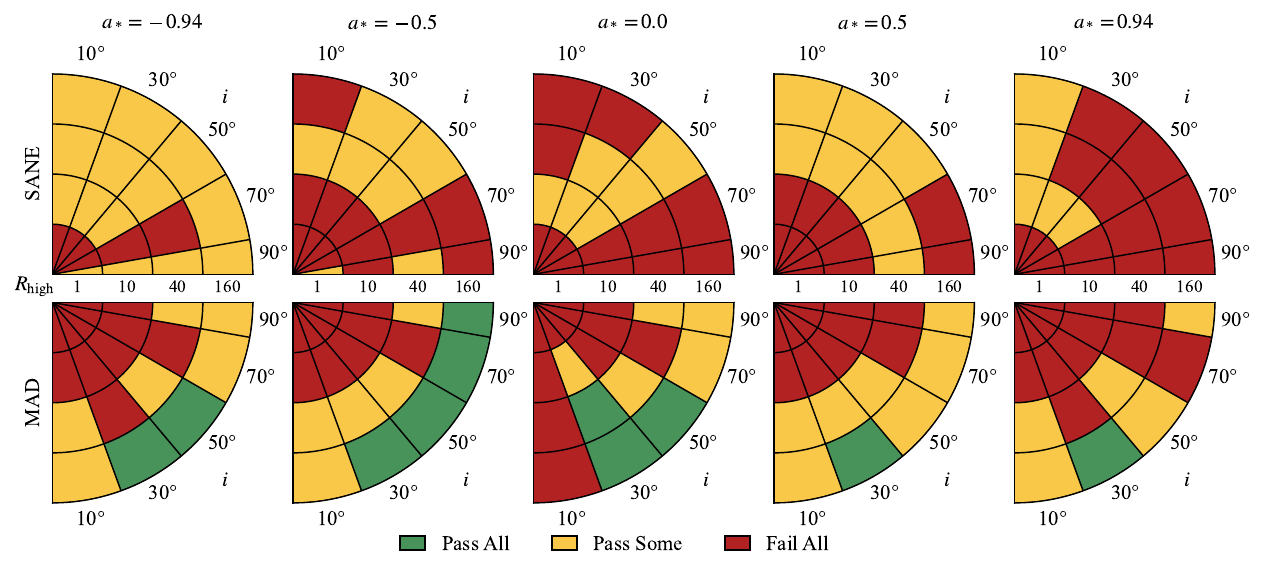}
  \caption{Combined non-EHT constraints (logical {\em and}).
    Green indicates that the \kharma, \bhac, and \hamr models pass, yellow that one or two of the fiducial models fail, and red that all three fail.}
  \label{fig:non_eht_cuts}
\end{figure*}

\subsubsubsection{86\,GHz Flux Density}

In a simplified picture \sgra's millimeter flux is produced in a photosphere that decreases in size as frequency increases.
Because optical depth is not large at 230\GHz ($\sim 0.4$ in the one-zone model) and the source structure is complicated (the optical depth varies across the image) the simplified picture is imprecise.
Nevertheless 86\GHz emission is on average produced at larger radius than 230\GHz emission, and the 86\GHz source size is larger than the 230\GHz source size.  The ratio of 86\GHz to 230\GHz flux density is therefore sensitive to the radial structure of the source plasma.

Figure~\ref{fig:86GHz_flux_pizza} records the results of applying this constraint.
Most $\Rh = 1$ models, both MAD and SANE, fail the 86\GHz flux density test.
The 86\GHz flux density is quite sensitive to $\Rh$.
For example, SANE $\abh = 0.5$, $i = 70\degree$, $90\degree$ models are too bright at $\Rh = 1$ and too dim at $\Rh = 10$.  This suggests that there are passing models in between, and that the parameter space is not sampled densely enough.
Finally, the 86\GHz flux constraint strongly favors MAD models over SANE models in all three fiducial model sets.

\subsubsubsection{86\,GHz Major Axis}

As for the 86\GHz flux, the 86\GHz size is sensitive to optical depth as a function of radius in the source plasma.
Figure~\ref{fig:86GHz_size_pizza} in Appendix~\ref{app:tables} shows the full results of applying this constraint.

The 86\GHz size is sensitive to inclination.
For example, the SANE, $\abh = 0$, $\Rh = 40$ models are too small at low inclination and too large when seen edge-on, because the edge-on models have prominent limb-brightened jet walls that are visible to $100\uas$.
The 86\GHz size constraint passes only $58\%$ of models and is therefore one of the tightest constraints.

The physical picture for 86\GHz source size is complicated, as is the extraction of the constraint itself from observations.
Notice that
\emph{i})~two different values for the 86\GHz intrinsic source size have been reported in the literature (see Section~\ref{sec:86size});
\emph{ii})~scattering is $7$ times stronger at 86\GHz than at 230\GHz;
\emph{iii})~scattering must be subtracted accurately to obtain the intrinsic source size; and
\emph{iv})~the error bars for the 86\GHz source size are narrow and this plays a key role in determining the strength of the constraint.

\subsubsubsection{\texorpdfstring{$2.2\um$}{2.2um} Median Flux Density}

$2.2\um$ photons are produced by the synchrotron process from electrons on the high energy end of the eDF.
For the one-zone model with $B = 30$\,G and $\Theta_e = 10$, the mean Lorentz factor is $\gamma = 30$ and the synchrotron critical frequency $\nu_\mathrm{crit} = \gamma^2 e B/(2 \pi m_e c) \simeq 80\GHz$.
Emission at $2.2\um$ is produced by electrons with Lorentz factor $\gamma \simeq 10^3$, so $2.2\um$ flux density is sensitive to $\Theta_e$ and $B$.
Both increase toward the horizon, and field strength is nearly independent of latitude, so $2.2\um$ photons are produced at small radius in regions where $\Theta_e$ is highest.

The sensitivity to $\Theta_e$ implies that $2.2\um$ flux density will be highest for models with higher temperatures.
For SANEs the midplane gas temperature, and therefore electron temperature in the $\Rh$ prescription, increases with $\abh$, so the highest $2.2\um$ flux density is at positive $\abh$.

The sensitivity to $B$ implies that $2.2\um$ flux density will be highest for parameters with stronger fields.  $B$ depends on the GRMHD flow configuration and also on the accretion rate, which is fixed by the observed $F_{230}$, so when all else is equal the $2.2\um$ flux density is highest when the accretion rate is largest.  The dependence of accretion rate on model parameters is discussed in Section~\ref{sec:accrate_outflowpower}.  In brief, for SANE models the accretion rate declines as $\abh$ increases and $\Rh$ decreases. For MAD models the accretion rate dependence on $\abh$ and $\Rh$ is relatively weak.

Finally, the $2.2\um$ flux density is also sensitive to inclination.  A combination of Doppler boosting and the rapid falloff in emissivity in the NIR means that at large inclination lower frequency emission from the approaching side of the accretion flow is boosted into the NIR and thus $2.2\um$ flux is higher at high inclination.

Figure~\ref{fig:passfail_sed} shows sample SEDs from our model library, where the left panel is a model that passes the $2.2\um$ flux limit and the right panel is a model that fails.
Models that pass the $2.2\um$ flux limit are shown in Appendix~\ref{app:tables} in Figure~\ref{fig:2um_flux_pizza}.
The rejected SANE models ($7\%$ rejected by all of \kharma, \bhac, and \hamr) tend to be at high inclination: their images are dominated by a bright spot on the approaching side of the disk.
The rejected MAD models ($53\%$) include nearly all models at $\Rh = 1$ and $\Rh = 10$, where $\Theta_e$ tends to be larger, and the majority of high-inclination models, where the effect of Doppler boosting is largest.

We find that some models are Compton dominated at $2.2\um$.
For example, $\abh = -0.94$ SANE models become optically thin at relatively low frequency as $\Rh$ goes to $1$, and thus synchrotron emission drops off rapidly as frequency increases.  When the synchrotron is weak enough the underlying bump of Comptonized millimeter photons dominates.

\subsubsubsection{X-ray Luminosity}

X-ray production in fiducial models is typically dominated by Compton upscattering of thermal synchrotron photons.
In the first Compton bump $\nu L_\nu$ is thus proportional to the y-parameter $y \sim 16 \Theta_e^2 \tau_e$ where $\tau_e$ is a characteristic electron-scattering optical depth and $\Theta_e$ is a typical dimensionless electron temperature.
At $\Rh = 1$ the X-ray band lies in the first Compton bump, while at larger $\Rh$ the bumps move to lower energy because the bulk of the Thomson depth is in the midplane where $\Theta_e \propto 1/\Rh$.

We find that in a few large $\Rh$ SANE models, however, X-ray emission is dominated by bremsstrahlung (synchrotron never dominates the X-ray in thermal models).  Bremsstrahlung emissivity $j_{\nu,b} \propto n^2$, so at fixed temperature bremsstrahlung increases rapidly with density. Notice that $j_{\nu,b} \propto \Theta_e^{1/2}$ for $\Theta_e > 1$ and $\Theta_e^{-1/2}$ for $\Theta_e < 1$, so cool disks enhance bremsstrahlung.  Bremsstrahlung therefore dominates Compton in models with high density and low temperature, i.e. some models with large $\Rh$ (see Section~\ref{sec:discussions}).

In models with bremsstrahlung-dominated X-ray emission the median radius of emission is $\simeq 20 \rg$.  Although the models are equilibrated at this radius the X-ray luminosity may be partially contaminated by emission from unequilibrated plasma at larger radii.  Because the fiducial models start with a torus of finite radial extent, however, they are also missing bremsstrahlung emission from outside the initial torus.  A full assessment of the associated uncertainty requires large, long runs.  Notice that because bremsstrahlung arises at large radii it varies more slowly than the synchrotron and Compton-upscattered X-ray emission and is therefore potentially distinguishable \citep{2013ApJ...774...42N}.

The left panel of ﻿Figure﻿~\ref{fig:passfail_sed} shows a model that passes the X-ray ﻿flux limit, while the right panel shows a model that fails.
The X-ray cuts are shown in Appendix~\ref{app:tables},  Figure~\ref{fig:xray_pizza}.  Some large $\Rh$ SANE models fail due to excess bremsstrahlung, although there is notable disagreement between \bhac and \kharma for SANE X-ray fluxes.
MAD models that fail have low $\Rh$ and are Compton-dominated in the X-ray.
Nearly all $\Rh = 1$ MAD models fail the X-ray constraint, as do many at $\Rh = 10$.  This is because the midplane $\Theta_e$ increases as $\Rh$ goes to $1$.
Since the midplane contributes most of the electron scattering optical depth, low $\Rh$ models have the largest $y$ parameter and are at greatest risk of overproducing X-rays.

\subsubsubsection{Summary of Non-EHT constraints}

Applying only non-EHT constraints leaves $6\%$ of models as shown in Figure~\ref{fig:non_eht_cuts}.
The surviving models are the result of applying a heterogeneous and noisy set of constraints using a hard cutoff, which somewhat obscures the underlying physical picture.
Nevertheless, the surviving 13 models are all MAD and all have $\Rh > 10$.  All but two have $i < 70\degree$.
This leaves a cluster of surviving MAD models at large $\Rh$ and low to moderate inclination.

\subsubsection{Variability}

\begin{figure}
  \centering
  \includegraphics[width=\columnwidth]{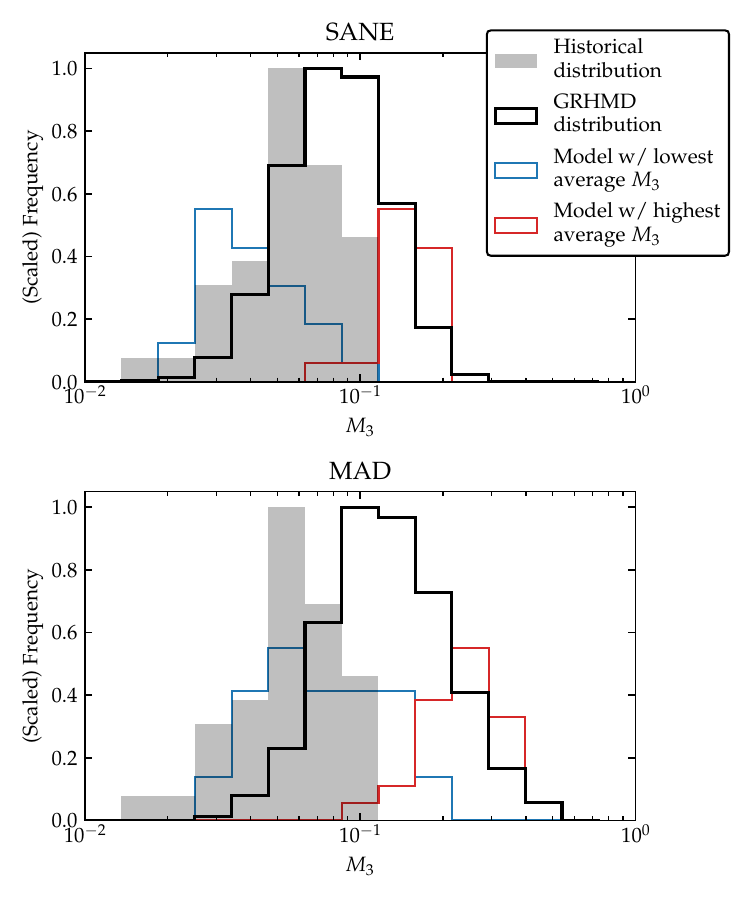}
  \caption{Distributions of 3-hour modulate index $\mi{3}$ for \bhac, \kharma, and \hamr models (black), compared to distributions from historical observations (gray).
The distributions for models with the lowest (blue) and highest (red) average $\mi{3}$ for SANEs and MADs are also shown.
The heights of these distributions have been scaled for clarity.}
  \label{fig:cmp_ALMA_var}
\end{figure}

Variability is central to the interpretation of EHT observations of \sgra: an $8\,\mathrm{hr}$ observation of \sgra lasts $\no{1400}\,\tg$, a timescale over which most models vary substantially.
In contrast, an $8\,\mathrm{hr}$ observation of M87* is $\sim \tg$ and on this timescale M87* hardly varies at all.

Recall that we consider two variability constraints, one on the 230\GHz light curve and the other on 230\GHz VAs.  We find that SANE models are less variable than MAD models. Only 3.5\% of models, all SANE, pass both variability constraints.  A possible interpretation of this result is that the models are missing a physical ingredient that would reduce variability, and this is discussed in Section~\ref{sec:discussions}.

\subsubsubsection{Modulation Index}

The distribution of 3-hour modulation index ($\mi{3}$) across all fiducial SANE models, across all fiducial MAD models, and across the historical dataset are shown in Figure~\ref{fig:cmp_ALMA_var}.
The plot also shows distributions for individual models with the lowest and highest median $\mi{3}$.

The $\mi{3}$ cuts are summarized in Appendix \ref{app:tables}, Figure~\ref{fig:m3_pizza}.
We find that:
\emph{i})~as a group, the fiducial models are more variable than the data;
\emph{ii})~the MAD models are more variable than SANE models;
\emph{iii})~eleven individual models  pass the constraint for all fiducial model sets, and these are exclusively SANE models;
\emph{iv})~there are some differences between variability in the fiducial model sets, with \hamr models notably more variable than \kharma and \bhac models; and
\emph{v})~the pass fractions for the fiducial model sets are 20\% for \kharma, 27\% for \bhac, and 7\% for \hamr.
The modulation index is the tightest single constraint on the models.

\subsubsubsection{4 \texorpdfstring{$G\lambda$}{Gl} Visibility Amplitude Variability}

\begin{figure}
  \centering
  \includegraphics[width=\columnwidth]{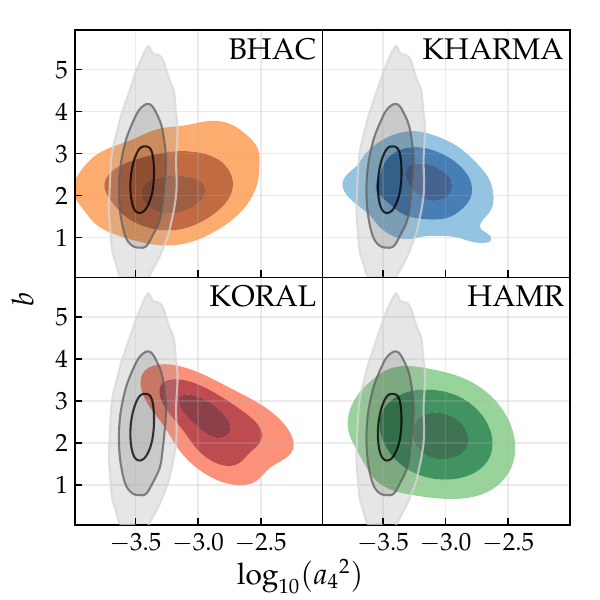}
  \caption{Distributions of $(\log_{10}(\afour^2), b)$ for \bhac, \kharma, \koral, and \hamr, compared against the 2017 EHT confidence regions (gray).}
  \label{fig:cmp_VLBI_var}
\end{figure}

The power-law index $b$ of the variance $\sigma_\text{var}^2 (|u|)$ at 2--6\,G$\lambda$ of the models is generally in good agreement with the value measured  from the 2017 EHT campaign (excluding April 11).
The amplitude $\afour^2$, however, varies depending on the model.

Figure~\ref{fig:cmp_VLBI_var} shows the distribution of $(\afour^2, b)$ from the EHT observation, along with the distributions across all fiducial models.
For a single model, the number of measurements of $(\afour^2, b)$ is equal to the number of windows for that model (three in most cases).  The \koral models appear more variable because they include only $\Rh = 20$ MAD models at various spins.

The models tend to be more variable than the observations, with face-on models performing better than edge-on models.
For SANE models, $\Rh = 10$ tends to be more variable than others.
For MAD models, there is a slight preference for lower $\Rh$.

\subsubsubsection{Long duration \koral models}

We have imaged a set of MAD models run with the \koral code out to $\sim \no{100000}\,\tg$.  These long duration  models have $\Rh = 20$, which lies off our fiducial model parameter grid.  They enable us to assess the importance of integration time for application of the constraints, and provide a more accurate distribution for, e.g. \mi{3}.

The \koral models are discussed in Appendix~\ref{app:variability}. In brief we find no evidence for significantly different variability when comparing the first and second half of the \koral runs, consistent with no long-term evolution of the variability.  We also find no significant differences when comparing the \koral runs to nearby models on the fiducial model parameter grid.  Notice that in Figure \ref{fig:cmp_VLBI_var} the \koral models are more variable than the other model sets only because the other model sets contain lower-variability SANE models.

\subsubsection{Summary of Constraints on Fiducial Models}
\label{sec:summarythermal}

\begin{deluxetable}{l|ccc}\label{tab:passfraction_thermal}
\tablecaption{Fiducial Model Pass Fractions}
\tablehead{
\colhead{constraint} &
\colhead{\kharma} &
\colhead{\bhac} &
\colhead{\hamr}
}
\startdata
230\GHz size            & 0.98 & 0.98 & 1.0  \\
VA morphology           & 0.84 & 0.83 & 0.80 \\
M-ring diameter         & 0.67 & 0.65 & 0.58 \\
M-ring width            & 0.35 & 0.21 & 0.29 \\
M-ring asym.            & 0.94 & 0.95 & 1.0 \\
\hline
86\GHz flux             & 0.74 & 0.68 & 0.62 \\
86\GHz size             & 0.65 & 0.59 & 0.46 \\
2.2\um flux             & 0.59 & 0.55 & 0.80 \\
X-ray flux              & 0.46 & 0.70 & 0.61 \\
\hline
lc varability           & 0.20 & 0.27 & 0.07 \\
4\,G$\lambda$ variability & 0.60 & 0.72 & 0.39 \\
\hline
EHT Constraints         & 0.25 & 0.19 & 0.22 \\
non-EHT Constraints     & 0.19 & 0.19 & 0.22 \\   
Variability Constraints & 0.16 & 0.27 & 0.03  
\enddata
\tablecomments{
Passing fractions for the fiducial \kharma, \bhac, and \hamr $\Rh$ thermal models, showing the consistency and relative power of the constraints. 
}
\end{deluxetable}

\begin{deluxetable*}{l|ccccc}\label{tab:fail_one_thermal}
\tablecaption{Fiducial models that fail only one constraint}
\tablehead{
\colhead{Code/Setup} &
\colhead{MAD/SANE} &
\colhead{Spin $\abh$} &
\colhead{Inclination $i$} &
\colhead{$\Rh$} &
\colhead{Failed Constraint}
}
\startdata
\kharma Thermal					&	SANE	&	0.94	&	10		&	40		&	86\,GHz size		\\
\kharma Thermal					&	SANE	&	0.94	&	30		&	40		&	86\,GHz size		\\
\kharma Thermal					&	SANE	&	0.94	&	50		&	1		&	86\,GHz size		\\
\kharma Thermal					&	MAD		&	0.5		&	30		&	40		&	\mi{3}				\\
\kharma Thermal					&	MAD		&	0.5		&	30		&	160		&	\mi{3}				\\
\kharma Thermal					&	MAD		&	0.94	&	10		&	160		&	\mi{3}				\\
\kharma Thermal					&	MAD		&	0.94	&	30		&	160		&	\mi{3}				\\
\hline
\bhac Thermal					&	SANE	&	-0.5	&	30		&	40		&	\Mring diameter   	\\
\bhac Thermal					&	SANE	&	0		&	30		&	40		&	\Mring diameter   	\\
\bhac Thermal					&	SANE	&	0.5		&	10		&	40		&   \mi{3}  			\\
\bhac Thermal					&	SANE	&	0.5		&	10		&	160		&   \mi{3}  			\\
\bhac Thermal					&	SANE	&	0.5		&	30		&	40		&   \mi{3}  			\\
\bhac Thermal					&	SANE	&	0.5		&	30		&	160		&   \mi{3}  			\\
\bhac Thermal					&	MAD		&	0.5		&	30		&	160		&   \mi{3}  			\\
\bhac Thermal					&	MAD		&	0.5		&	50		&	160		&   \mi{3}  			\\
\bhac Thermal					&	MAD		&	0.94	&	10		&	160		&   \mi{3}  			\\
\bhac Thermal					&	MAD		&	0.94	&	30		&	160		&   \mi{3}  			\\
\enddata
\tablecomments{Models which pass all but one  constraint. Since no model passes all constraints, these represent the parameters that are closest to being consistent with observations.}

\end{deluxetable*}

None of the fiducial models  survive the full gauntlet of constraints. The pass fractions for individual constraints for the \bhac, \kharma, and \hamr fiducial models are listed in Table~\ref{tab:passfraction_thermal}.
\mi{3} is the most severe constraint, followed by the \mring width constraint.
Together the variability constraints pass only $4\%$ of fiducial models and prefers SANEs, which are less variable than MADs, while the remaining constraints prefer MAD models.

It is likely that the models are physically incomplete.
It is also possible, however, that one of the constraints is measured incorrectly, that one of the constraints is applied incorrectly, or that one of the constraints is poorly predicted for numerical reasons.
To investigate this, we identify all models that fail only one constraint in Table~\ref{tab:fail_one_thermal}.
We find that the critical constraints are 86\GHz size, \mring diameter, and $\mi{3}$.
Notice that there is overlap between \kharma and \bhac in MAD models that fail the $\mi{3}$ constraint.
The \hamr models fare significantly worse than the \kharma and \bhac models in the $\mi{3}$ constraint: only 7\% of models pass.
The remaining models all fail at least one additional constraint, leading to their exclusion from Table~\ref{tab:fail_one_thermal}.

\subsection{Exploratory Models}\label{sec:explore}

Next we go beyond the fiducial models and consider the exploratory models, which include: aligned models that use an alternative scheme for assigning temperatures to a thermal eDF; aligned models with a power-law component or $\kappa$ component in the eDF; tilted models; and stellar wind-fed models.
Unless stated otherwise, exploratory models are imaged over only $5 \times 10^3\,\tg$, yielding weaker constraints.  In all cases we focus on how the exploratory models differ from the fiducial models.

\subsubsection{Critical Beta Model}

The $\Rh$ prescription provides a convenient, one-parameter model for assigning electron temperatures, but here is a vast function space of possible alternative parameterizations.
One well-motivated choice is the critical beta model \citep{2020MNRAS.493.1404A}, which sets $T_e = T_e(R)$ and $R = f \exp(-\beta/\beta_c)$ (see Equation~\ref{eq:thermaleDF}).
This ``critical beta'' model has two parameters, $f$ and $\beta_c$.
We consider a single point in the parameter space: $f = 0.5$, $\beta_c = 1$.
Compared to the \Rh temperature prescription, the main new characteristic of the critical beta models is that the electron to ion temperature ratio approaches 0 at high $\beta$ instead of $1/\Rh$.

We have run all tests except X-ray for the critical beta models.
The $2.2\um$ flux is calculated by imaging only and therefore does not include Compton scattering.

All critical beta models fail the non-EHT constraints, with the 86\GHz size constraint rejecting most models as too small.
The variability constraints pass $23\%$ of the models.
No models survive the combined EHT and non-EHT constraints even if variability constraints are excluded.
Notice that this does not imply that critical beta models are ruled out, since we have only tested a single point in the $f,\beta_{\rm crit}$ parameter space.

\subsubsection{Thermal Plus \texorpdfstring{$p = 4$}{p=4} Power-law Models}

So far we have assumed a thermal eDF (Equation~\ref{eq:thermaleDF}).
Fully kinetic simulations as well as resistive MHD predict that reconnection in current sheets within the accretion flow and in the jet sheath leads to the acceleration of particles to higher energies, resulting in the emergence of a power-law tail  \citep[e.g.,][and references therein]{Sironi2021}.
Such acceleration events are thought to be the origin of near-infrared and X-ray flares detected in \sgra.
Here we do not address flare mechanisms but seek to constrain the contribution of non-thermal electrons to the quiescent emission of \sgra.

Below, we assume different forms of the eDF assuming that a fraction of the electron population is accelerated into a non-thermal tail.
There are multiple ways of doing this, but we will continue to assume that the eDF depends instantaneously on local conditions and set the accretion rate so that the 230\GHz time-averaged compact flux is 2.4\Jy.

First we consider a hybrid thermal/power-law distribution using \hamr/\bhoss.
Since we are modeling quiescent emission, we assume a steep power-law index of $p=4$ with a constant non-thermal acceleration efficiency $\epsilon=n_{\rm e, power-law}/n_{\rm e, thermal}=0.1$, typical of PIC simulations \citep[e.g.,][]{Sironi2015,Crumley2019}.
Following \citet{Chatterjee2021}, the power-law tail is stitched to the thermal core by choosing the minimum Lorentz factor limit of the power-law, $\gamma_{\rm min}$, to be at the peak of the Maxwellian component.
The upper end of the power-law is set to $10^5 \gamma_{\rm min}$ (see Equation~\ref{eq:non-thermaleDF}).
The temperature of the thermal component is set by the $\Rh$ prescription (Equation~\ref{eq:rhigh_prescription}).
We find that the accretion rate is slightly smaller than for corresponding thermal models, consistent with a small contribution from the power-law component to the 230\GHz total intensity.

\subsubsubsection{230\GHz VLBI pre-image size}

Hybrid thermal/power-law models have larger 230\GHz VLBI pre-image sizes compared to their purely thermal counterparts.
This is because the power-law component of the eDF allows high energy electrons in weak magnetic fields at distances more than a few gravitational radii (i.e., larger than the typical emission radius of the 230\GHz images) to contribute to the total image.
However, the extension in the images is much smaller for MAD models, with most MAD images displaying an increase in size of $<10\%$.

\subsubsubsection{86\GHz flux and image size}

In general, the $R_{\rm high}=1$ models produce too much 86\GHz flux.
Since the lower limit of the power-law $\gamma_{\rm min}$ is directly affected by the local electron temperature, the highest energy electrons are located in the jet sheath where $T_i \approx T_e$.
Indeed this is why SANE models produce more 86\GHz flux when non-thermal electrons are introduced, especially at larger $\Rh$ values.
On the other hand, MAD thermal and mixed thermal/non-thermal models behave similarly as the bulk of the emission is produced in the inner disk.

The 86\GHz image sizes for the hybrid \hamr models are, on average, larger than their thermal-only counterparts, similar to the 230\GHz image sizes.
The higher energy electrons of a hybrid thermal/power-law population emit at higher frequencies than their thermal core, thereby extending the image size.
This effect increases the image size of MAD models by only a few percent.

\subsubsubsection{$2.2\um$ constraint}

The addition of the power-law tail increases the flux at $2.2\um$ and thus the GRAVITY-based $2.2\um$ median flux density threshold of $1.0\,\mathrm{mJy}$ provides a strong constraint on the power-law index and the acceleration efficiency.
In brief, 59\% of the power-law models, especially $\Rh=1$ and $40$ MAD models, are ruled out by the $2.2\um$ constraint.

\subsubsubsection{Summary}

Overall, \hamr hybrid thermal/power-law models behave quite differently from their thermal counterparts.
For the thermal models, both EHT and non-EHT constraints are equally successful in ruling out models, with 22\% passing for each constraint set. For the power-law model set non-EHT constraints pass 39\% of models while EHT constraints pass 10\% of models.
This disparity occurs for two reasons:
\emph{i}) introducing non-thermal electrons pushes the 86 GHz image size to the acceptable range as thermal models typically exhibit small image sizes; and
\emph{ii}) the m-ring width is found to be smaller for the hybrid models.
This could be due to a change in the gas density scaling that is required to match the 230\GHz flux.
Nonthermal models require a smaller normalization value, meaning a smaller electron number density as compared to the corresponding thermal models.
A decrease in the number density lowers the optical depth, leading to a thinner photon ring.
For the initial $\no{5000}\,\tg$ survey, two mid-inclination power-law models survive: a SANE $\abh=0.94$ model and a MAD $\abh=0.5$, $\Rh=1$ model (see Table~\ref{tab:passfraction}), although ultimately both models are ruled out when extended to $\no{15000}\,\tg$.

\subsubsection{Constant \texorpdfstring{$\kappa$}{kappa} models with \texorpdfstring{$\kappa = 5$}{kappa = 5}}
\label{sec:constant_kappa}

Next we consider a model in which all electrons are in a $\kappa$ eDF, which has a thermal core and a power-law tail.
We set $\kappa = 5$ everywhere,  motivated by \citet{2016PhRvL.117w5101K}, who found $\kappa = 5$ to be a good fit to the ion DF in a 3D hybrid simulation of MHD turbulence.
A similar application of $\kappa$ eDFs with fixed $\kappa$ values for \sgra can be found in \citet{2018A&A...612A..34D}.
The power-law tail has $p = \kappa - 1 = 4$, and at high frequency $\nu L_\nu \sim \nu^s$ where $s = 2 - \kappa/2 = -1/2$.\footnote{Unless stated otherwise, the width parameter $w$ of the $\kappa$ distribution (see Equation~\ref{eq:kappa}) is set by $w = (\kappa - 3) \Theta_e/\kappa$, where the dimensionless electron temperature $\Theta_e$ is computed according to Equation~(\ref{eq:rhigh_prescription}).}
We image \bhac GRMHD simulations from 25\,kM to 30\,kM using \bhoss \citep{Younsi2012,Younsi2020}.
The accretion rate required to obtain 2.4\Jy is smaller than for the thermal models.
This implies that many of the $\kappa=5$ models are optically thin at 230\GHz and show thinner rings than their thermal counter-part (see first and second row in Figure~\ref{fig:SANE_edfs}).

\subsubsubsection{230\GHz size and light curve variability}

We find that the $\kappa=5$ models produce results that are generally consistent with the \bhac thermal models.
Especially at 230\GHz we find similar passing fractions for the 230\GHz source sizes.
92\% of the $\kappa=5$ models pass the size constraint compared to 98\% for the thermal models.
This can be explained mainly by SANE models at low $\Rh$ which are larger than the thermal models.
Variability is almost completely unaffected by the $\kappa$ distribution.
We find 29\% are in agreement with the \mi{3} constraint compared to 27\% for the thermal models.
The $\kappa=5$ models have a higher $\mi{3}$ for a small number of SANE $\Rh \geq 40$ models.
However, since the $\mi{3}$ constraint is computed for a time window of length only $\no{5000}\,\tg$, a factor three shorter than for the thermal models, this increase does not increase the fraction of models ruled out by this constraint.

\subsubsubsection{Visibility Amplitude Morphology }

The $\kappa=5$ models are optically thinner than the corresponding thermal models and typically show a thin, bright ring feature.
In consequence only 59\% of the $\kappa=5$ models pass the \vam constraint while the passing fraction for the thermal models is 84\%.
Similarly, due to the change in optical depth, only 55\% of the non-thermal models are in agreement with the \vam in contrast to 72\% of the thermal models.

\subsubsubsection{M-ring fits}

The m-ring constraints on diameter, width and asymmetry are passed by 71\%, 3\% and 73\% of the $\kappa=5$ models.
Except for the diameter all pass fractions are smaller than for the thermal models (65\%, 21\%, and 95\% for diameter, width and asymmetry).
The slightly larger pass fraction for the diameter could be affected by the shorter time window used for the $\kappa$ models as compared to the thermal ones.
However, the low fraction for the m-ring width can be explained by the optical depth of the $\kappa$ models.
Most of the $\kappa$ models are optically thinner than their thermal counter-part which leads to a finer, brighter ring-structure, and this is picked up by the m-ring fitting (see Figure~\ref{fig:SANE_edfs}).

\subsubsubsection{86\GHz source size}

For MAD models the size of the $\kappa=5$ models does not change.
This can be explained by the fact that most of the emission is produced in the midplane.
For the SANE models we find two different behaviors: the source size increases for $\Rh < 40$ and decreases for $\Rh \geq 40$, especially for positive black hole spins and high inclinations (compare first and last panel in the bottom row of Figure~\ref{fig:SANE_edfs}).
This change in size is consistent between the images at 230\GHz and at 86\GHz.
The passing fraction for the $\kappa=5$ models drops to 29\% as compared to 59\% for the thermal models.

\subsubsubsection{86\GHz flux}

The $\kappa=5$ models are relatively optically thin at 86\,GHz.
Together with the spectral slope $p=\kappa-1$ the flux at lower frequencies can be approximated as: $2.4\times \left(\nu/230\GHz\right)^{-(p-1)/2}$.
This leads to an 86\GHz flux density $\sim$ 10\Jy which is far above the 86\,GHz flux constraint of $2\pm0.2\Jy$.
In consequence the passing fraction for the $\kappa=5$ models drops to 12\% compared to 68\% for the thermal models.

\subsubsubsection{$2.2\um$ Constraint}

All MAD models fail the $2.2\um$ constraint, as do SANE models with $\Rh>1$.
This can be explained by the power-law tail of the $\kappa$ eDF (see Equation~\ref{eq:kappaeDF}) as compared to the exponential behaviour in the thermal eDF (see Equation~\ref{eq:thermaleDF}).
Only 14\% of the $\kappa$ models pass in contrast to 70\% of the thermal models.  Evidently the NIR flux density provides a powerful constraint on any non-thermal component in the eDF.

\subsubsection{Mixed Thermal/\texorpdfstring{$\kappa$}{kappa} Model}

\begin{figure*}
  \centering
  \includegraphics[width=\textwidth]{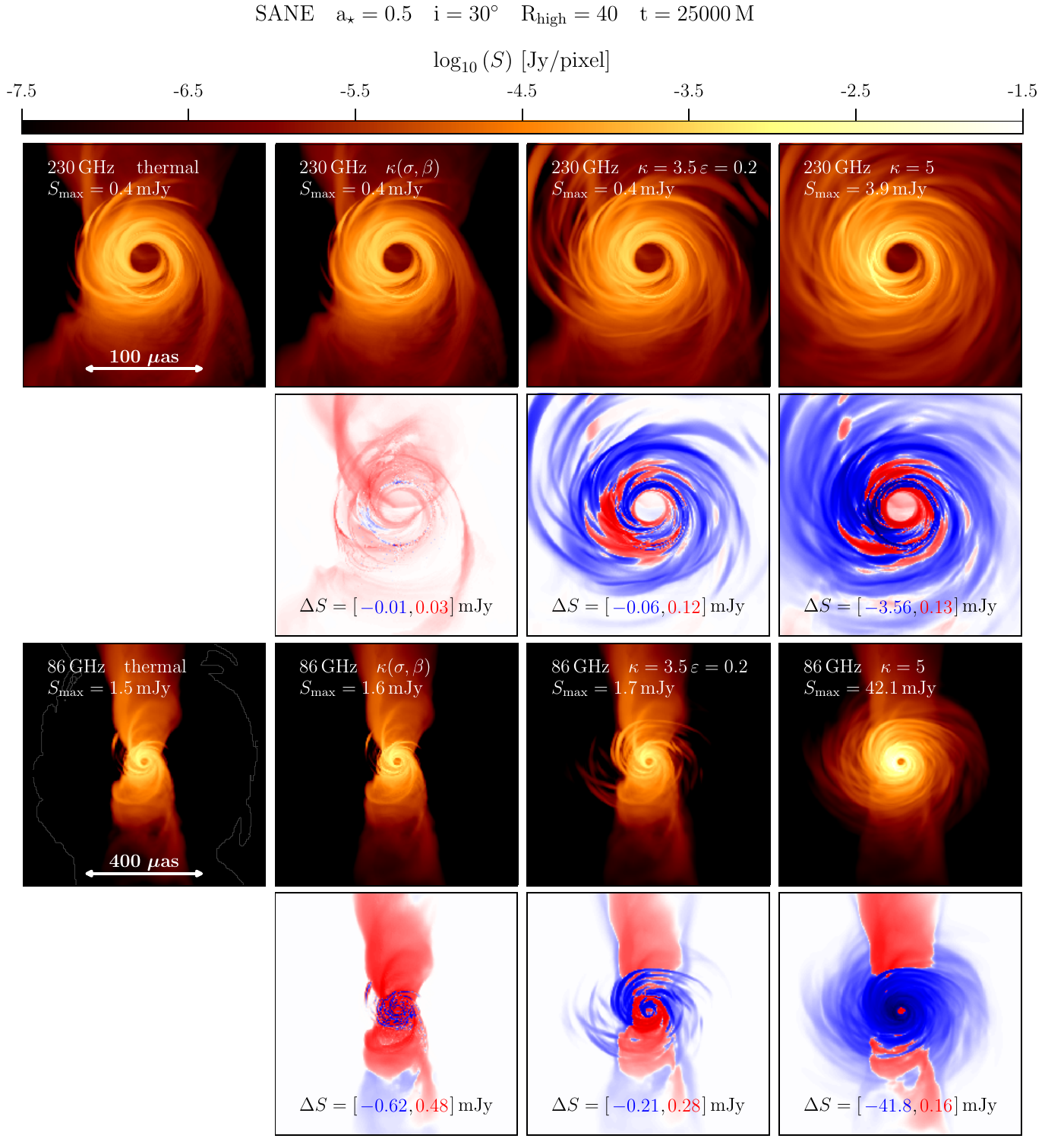}
  \caption{Influence of the eDF on the image structure for a SANE model with spin $a_{\star}=0.5$ seen under a viewing angle of $30\degree$ using $\Rh = 40$ at $t = \no{25000}\,\tg$.
    In the first and third row the panels show the image structure from left to right a thermal, variable kappa $\kappa(\sigma,\beta)$, $\kappa=3.5$ (fixed) with efficiency $\varepsilon=0.2$ and $\kappa=5$ (fixed) everywhere eDF at 230\GHz and at 86\GHz.
    Notice the increased field of view for the 86\GHz images.
    The second and fourth row show the difference between thermal image and the different non-thermal eDFs.}
  \label{fig:SANE_edfs}
\end{figure*}

Next we consider a mixed thermal/non-thermal eDF, with the non-thermal component following the $\kappa$ DF with $\kappa = 3.5$.
At high frequency $\nu L_\nu \sim \nu^s$ with $s = 2 - \kappa/2 = 1/4$, similar to what is seen in $2.2\um$ flares \citep{2007ApJ...667..900H}.
For this model set the GRMHD simulations use \bhac and the imaging uses \bhoss.

The fraction of non-thermal electrons is assumed to depend on $\sigma$ and $\beta$.  The emissivity
\begin{equation}
  j_{\nu,\rm{tot}}=(1-\epsilon) j_{\nu,\rm{thermal}} + \epsilon j_{\nu, \kappa},
  \label{eq:kappaeff}
\end{equation}
where the non-thermal efficiency
\begin{equation}
  \epsilon(\varepsilon,\beta,\sigma)=\varepsilon\,
  \left[1 - e^{-\beta^{-2}}\right]
  \left[1-e^{-(\sigma/\sigma_{\rm min})^2}\right].
  \label{eq:efficiencybetasigma}
\end{equation}
Evidently $\epsilon \rightarrow 0$ in the disk while $\epsilon \rightarrow \varepsilon$ in the jet.
Since we remove emission at $\sigma > \sigma_{\rm cut} = 1$ the non-thermal electrons are confined to the jet sheath.

We set $\sigma_{\rm min}=0.01$ and vary the base efficiency, $\varepsilon$ over 0.05, 0.1 and 0.2.
At each $\varepsilon$ we generate a model set spanning the same parameter space as the thermal models (see Table~\ref{tab:radiativemodels}) and normalize the accretion rate using the standard procedure (see Section~\ref{sec:models}).

The mass accretion rate required to obtain 2.4\Jy at 230\GHz only changes on average around 1.5\% as compared to the thermal models.
This small variation in the mass accretion rate reveals the fact that most of the emission at 230\,GHz is created from the thermal part of the hybrid eDF, consistent with the small fraction of non-thermal particles added ($\varepsilon=0.05$, $0.1$, and $0.2$).

\subsubsubsection{230\GHz size and light curve Variability}
\label{varkappa230}

The addition of non-thermal particles does not substantially affect the flux or size of the image at 230\GHz.

For MAD models, 230\GHz emission is mostly produced in the disk region (see \citetalias{M87PaperV} and Figure~8 in \citealt{2022ApJS..259...64W} for a 3D rendering).
Thus, the images are unaffected by the non-thermal particles, which are located in the jet.

For SANE models, increasing $\Rh$ pushes the emission towards the jet sheath which increases the source size for high spins and large $\Rh$.
However, the effect of non-thermal particles on the image is minor, because most of the emission is still produced by thermal electrons with temperature set by the $\Rh$ prescription (compare first and third panel in top row of Figure~\ref{fig:SANE_edfs}).

The passing fraction for the 230\GHz image size is 98\%, independent of $\varepsilon$, consistent with the thermal models.
We find that the 47\% of the models are in agreement with the 230\GHz variability constraint.
This passing fraction is larger than for the thermal models (27\%), due to the shorter time window ($\no{5000}\,\tg$) considered for the non-thermal models in contrast to $\no{15000}\,\tg$ for the thermal ones.

\subsubsubsection{Visibility Amplitude Morphology and Variability}

Since 230\GHz images of the $\kappa=3.5$ models with variable efficiency are similar to the thermal models (see previous paragraph) the fraction of passing models for the VA morphology are comparable.
The three non-thermal models have an average passing fraction for the \vam of 80\% whereas 84\% of the thermal models pass.
The 4\,$G\lambda$ visibility amplitude variability constraint passes 72\% of the models for both thermal and non-thermal eDF.

\subsubsubsection{M-ring fits}

Given that including non-thermal particles via the equations presented in Equation~\ref{eq:efficiencybetasigma} does not change the image structure and variability properties of the 230\GHz images the M-ring fits provide the same passing fractions for the diameter (65\%), width (22\%) and asymmetry (95\%).

\subsubsubsection{86\GHz source size and flux}

The 86\GHz source is only slightly affected by the addition of non-thermal particles as compared to the thermal models.
Only the SANE models with $\Rh \geq 40$ and $\abh > 0$ produce 86\GHz image sizes larger than the thermal SANE models.
This effect can be seen in  first and third panel in the bottom row of Figure~\ref{fig:SANE_edfs}.
Notice the increased flux density in the jet sheath in the difference image (blue color).
This trend increases with the efficiency and is reflected in the decreasing pass fraction: 56\% (for $\varepsilon=0.05,0.1$) and 55\% ($\varepsilon=0.2$) as compared to thermal models (59\%).
A similar trend is found for the 86\GHz flux density.
The non-thermal particles are mainly located in the jet and thus contribute to the 86\GHz flux.
Again, jet dominated high spin SANE models typically fail the 86\GHz flux constraint.
With increasing efficiency i.e.
adding more non-thermal particles the pass fraction decreases, with 67\%  at $\varepsilon=0.05$, 66\% passing at $\varepsilon=0.1$, and 63\% passing at  $\varepsilon=0.2$ compared to a pass fraction of
68\% for the thermal models.

\subsubsubsection{$2.2\um$ Constraint}

The $2.2\um$ flux density increases for all models.
For SANE models, except $\Rh=1$, the addition of non-thermal particles leads to over-production of $2.2\um$ photons.
For MAD models, all models over-produce at $2.2\um$ for $\varepsilon \ge 0.05$.
As noted above, $2.2\um$ emission is produced from the tail of the eDF.
The thermal eDF tail decreases exponentially, while the $\kappa$ eDF tail decreases as a power law, so the increase in $2.2\um$ flux density is unsurprising.
This is a general feature of the non-thermal models: $2.2\um$ observations sharply limit the allowed population of non-thermal electrons.

\subsubsection{Variable \texorpdfstring{$\kappa$}{kappa} Model}

The high-energy variability observed in many astrophysical sources including the galactic center may be associated with
magnetic reconnection.
Particle-in-cell (PIC) simulations have found that the slope of the non-thermal tail depends on  $\sigma$ and $\beta$ \citep[see, e.g.,][]{2018ApJ...862...80B}.  Here we consider a $\kappa$ eDF model in which $\kappa$ and $w$ vary  following the prescription of \cite{2018ApJ...862...80B}:
\begin{align}
  \kappa &= 2.8 +0.7\sigma^{-1/2} + 3.7\sigma^{-0.19}\tanh{(23.4\,\sigma^{0.26}\beta)}, \label{eq:kappa}\\
  w      &= \frac{ \kappa -3 }{\kappa} \Theta_e.
\end{align}

We use emissivities and absorptivities from  \cite{2016ApJ...822...34P}, computed numerically for the interval $3 < \kappa \le 8$.
For $\kappa > 8$ we substitute a thermal eDF.
As in the fiducial models we turn off emission at $\sigma > 1$.

The variable $\kappa$ models are computed from \hamr and \bhac GRMHD models where the time windows
$(\no{30000}$--$\no{35000})\,\tg$ (\hamr) and
$(\no{25000}$--$\no{30000})\,\tg$ (\bhac) are used.

We find that the mass accretion rate needed to obtain $\langle F_{230} \rangle = 230\GHz$ is on average 4\% larger than for the thermal models, and thus the variable $\kappa$ models have slightly higher optical depth.

\subsubsubsection{230\GHz size}

The disk region is dominated by thermal electrons (i.e. large $\kappa$) while the jet sheath has the lowest $\kappa$.
Therefore, the 230\GHz source size of the variable $\kappa$ models is similar to the thermal ones and no difference in pass fraction is found.
98\% of both models are in agreement with the 230\GHz size estimate (see first and second panel in the top row of Figure~\ref{fig:SANE_edfs}).

\subsubsubsection{Visibility Amplitude Morphology and Variability}

For the null location of the variable $\kappa$ models we find no difference to their thermal counterparts and for both $\sim$80\% pass this constraint.
However, there is a clear discrepancy between the variable $\kappa$ and thermal models regarding the VA morphology.
Only 60\% of the $\kappa$ models pass the 4\,$G\lambda$ visibility amplitude variability constraint, in contrast to 72\% of the thermal models.

\subsubsubsection{M-ring fits}

The thermal and variable $\kappa$ models agree in the passing fraction for the m-ring diameter (66\%), m-ring width (22\%) and asymmetry (95\%).

\subsubsubsection{86\GHz source size and flux}

The pass fraction for the 86\GHz source size constraint are comparable for thermal (59\%) and variable $\kappa$ models (55\%).
Given that most of the variable $\kappa$ models are optically thicker than their thermal counterparts the 86\GHz flux is on average lower which increases the passing fraction from 68\% (thermal eDF) to 75\% (variable $\kappa$ eDF).

\subsubsubsection{$2.2\um$ Constraint}

Including non-thermal particles via the $\kappa$ eDF increases the 2.2\um flux as compared to the thermal eDF.  In our prescription $\kappa$ decreases (more high energy electrons) as $\sigma$ increases.  In MAD models $\sigma$ is systematically larger than in SANE models.  Consistent with this, we find that most of the MAD models fail the 2.2\um constraint whereas in SANE models $\kappa$ is large and the passing fraction for SANE models is almost indistinguishable from their thermal counterparts.
In total 35\% of the variable $\kappa$ models pass the 2.2\um constraint compared to 55\% for models including only thermal particles.

\subsubsubsection{X-ray constraint}

Including non-thermal particles via the variable $\kappa$ eDF reduces the pass fraction from 61\% (for thermal eDF) to 35\%.  The $\kappa$ eDF provides a larger population of seed photons in the NIR and at higher energies that can be boosted into the X-ray by a single scattering event.

\subsubsubsection{Trends across \bhac and \hamr models for variable $\kappa$ models}

For the variable $\kappa$ models we have redundant models from \bhac and \hamr (see Table ~\ref{tab:GRMHDmodels}).
Both models sets show similar trends for all constraints (see Table~\ref{tab:passfraction}).

\subsubsection{Summary of Constraints on Non-Thermal Models}

\begin{deluxetable*}{l|ccccccc}\label{tab:passfraction}
\tablecaption{Exploratory Model Pass Fractions}
\tablehead{
\colhead{constraint/model} &
\colhead{}                 &
\colhead{\bhac}            &
\colhead{}                 &
\colhead{}                 &
\colhead{\hamr}            &
\colhead{}                 &
\colhead{}
}
\startdata
& Thermal & $\kappa(\sigma,\beta)$& $\kappa=3.5\,\varepsilon=0.05,0.1,0.2$& $\kappa=5$ & Thermal & $\kappa(\sigma,\beta)$ & $p = 4$ \\
\hline
230\,GHz size           & 0.98 & 0.99 & 0.98, 0.98, 0.98 & 0.92 & 1.0  & 0.99 & 0.94 \\
VA Morphology           & 0.83 & 0.80 & 0.81, 0.81, 0.78 & 0.59 & 0.80 & 0.88 & --   \\
M-ring diameter         & 0.65 & 0.69 & 0.66, 0.66, 0.67 & 0.71 & 0.58 & 0.67 & 0.89 \\
M-ring width            & 0.21 & 0.21 & 0.24, 0.23, 0.23 & 0.03 & 0.29 & 0.40 & 0.13 \\
M-ring asym.            & 0.95 & 0.97 & 0.95, 0.95, 0.94 & 0.73 & 1.0  & 0.97 & 0.98 \\
\hline
86\GHz flux             & 0.68 & 0.75 & 0.67, 0.66, 0.63 & 0.12 & 0.62 & 0.65 & 0.72 \\
86\GHz size             & 0.59 & 0.57 & 0.56, 0.56, 0.55 & 0.38 & 0.46  & 0.45 & 0.47 \\
2.2\um flux             & 0.55 & 0.35 & 0.14, 0.12, 0.12 & 0.14 & 0.80 & 0.2  & 0.41 \\
X-ray flux              & 0.70 & --   & --               & --   & 0.61 & 0.35 & --   \\
\hline
lc variability          & 0.27 & 0.30 & 0.47, 0.47, 0.46 & 0.29 & 0.07 & 0.28 & 0.41 \\
4\,G$\lambda$ variability & 0.72 & 0.60 & 0.74, 0.73, 0.71 & 0.55 & 0.39 & 0.46 & 0.63 \\
\hline
EHT Constraints         & 0.19 & 0.17 & 0.17, 0.16, 0.15 & 0.01 & 0.22 & 0.27 & 0.10 \\
non EHT Constraints     & 0.19 & 0.12 & 0.01, 0.0,  0.0  & 0.14 & 0.22 & 0.09 & 0.39 \\
Variability Constraints & 0.27 & 0.28 & 0.42, 0.42, 0.42 & 0.27 & 0.03 & 0.24 & 0.38
\enddata
\tablecomments{Pass fractions for \bhac and \hamr models for various eDFs. Note that the thermal models are run for $\no{10000}\,\tg$ and $\no{15000}\,\tg$ for \bhac and \hamr respectively, while the non-thermal models are only run for $\no{5000}\,\tg$. This results in a much lower pass fraction in the thermal models for the lightcurve variability constraint and all three \mring constraints.}
\end{deluxetable*}

In Table~\ref{tab:passfraction} we list the pass fractions for \bhac and \hamr models using different eDFs.
Most non-thermal eDF models produce little change compared to the thermal models for most constraints.
The 86\GHz size and flux, which are the most important non-EHT constraints, are only marginally affected by the addition of non-thermal electrons.
This behaviour is obtained especially for eDFs which mainly add non-thermal particles in the jet while the disk is populated by thermal ones.
In our case this setup is given for variable $\kappa$ and $\kappa=3.5$ with variable efficiency and is consistent between \bhac and \hamr models (see Table~\ref{tab:passfraction}).

If non-thermal particles are included also in the disk, either via a power-law with slope p=4 stitched to a thermal distribution or via a $\kappa=5$ distribution, then there are some variations in pass fractions as compared to the above-mentioned eDFs.  For the power-law models the addition of non-thermal electrons increase the 86\GHz size by an average of 50\%.
However the pass fractions with respect to the thermal models is not changed.
In contrast the $\kappa=5$ model pass fractions decrease by 20\% compared to the thermal models.
For $p=4$ and $\kappa=5$, fewer models pass the 230\GHz m-ring width, with a consistent decrease by $\sim20\%$ for both models.
Interestingly the other m-ring constraints i.e., the diameter and asymmetry, are not affected by the addition of non-thermal particles.
This can be explained by the finer and brighter ring feature found in $p=4$ and $\kappa=5$ models connected to their smaller optical depth compared to their thermal counter-parts (see Table~\ref{tab:passfraction}).

In general the fraction of models passing $\mi{3}$ increases with the addition of non-thermal particles, independent of the prescriptions of the eDF.  This is due to the shorter duration of the exploratory runs and not an actual reduction in variability.   The main characteristic of the non-thermal models is the increase of 2.2\um and X-ray flux densities. However, in a large fraction of models this leads to overproduction of 2.2\um or X-ray flux and the pass fractions are reduced.

\begin{deluxetable*}{l|ccccc}\label{tab:fail_none}
\tablecaption{Exploratory models that pass all constraints}
\tablehead{
\colhead{Code/Setup} &
\colhead{MAD/SANE} &
\colhead{spin} &
\colhead{inc} &
\colhead{$\Rh$} &
\colhead{Constraint(s) failed at $\no{15000}\,\tg$}
}
\startdata
\bhac $\kappa(\sigma, \beta)$	&	MAD		&	0.5		&	10		&	80 		& 	Mring width, \mi{3}		\\ 
\bhac $\kappa(\sigma, \beta)$	&	MAD		&	0.5		&	10		&	160		&	Mring width, \mi{3}		\\ 
\bhac $\kappa(\sigma, \beta)$	&	MAD		&	0.5		&	30		&	160		&	\mi{3}					\\ 
\hline
\bhac $\epsilon = 0.05$			&	SANE	&	0.94	&	10		&	10 		& 	\mi{3}					\\	
\hline
\hamr $p = 4$	  				&	SANE	&	0.94	&	50		&	40 		& 	2.2\,$\mu$m flux		\\	
\hamr $p = 4$	  				&	MAD		&	0.5		&	50		&	1 		&  	Mring diameter, \mi{3}
\enddata
\tablecomments{Exploratory models which pass all constraints when computed to $\no{5000}\,\tg$. When extended to $\no{15000}\,\tg$, each model fails one or more constraints.}

\end{deluxetable*}

Six non-thermal models pass all 11 constraints (see Table~\ref{tab:fail_none}).
These models are:
one \hamr MAD $p=4$ model with $\abh=0.5$ seen under an inclination $i=50\degree$ and
$\Rh=1$ and a high spinning SANE model with $\abh=0.94$ at an inclination of $i=50\degree$ with $\Rh=40$.
From the \bhac variable $\kappa$ three models are in agreement with all constraints namely: spin $\abh=0.5$ at inclination $i=10\degree$ at $\Rh=80$ and 160 and a model with the same spin seen under a slightly larger angle of $i=30\degree$ with $\Rh=160$.
The last of the six survivors is a \bhac SANE model with variable efficiency of $\varepsilon=0.05$ with an inclination of 10$^\degree$ and a $\Rh=10$.
These models share a common low inclination angle $i \leq 60\degree$ and positive spin.
We note that the MAD models coincide with the cluster of thermal models found for both \bhac and \kharma models (see Section~\ref{sec:summarythermal}).
We also show the nonthermal models that fail only one constraint in Table~\ref{tab:fail_one_nonthermal}.

\begin{deluxetable*}{l|ccccc}\label{tab:fail_one_nonthermal}
\tablecaption{Exploratory models that fail only one constraint}
\tablehead{
\colhead{Code/Setup} &
\colhead{MAD/SANE} &
\colhead{spin} &
\colhead{inc} &
\colhead{$\Rh$} &
\colhead{Constraint Failed}
}
\startdata
\bhac $\kappa(\sigma, \beta)$	&	SANE	&	0		&	70		&	1		&	86\,GHz flux		\\
\bhac $\kappa(\sigma, \beta)$	&	MAD		&	0.5		&	10		&	40		&	2.2\,$\mu$m flux	\\
\bhac $\kappa(\sigma, \beta)$	&	SANE	&	0.5		&	10		&	40		&	\mi{3}   			\\
\bhac $\kappa(\sigma, \beta)$	&	SANE	&	0.5		&	30		&	40		&	\mi{3}   			\\
\bhac $\kappa(\sigma, \beta)$	&	SANE	&	0.5		&	30		&	80		&	\mi{3}   			\\
\bhac $\kappa(\sigma, \beta)$	&	SANE	&	0.5		&	30		&	160		&	\mi{3}   			\\
\bhac $\kappa(\sigma, \beta)$	&	MAD		&	0.5		&	30		&	80		&	\mi{3}   			\\
\bhac $\kappa(\sigma, \beta)$	&	MAD		&	0.5		&	50		&	160		&	\mi{3}   			\\
\bhac $\kappa(\sigma, \beta)$	&	MAD		&	0.94	&	10		&	160		&	\mi{3}   			\\
\hline
\bhac $\epsilon = 0.05$			&	SANE	&	0.94	&	30		&	10		&	86\,GHz flux   		\\
\bhac $\epsilon = 0.05$			&	MAD		&	0.5		&	30		&	160		&	2.2\,$\mu$m flux   	\\
\bhac $\epsilon = 0.05$			&	MAD		&	0.5		&	50		&	160		&	2.2\,$\mu$m flux   	\\
\bhac $\epsilon = 0.05$			&	MAD		&	0.94	&	10		&	40		&	2.2\,$\mu$m flux   	\\
\hline
\bhac $\epsilon = 0.10$			&	SANE	&	0.94	&	10		&	10		&	2.2\,$\mu$m flux   	\\
\bhac $\epsilon = 0.10$			&	MAD		&	0.5		&	30		&	160		&	2.2\,$\mu$m flux   	\\
\bhac $\epsilon = 0.10$			&	MAD		&	0.5		&	50		&	160		&	2.2\,$\mu$m flux   	\\
\bhac $\epsilon = 0.10$			&	MAD		&	0.94	&	10		&	40		&	2.2\,$\mu$m flux   	\\
\hline
\bhac $\epsilon = 0.20$			&	SANE	&	0.94	&	10		&	10		&	2.2\,$\mu$m flux   	\\
\bhac $\epsilon = 0.20$			&	MAD		&	0.94	&	10		&	40		&	2.2\,$\mu$m flux   	\\
\bhac $\epsilon = 0.20$			&	MAD		&	0.94	&	10		&	160		&	2.2\,$\mu$m flux   	\\
\hline
\bhac $\kappa = 5$				&	MAD		&	0.94	&	50		&	1		&	2.2\,$\mu$m flux   	\\
\hline
\hamr 30$^\circ$ tilt  			&	SANE$^a$	&	0.94	&	10		&	160		&	86\,GHz size
\enddata
\tablecomments{Models which pass all of the constraints except for one. Since no model passes all constraints, these represent the parameters that are closest to being consistent with observations. $^a$ For the tilted model, $\phi/\phi_{crit} \simeq 0.8$.}

\end{deluxetable*}

\subsection{Tilted Models}

Aligned models are a special case: in general one expects that the spin angular momentum of the black hole and the orbital angular momentum of the accretion flow are misaligned.
Here we consider misaligned flows around an $a_*=15/16$ black hole from \citet{Liska2018} and \citet{Chatterjee2020}.

All aligned models considered so far produce either a SANE or MAD accretion flow.
The tilted disk model initial conditions, however, produce a strongly magnetized near-MAD outcome with dimensionless magnetic fluxes between 25--50, a state we describe as  IN-SANE.
We consider three GRMHD simulations with tilt $0\degree$, $30\degree$ and $60\degree$.

The tilted models exhibit a warped disk due to  Lense-Thirring precession.
The time-averaged disk and jet are therefore non-axisymmetric.
Since the inner and the outer disk have different orientations, it is necessary to specify the coordinate axis of the observer.
We consider three  observer inclinations with respect to the {\em outer} disk at a single azimuth of $0\degree$ \citep[for more details, see][]{Chatterjee2020}.\footnote{A full parameter survey would run over azimuth angle as well.}

The 230\GHz pre-image size of edge-on large $\Rh$ models  increases slightly for the tilt-$60\degree$ compared to the aligned case.
This occurs because the inner jet is warped and creates an extended image.
This effect is also seen in the 86\GHz image size.
On the other hand, the 86\GHz flux varies little with tilt despite the presence of a boosted jet component at large tilt angles.

Variability increases with tilt.
In tilted disks, accretion occurs via thin plunging streams \citep[e.g.,][]{Fragile2007} where electrons in the shocked flow can be heated to relativistic temperatures \citep[e.g.,][]{Dexter2013, 2014ApJ...780...81G, White2019}, forming localized, fluctuating hotspots more easily than in aligned disks and increasing flux variability \citep{Chatterjee2020, 2022ApJ...925..119B}.  Nevertheless, 20/27 models pass the \mi{3} constraint because of the short duration of the tilted models, which provide fewer \mi{3} samples than the fiducial models.

The $2.2\um$ flux density also increases with tilt.
The $2.2\um$ flux exceeds the 1.0\,mJy limit for all 3 tilts, with a few exceptions, e.g., $\Rh=160$ models at $10\degree$ inclination, which makes it difficult to favor the aligned case over the tilted one.
Furthermore misalignment destroys the axisymmetric nature of the accretion flow.
The current model set covers a small parameter space in inclination and $\Rh$.
A thorough exploration of the source azimuthal angle with respect to the observer is left to future studies.

To summarize: for the model set considered here tilt primarily affects variability and the $2.2\um$ flux density, tending to increase both and thus shifting acceptable aligned models into rejected models as tilt angle increases.
These trends are consistent with those observed by \cite{2022ApJ...926..136W}.

\subsection{Stellar Wind Fed Models}

The accretion models of \cite{2020ApJ...896L...6R, 2020MNRAS.492.3272R, 2018MNRAS.478.3544R} track plasma from  magnetized stellar winds down to the event horizon and provide a self-consistent picture of the origin of both gas and magnetic fields in the accreting plasma in \sgra.
The resulting inflow does not fully circularize, so the models provide a distinct alternative to the fiducial models, which {\em assume} that the torus initial conditions relax to an astrophysically accessible state for the inner accretion flow.
In the wind-fed models the density of the wind is fixed, so the 230\GHz flux density is matched to observations by varying $\Rh$ instead.

We use two versions of the model: one in which the stellar wind magnetization is low ($\beta = 10^6$) and a second in which the magnetization is high ($\beta = 10^2$).
$\Rh$ is adjusted until each model has the observed time-averaged 230\GHz flux density, with $\Rh = 13$ ($\beta = 10^6$) and $\Rh = 28$ ($\beta = 10^2$).

Both wind-fed models produce rings that are too narrow, failing the \mring width test.
In addition both are too bright at 86\GHz and fail the $\mi{3}$ test, although they are quieter than MAD models and close to the cutoff.

Both non-EHT and EHT constraints have the power to test wind-fed models.
It is {\em not} possible to draw broad conclusions about the viability of the wind-fed models in general, however, since the two models tested here contain only a single spin ($\abh=0$) and all use the $\Rh$ thermal eDF model.

\section{Discussion}
\label{sec:discussions}

\subsection{The Goldilocks Model and Polarization}\label{sec:goldilocks}

The fiducial models cover a regular grid in parameter space.
In one instance, for \kharma models, adjacent points in parameter space fail only one constraint for opposite reasons: the SANE, $\abh = 0.5$, $\Rh = 40$ $i = 10\degree$ models fails because the 86\GHz image is too small, while the $i = 30\degree$ models fail because the 86\GHz image is too large, suggesting  that a model with intermediate inclination would pass all constraints.

We analyzed an intermediate inclination model at $i = 20\degree$.
This ``goldilocks'' model passes {\em all} constraints (we did not compute the X-ray luminosity but the neighboring $i=10\degree$ and $i=30\degree$ models pass the X-ray constraint).

We have imaged a series of \kharma models with inclinations between $10\degree$ and $30\degree$.
The cause of the inclination sensitivity is an extended, 86\GHz-bright, jet.
At $i = 10\degree$ the jet is nearly parallel to the line of sight and the source size is dominated by the accretion flow, but as $i$ increases the jet, which is radially extended, begins to extend past the accretion flow and dominate the source size.

Despite this success we regard the goldilocks model as unpromising for two reasons.
First, the $10\degree$ and $30\degree$ \bhac models fail the X-ray constraint, and the neighboring $10\degree$ and $50\degree$ \hamr models fail several constraints.

In addition, the goldilocks model is likely underpolarized at 230\GHz.
The source-integrated linear polarization of \sgra at the time of the 2017 campaign was between $6.9\%$ and $7.5$\% \citep{2021ApJ...910L..14G}, consistent with historical measurements.
We will consider linear and circular polarizations in future papers, but our preliminary finding is that the goldilocks model has linear polarization $\approx 1$\%, which is far too low. We find that the best-bet region MAD models, considered below, have linear polarization compatible with observations.

\subsection{Testing Exploratory Models that Pass}

Six exploratory models, considered in Section~\ref{sec:explore},  pass all constraints.
Although they appear promising these six models are imaged for only $\no{5000}\,\tg$ and thus have been tested more weakly than the fiducial thermal models.

To evaluate the effect of run duration we imaged the six passing exploratory models for  $\no{15000}\,\tg$.
All failed one or more constraints, which are listed in Table \ref{tab:fail_none}.
Evidently adding a population of non-thermal electrons does not provide a consistent way of transforming failing fiducial models to passing models.

The pass/fail status of the model can be sensitive to the length of integration, and it is important to image the models for at least $\no{15000}\,\tg$.
In connection with this we note that the \koral models, which were imaged at 230\GHz for $\sim 10^5 \tg$, were generally consistent with the fiducial models but provide tighter $\mi{3}$ constraints (see Appendix~\ref{app:variability}).
The constraints that are most sensitive to model duration are $\mi{3}$ (all models failed $\mi{3}$ after being extended) and the \mring fits (two failed \mring width, one failed \mring diameter).

\subsection{Origin of Variability Excess}

Approximately 84\% (\kharma), 73\% (\bhac), and 97\% (\hamr) of the fiducial models fail one or both variability constraints.
This naturally leads one to ask whether there is an observational or modeling problem with these constraints.

For example, it is possible that a fraction $f_\mathrm{ext}$ of the $230\GHz$ flux density is in an extended structure (e.g. a jet) that is slowly varying, unresolved by ALMA, and resolved out by EHT.
The observed $\mi{3}$ would then be smaller than the true $\mi{3}$ for the compact source by a factor of $1/(1 + f_\mathrm{ext})$.
The $4\,\mathrm{G}\lambda$ amplitude variability is normalized by the light curve and thus $a_4$ would be suppressed by a similar factor.

Diffuse emission on scales larger than the VLBI images ($\sim 100\uas$) and smaller than connected element interferometer measurements ($\sim 100\,\mathrm{mas}$) is difficult to constrain.
The EHTC imaging strategy involves a self-calibration step that assumes no diffuse structure on scales between the zero-baseline and the shortest VLBI baselines (\citetalias{PaperIII}).
However, longer wavelength VLBI observations place constraints on any emission on these scales under the assumption of flat or steep spectra.
For instance, $230$ and $43\GHz$ VLBI observations that use the Los Alamos-Pie Town-VLA baselines probe scales of $\sim 1$ to $10\,\mathrm{mas}$ and demonstrate no inconsistency in closure amplitudes and closure phases with a symmetric, two-dimensional Gaussian model \citep{2004Sci...304..704B}.
VLBI observations at 3mm wavelength are also fully consistent with a two-dimensional Gaussian model with an upper limit of $\sim 10\,\mathrm{mJy}$, or approximately 1\% of the total flux density \citep{2019A&A...621A.119B}.
Additionally, a dust contribution is constrained by shorter wavelength ALMA observations that find a flat or slightly falling spectrum up to 900 GHz \citep{2019ApJ...881L...2B}.
A substantial diffuse dust contribution would only be consistent if its properties were tuned to match a steeply falling compact synchrotron spectrum, which would likely be inconsistent with the substantially variable far infrared component of \sgra emission \citep{2016ApJ...825...32S, 2018ApJ...862..129V}.

Future observations with short-baseline coverage such as that provided by the Kitt Peak to SMT baseline will enable tighter constraints on any diffuse component.
If confirmed, the presence of diffuse flux would require re-normalization of the models and re-evaluation of the constraints.
A reduction of $\sim 30\%$ in the compact flux would lead many SANE models to fail the \mi{3} constraint because they would be not variable enough and would lead most MAD models to pass the \mi{3} constraint.
Since $F_{230} \sim \dot{M}^2$ a $15\%$ change in model density normalization, and therefore in optical depth, would suffice.

The variability excess might also be caused by physical incompleteness of the models.
Collisionless effects and radiative cooling could both reduce variability.

We model the accreting plasma as an ideal fluid when in fact it is collisionless, with Coulomb mean free path large compared to $\rg$.
This is less worrisome than one might think: electrons and ions are confined to helical orbits around field lines, implying an effective mean free path perpendicular to the field lines of order the gyroradius $\sim 56 \Theta_e [B/(30\,\mathrm{G})]^{-1} \cm \ll \rg$.
The mean free path parallel to field lines is still long, but may be limited by scattering off electromagnetic field fluctuations excited by kinetic instabilities rather than Coulomb scattering.

Future global kinetic general relativistic particle-in-cell simulations may be able to test how well the ideal fluid model describes collisionless accretion flows.
Meanwhile, a relativistic fluid model incorporating small collisionless corrections was developed by \citet{2015ApJ...810..162C} and studied numerically by \citet{2017MNRAS.470.2240F}.
The leading order corrections are conduction and pressure anisotropy (i.e. viscosity).
The effect of conduction and viscosity on our constraints is not known, but it is known that viscosity reduces turbulent stress in SANE models \citep{2017MNRAS.470.2240F}, consistent with a reduction in variability.
It is also plausible that conduction smoothes out temperature maxima, possibly also leading to a reduction in variability.

Our models neglect radiative cooling.
Cooling is fastest where the electron temperature is highest, so cooling has the potential to blunt local maxima in temperature~\citep{2020MNRAS.499.3178Y}.
If local maxima with short cooling times contribute significantly to variability then cooling could reduce $\mi{3}$.
Self-consistent cooling requires integration of an electron energy equation \citep[e.g.][]{2015MNRAS.454.1848R}, and assignment of a density scale (mass unit or accretion rate) when the GRMHD simulation is run, vastly increasing computational cost.

Another possibility is that self-consistent electron heating reduces variability.
In Appendix~\ref{app:variability} we consider a set of such models from \citet{2020MNRAS.494.4168D}.
These models show a variability excess that is similar to the fiducial models.

The $\Rh$ prescription contains a parameter $\Rl$ that determines the ion-electron temperature ratio at low $\beta$, which we have consistently set to $1$.
Increasing $\Rl$ models the effect of rapid cooling in low $\beta$ regions.
In Appendix~\ref{app:variability} we consider $\Rl$ up to 10 in a sample of four models and show that it does not systematically reduce variability.

The variability excess might also be caused by numerical inaccuracies in radiative transfer, truncation error in the GRMHD integrations (limited resolution), limited simulation duration, or misspecification of the adiabatic index.
These are considered in Appendix~\ref{app:numerical} and~\ref{app:variability}.
We find no evidence that any of these effects is producing the variability excess.  A more extensive study on how these effect could alter the variability is warranted but
outside the scope of this paper.

\subsection{Best-Bet Region without Variability}

From the preceding discussion it is possible that a combination of extended flux, viscosity, cooling, and numerical limitations affect model variability enough to compromise the variability constraints.
Notice that a relatively small change in variability, $\sim$ 30\% in \mi{3}, is sufficient to promote many of the MAD models from failing to passing.
If we exclude the variability constraints, then, which models are favored?

\begin{figure*}
  \centering
  \includegraphics[width=0.8\textwidth]{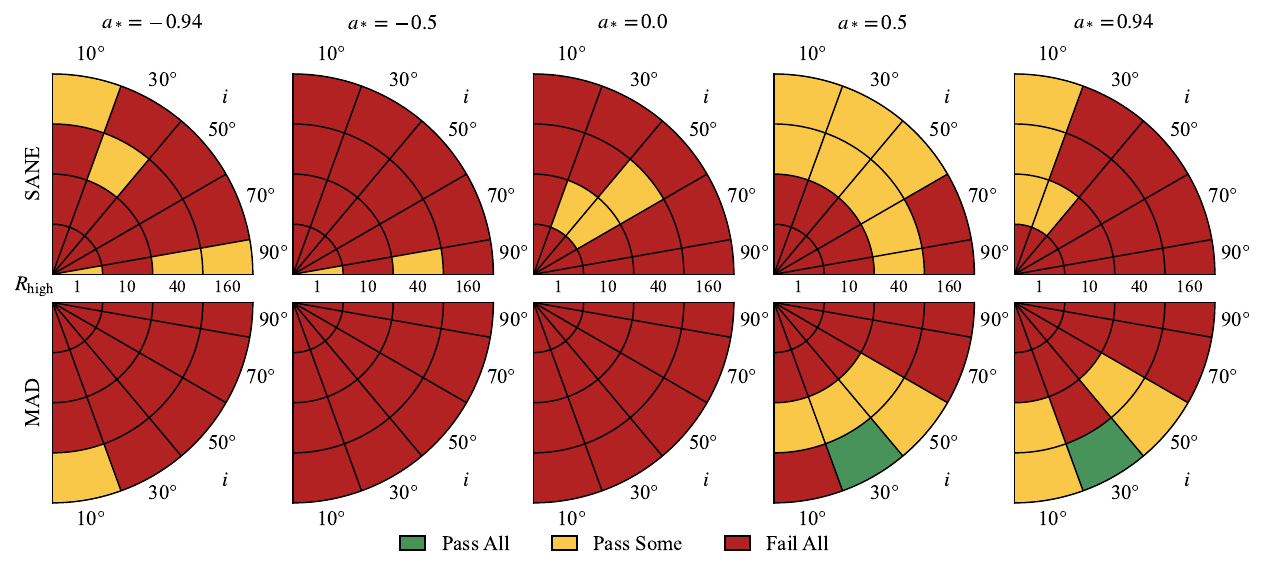}
  \caption{Combined constraints without structural or flux variability.
Green indicates that the \kharma, \bhac, and (for $i = 10\degree, 50\degree, 90\degree$) \hamr models pass, yellow that one or two of the fiducial models fail, and red that all models fail.}
  \label{fig:all_pizza}
\end{figure*}

Figure~\ref{fig:all_pizza} (which also appears as  Figure~\ref{fig:all_pizza_app} in Appendix~\ref{app:tables}) shows the result of applying all constraints except structural and flux variability to the fiducial models.
Most negative spin models are ruled out, and two MAD, positive spin, low inclination, large $\Rh$ models pass all constraints for \kharma, and \bhac (the \hamr simulations do not include $i = 30\degree$).
Nearby models in parameter space are close to passing in the sense that they pass for one or more of \kharma, \bhac, and \hamr. \footnote{If we use the more permissive 86GHz size constraint from \cite{2019ApJ...871...30I}, we find two new best-bet models, one MAD model close to the best-bet region at  $\abh = 0.94$, $i = 10 \degree$, and $\Rh = 160$; and one SANE model at $\abh = 0.5$, $i = 30 \degree$, and $\Rh = 10$.}

We will call the green region of parameter space in Figure~\ref{fig:all_pizza} the {\em best-bet region}.
The models in this part of parameter space perform well and explain nearly all the data.

Given the uncertainty associated with the variability excess and the possibility of missing physical ingredients in the model the existence of the best-bet region cannot be regarded as evidence that \sgra has positive spin and low inclination.
Given the large and uncertain set of constraints, however, it is remarkable that {\em any} models perform as well as these do.
The models in the best-bet region merit additional analysis, and it is interesting to ask what they predict for future observations.

\subsection{Fiducial Models Accretion Rate and Outflow Power}
\label{sec:accrate_outflowpower}

\begin{figure*}
  \centering
  \includegraphics[width=0.95\textwidth]{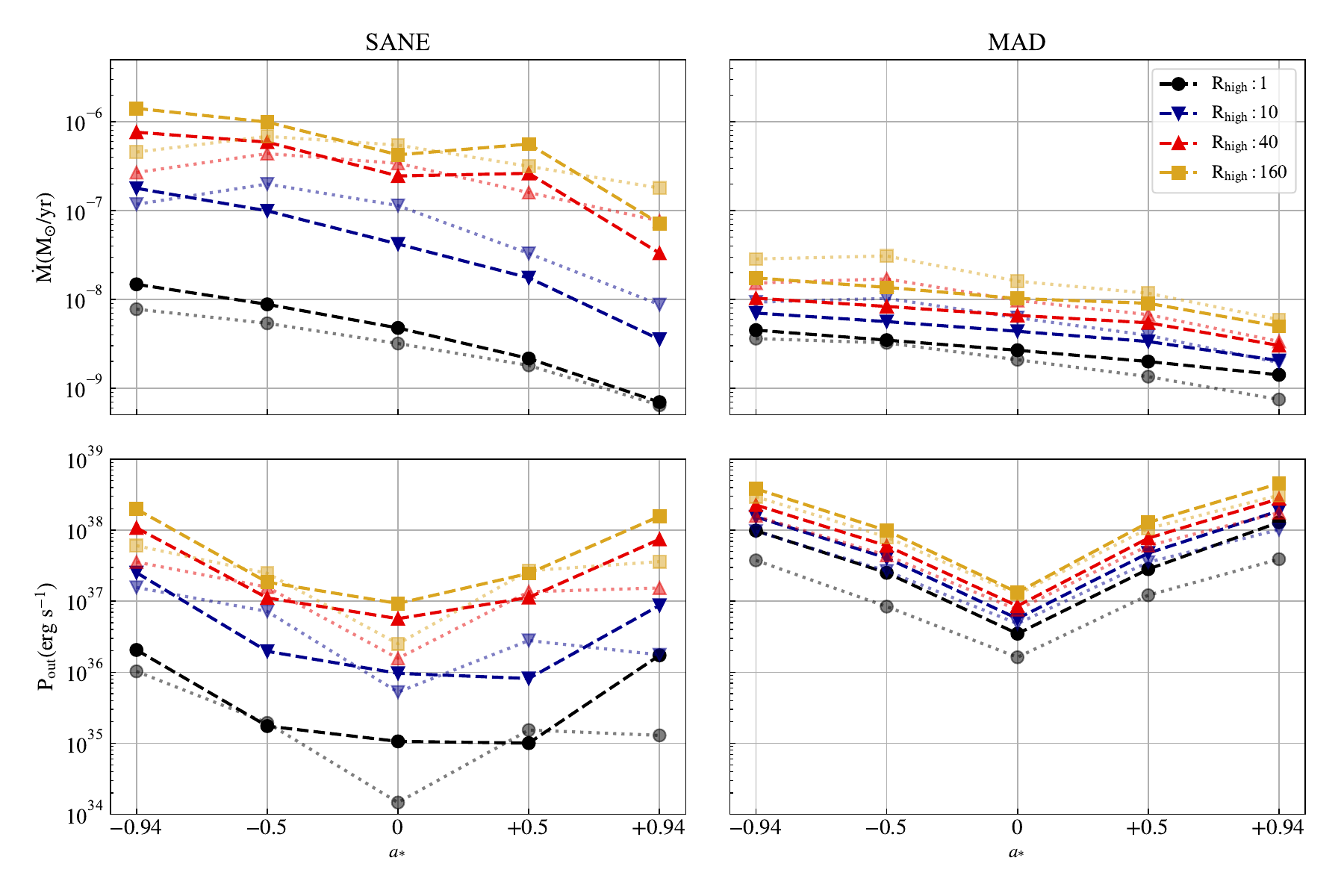}
  \caption{Comparison of the dependence on spin $\abh$ and high end of the temperature ratio $\Rh$ for the accretion rate $\dot{M}$ and outflow power $P_{\rm out}$ in the SANE and MAD cases.
    {\it Top:} Accretion rate $\dot{M}$ for fiducial models.
    Since $\dot{M}$ depends weakly on inclination $i$, we show only $i=50\degree$.
    {\it Bottom:} Outflow power $P_\mathrm{out}$ for fiducial models at $i = 50\degree$.
    The colors and markers vary with $\Rh$.
    The dashed lines correspond to \kharma thermal models while the dotted lines indicate \bhac thermal models.}
  \label{fig:accretion_outflow_power_illinois_thermal}
\end{figure*}

What is the accretion rate in \sgra?  The time-averaged
\begin{equation}
  \dot{M} \equiv \frac{1}{\Delta t}\int dt\int d\theta d\phi\hspace{0.1cm}\sqrt{-g}\big(-\rho u^{r}\big),
\end{equation}
at the event horizon; the quantity in parentheses is the inward rest-mass flux density.

Figure~\ref{fig:accretion_outflow_power_illinois_thermal} (top panels) shows $\dot{M}$ in solar masses per year for the \kharma and \bhac thermal models.
The accretion rates follow immediately once the models are normalized so that $\langle F_{230}\rangle = 2.4\,\mathrm{Jy}$.

MAD models accrete at $10^{-9}$ to $10^{-8} M_{\odot}$yr$^{-1}$ while SANE models have a broader range, $10^{-9}$ to $10^{-6} M_{\odot}$yr$^{-1}$.
$\dot{M}$ is an increasing function of $\Rh$, which is a result of the thermal synchrotron emissivity increasing with increasing electron temperature, so that at fixed flux density lower electron temperature models (higher $\Rh$) have higher $\dot{M}$.
SANE models exhibit a stronger dependence on $\Rh$ than MAD models.
The emission in the SANE models is predominantly from regions with large $\beta$, where the electron temperature is regulated by $\Rh$.
MAD models produce more of their emission in regions with $\beta\sim 1$, where $\Rh$ only has a weak effect on electron temperature  (see \citetalias{M87PaperV}, Figure~4).
For both SANE and MAD models, $\dot{M}$ decreases with increasing spin.
This follows from the dependence of temperature on spin shown in Figure~\ref{fig:grmhd_temp}---higher spin models have higher gas temperature, which in the $\Rh$ model implies higher electron temperature.

Retrograde SANE, $\Rh=40$ and $160$ models produce the largest $\dot{M} \sim 10^{-6}M_{\odot}\yr^{-1}$.
These models have a high midplane density of cool electrons that, in \kharma models, overproduce X-ray emission through bremsstrahlung and are therefore ruled out.
Critical beta models have accretion rates that lie between the fiducial model values for $\Rh = 10$ and $\Rh = 40$ for the selected critical beta parameter values ($f=0.5$, $\beta_\mathrm{crit}=1$).

How do our $\dot{M}$ compare with earlier estimates?
Linear polarization and Faraday rotation measurements at millimeter and sub-millimeter (submm) wavelengths \citep{2000ApJ...538L.121A, 2000ApJ...545..842Q, 2003ApJ...588..331B, 2006ApJ...640..308M, 2006JPhCS..54..354M, 2006ApJ...646L.111M} and X-ray emission \citep{2003ApJ...591..891B, doi:10.1126/science.1240755} combined  with semi-analytic models predict $\dot{M} \sim (10^{-9}$--$10^{-7} M_{\odot}\yr^{-1}$.
The broad range of values is due to the differences in regions of radio emission in the theoretical models that are considered (ADAFs: \citealt{1998ApJ...492..554N, Yuan_2003}; Jet models: \citealt{1993A&A...278L...1F, 2000A&A...362..113F}; ADAF+jet models: \citealt{2002A&A...383..854Y}).
All our MAD model $\dot{M}$s fall within the range of these historical observational estimates.
In contrast almost all SANE models with larger $\Rh$ parameter (except SANE $\abh=0.94$, which is one of our best models) have an $\dot{M}$ that is inconsistent with earlier estimates.

All fiducial models produce outflows in the polar regions.
In many cases the outflows can be divided into a slower, denser disk wind and a relativistic, high $\sigma$ Poynting jet.
The outflows have a power that is comparable to or larger than the bolometric luminosity.
What is the outflow power $P_\mathrm{out}$?

First we must define $P_\mathrm{out}$.
There are a number of competing definitions in the literature; we set
\begin{equation}
  P_\mathrm{out} \equiv \frac{1}{\Delta t}\int dt \int d\phi \int_\mathrm{poles} d\theta \sqrt{-g}\left(-T^{r}_{t}-\rho u^{r}\right),
\end{equation}
where ``poles'' indicates $\theta < 1\,\mathrm{rad}$ or $\theta > (\pi-1)\,\mathrm{rad}$.  We include only those $\theta$ where the time- and azimuth-averaged energy flux is outward.
The integral is evaluated at $r = 100\,\rg$.
$P_\mathrm{out}$ includes power in the relativistic Poynting jet, if present, and in the slower, denser disk wind.

Figure~\ref{fig:accretion_outflow_power_illinois_thermal} (bottom panels) shows $P_\mathrm{out}$ for the fiducial \kharma and \bhac models.
As expected, the outflow power increases with the magnitude of black hole spin.
We find that SANE models have $10^{35} \lesssim P_\mathrm{out} \lesssim 10^{38} \ergps$.
Evidently $P_\mathrm{out}$ increases with $\Rh$, as expected from the behavior of $\dot{M}$: higher $\dot{M}$ models have stronger magnetic fields.
We find that MAD models have $10^{37} \lesssim P_\mathrm{out} \lesssim 10^{39}\ergps$ and are, on average, more powerful and less sensitive to $\Rh$.

Many high spin or large $\Rh$ models have $P_\mathrm{out} \sim 10^{38} \ergps$.
An outflow with this power could produce dramatic observable effects in the dense interstellar medium of the Galactic Center.
For instance, the X-ray transient CXOGC J174540.0-290031, located only 0.1\pc from \sgra and with an estimated jet power of $\sim 10^{36}\ergps$ produced a compact bipolar lobe at radio wavelengths with peak flux densities near 100$\,\mathrm{mJy}$ \citep{2005ApJ...633..218B}.
A more continuous  outflow, however, might clear out a substantial volume of space, making identification of any interaction with the ISM less certain.
Nevertheless, there have been a number of large scale features that have been suggested as the result of interaction of a jet with the ISM \citep[e.g.,][]{2013ApJ...779..154L,2021ApJ...922..254C}.

\subsection{Caveats and Limitations}\label{sec:limits}

\subsubsection{Electron Distribution Function}

One of the central uncertainties in modeling \sgra is electron distribution function (eDF) assignment.
Do the surviving models have anything to say about the eDF?

In our fiducial models, thermal eDFs with equal ion and electron temperature ($\Rh = 1$) are ruled out, in most cases by more than one constraint.
For MAD models this is easily explained: Comptonization is strong at $\Rh = 1$ and X-rays are overproduced.
For MAD models, $\dot{M}$ and therefore the electron scattering optical depth $\tau_\mathrm{es}$ is insensitive to $\Rh$.
Since the amplitude of the first Compton bump is $\propto y = 16 \Theta_e^2 \tau_\mathrm{es}$, the high $\Theta_e$ at $\Rh = 1$ produces a large X-ray flux.
For SANE models the situation is more complicated (see Appendix~\ref{app:tables}), with \mring width rejecting many $\Rh = 1$ models at $\abh \le 0$, and 86\GHz flux and size rejecting the rest.
The latter is a consequence of model electron temperature reaching a maximum at large $\abh$ and low $\Rh$, so that optical depth is a minimum and therefore so is the 86\GHz image size.

In some fiducial SANE models the sense of the X-ray constraint is reversed: bremsstrahlung is strong and X-rays are overproduced where $\Rh$ is {\em large}.
Again this is easily understood: when $\Rh$ is large the midplane electrons are cold, the accretion rates (and therefore $n^2$) are high, and the bremsstrahlung emissivity is large.
More generally the X-rays provide a strong constraint on the presence of dark (subrelativistic) electrons, which are otherwise undetectable in millimeter wavelength emission or absorption, although they can produce strong Faraday rotation.

We have tested a large set of non-thermal models, which have a power-law tail on the eDF.
Although integration times for the non-thermal models are too short to provide strong model constraints, there are trends that emerge from the existing data.

First, the 230~GHz images are relatively insensitive to the presence of non-thermal electrons for models in which most of the non-thermal electrons are introduced in and near the outflow region.
This is encouraging: the 230~GHz image is generated by electrons in an approximately thermal core of the eDF, and is relatively insensitive to the behavior of the tails.

Second, as one might expect, the 2.2\um flux density is an increasing function of non-thermal electron density.
In many models (e.g. the variable efficiency models of Section~\ref{sec:constant_kappa}) the addition of a power-law tail changes a thermal model that passes the $2.2\um$ test into a non-thermal model that fails.

Third, it is important to understand that many of the non-thermal models we use are linked to the $\Rh$ prescription in some way.
For example, the $\kappa$ distribution function contains a width parameter $w$, and this is set using an $\Rh$-like prescription with width depending on $\beta$.
Our non-thermal models are only a few points in a vast function space of possible non-thermal parameterizations, with none of the models considered allowing for an electron energy density that depends on plasma history as well as instantaneous plasma state.

\subsubsection{Collisionless plasma effects}

{The mean free path to Coulomb scattering for particles is typically larger than or comparable to the system size in \sgra, rendering its plasma collisionless.
The GRMHD simulations used in this work describe a collisional system, whereas a first-principles modeling of the collisionless plasma requires a fully kinetic treatment.
General relativistic (radiative) kinetic simulations are crucial for dynamically probing the electron temperature, effects of non-thermal distribution functions, and pressure anisotropy and their interplay with radiation in collisionless plasma in the accretion disk and jet.
While global general relativistic kinetic simulations cannot be performed with full physical separation between microscopic plasma scales (the particle's Larmor radius $r_{\rm L}$, and plasma skin depth $d_{\rm e}$) and macroscopic scales (the gravitational radius $r_{\rm g}$), they can achieve the right hierarchy of scales ($r_{\rm g} \gg d_{\rm e} \gg r_{\rm L}$) for magnetized plasmas \citep{2018A&A...616A.184L,2018ApJ...863L..31C,2019PhRvL.122c5101P,2020PhRvL.124n5101C,2020ApJ...895..121C,2020ApJ...902...80K,2021A&A...650A.163C,2021PhRvL.127e5101B}.
Even in GRMHD, it is computationally challenging to resolve plasma heating processes powering the observed radiation in a converged manner.
It is not yet feasible to resolve dissipation at the smallest scales of the turbulent cascade or the interplay between turbulence and reconnection at a similar level as in local box simulations \citep{2012ApJ...755...50R,2013ApJ...773..118H,2015PhRvL.114f1101H,2016PhRvL.117w5101K,2017PhRvL.118e5103Z,2018PhRvL.121y5101C,2018ApJ...859..149I,2019PhRvL.122e5101Z,2021ApJ...921...87N,2021ApJ...923L..13C}.
However, \citet{2019ApJS..243...26P} and \citet[in prep.]{Olivares_et_al} show that the global accretion dynamics (mass accretion rate, magnetic flux on the horizon, and MRI quality factor) are converging between the different simulations in this work.
Kinetic processes in the (near-)collisionless plasma may increase the effective particle collision rate \citep[see, e.g.,][]{2016PhRvL.117w5101K}.
Deviations from the infinitely conducting ideal fluid approximation may alter the thermodynamics of the flow \citep[see, e.g., ][and Appendix~C1]{2017MNRAS.470.2240F}.
Some aspects of (near-)collisionless plasma dynamics can be described with non-ideal effects (e.g., viscosity, resistivity, heat conduction, pressure anisotropy) in GRMHD simulations of black hole accretion, e.g.,  \cite{2014MNRAS.440L..41B,2015ApJ...810..162C,2016MNRAS.456.1332F,2017ApJ...837...92C,2017MNRAS.470.2240F,2018ApJ...859...28Q,2019ApJS..244...10R,2019ApJ...882....2V,2020ApJ...900..100R,2021PhRvD.104j3028M,2021arXiv211103689N,2021arXiv211105752M}.
For example, the first efforts have recently been made with high-resolution global GRMHD simulations to capture heating through magnetic reconnection in the largest current sheets in the system \citep{2020MNRAS.495.1549N,2020ApJ...900..100R,2021MNRAS.508.1241C,2022ApJ...924L..32R,2021arXiv211103689N}.}

\subsubsection{Positrons}\label{sec:pair}

So far we have considered only ion-electron plasmas, but pairs can be produced in the 230\GHz emission region through pair discharges or through so-called pair drizzle.

The importance of pairs has been assessed using phenomenological models \citep{2020ApJ...896...30A, 2021ApJ...923..272E} and depends sensitively on the efficiency with which a reservoir of magnetic energy can be converted into pairs.
If this efficiency is large then pairs can substantially increase intensity in the jet region.

Production of pairs through the drizzle process is weak in \sgra because its luminosity is low.
\cite[][see also \citealt{2021ApJ...907...73W}]{2011ApJ...735....9M} estimate the drizzle pair density of \sgra to be $10^{-8}\cm^{-3}$.
This is well below the Goldreich-Julian density $\sim \abh B c^2/(4 \pi e G M) \sim 10^{-2} \abh [B/(30\,{\rm G})] \cm^{-3}$ required to screen electric fields, suggesting pair discharges are likely.
If pair discharges serve only to raise the pair density to the Goldreich-Julian density, however, then the jet is unlikely to outshine the accretion flow, where the magnetic field strength is similar to that in the jet but the characteristic number density is $\sim 10^6 \cm^{-3}$.
The maximum conceivable impact of pairs on EHT observations is obtained by converting about half of the magnetic energy density into pairs with Lorentz factor $\sim 30$ so that in \sgra's $\sim 30$\,G magnetic field the emissivity of the resulting pairs peaks close to 230\GHz.
Then the pair density $\sim 10^{6} \cm^{-3}$ and the jet might compete with the accretion flow.

\subsection{Outlook}\label{sec:future}

Except for a brief discussion in Section \ref{sec:goldilocks} our analysis omits polarization.
Future analyses will test our models against integrated polarization of \sgra \citep{2021ApJ...910L..14G} and against polarized imaging, as was done for M87* \citepalias{M87PaperVII, M87PaperVIII}.

Our analysis also omits discussion of one of the main observational features of \sgra: the near infrared and X-ray flares, for which there is as yet no consensus model.
Our analysis is built on the notion that the near infrared flares, at least, could be produced by accelerating a small fraction of the electron population into a non-thermal tail.
The increase in $2.2\um$ flux density with non-thermal population seen in Section~\ref{sec:comparisons} is consistent with this notion.

The agreement between our results on the source size and orientation and those of the GRAVITY collaboration \citep{2018A&A...618L..10G} also support the similarity in the spatial distribution of the electrons producing near-infrared flares and those responsible for the 230\GHz emission.
\citealt{2018A&A...618L..10G} find that the near-infrared flares originate from a region $\approx 40$--$50\uas$ from the black hole, only slightly larger than the diameter of the \mring fit to the 230\GHz image.
Moreover, the combined results of our analysis point towards a low observer inclination, again consistent with the GRAVITY results , although the GRAVITY results are based on a model with a hotspot orbiting at $\lesssim 40\degree$ from the plane of the sky.

In M87* we were able to identify a sense of rotation of the source from the asymmetry of the observed ring and the orientation of the large scale jet.
For \sgra we have not yet been able to identify a preferred position angle for the source or measure the amplitude of source asymmetry, perhaps because it is small (\citetalias{PaperIV}).
The sparse baseline coverage from 2017 sharply limits our ability to detect asymmetry, but the 2021 observation already had better coverage, and future EHT campaigns will add even more stations.
Unless the source is aligned and $i \approx 0\degree$, all our models (which have a definite sense of rotation) predict that the brightest point on the ring is produced by Doppler boosting and should lie on the approaching side of the accretion flow.  The sense of rotation, determined from either the helicity of spiral features in the flow or from tracking rotation of bright spots, could be compared to the clockwise motion of near infrared flares observed by GRAVITY \citep{2018A&A...618L..10G}.

Our analysis shows the value of simultaneous measurements.
Simultaneous or near-simultaneous GMVA observations, in particular, provide a powerful constraint on the models.
The eDF is a major source of uncertainty in our analysis, and since the submillimeter (submm) and 2.2\um flux density are most sensitive to the eDF, future analyses should incorporate submm data and future EHT campaigns should seek near-simultaneous submm observations.

Our analysis provides some guidance for future numerical modeling of \sgra.
It is clear that models require long integrations ($\gtrsim \no{30000}\,\tg$) to provide a converged characterization of the source.
We also note that there are regions in parameter space where we may not have sampled densely enough---where, for example, the model is too large and too small at adjacent points in parameter space.
An example of this is the 86GHz size measurement, which is sensitive to inclination, as seen for the goldilocks model.
We found that being able to compare three simulation pipelines helped us identify numerical sensitivities and saved us from error on a number of occasions.
One point raised in these comparisons is that the $2.2\um$ flux density is sensitive to where emission is cut off in high $\sigma$ regions of the flow---the so-called $\sigma_\mathrm{cut}$ parameter.
This point merits future investigation.

Throughout this work we have assumed that the mass and distance to \sgra are known.
This assumption could be relaxed and the models checked for consistency with the stellar orbit measurements of mass and distance.
Phenomenological accretion flow models are particularly well suited to this type of study since they are inexpensive to compute \citep[e.g.,][]{2009ApJ...697...45B}.

\section{Conclusions}
\label{sec:conclusions}

\begin{figure*}
  \centering
  \includegraphics[width=\textwidth]{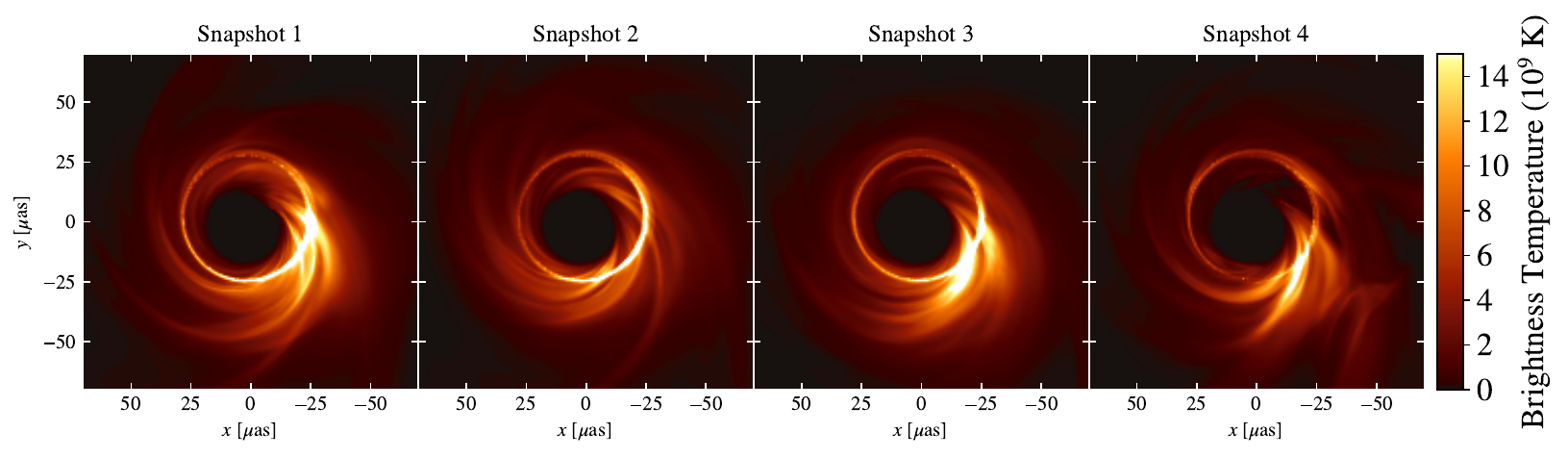}
  \caption{%
    230\GHz full resolution snapshots from a fiducial model in the
    best-bet region.
    This model passes 10/11 constraints.
    The different panels are snapshots taken from the ``best time'',
    when the synthetic observation has good \uv coverage.%
  }
  \label{fig:bestbet_imgs_snapshot}
\end{figure*}

\begin{figure*}
  \centering
  \includegraphics[width=\textwidth]{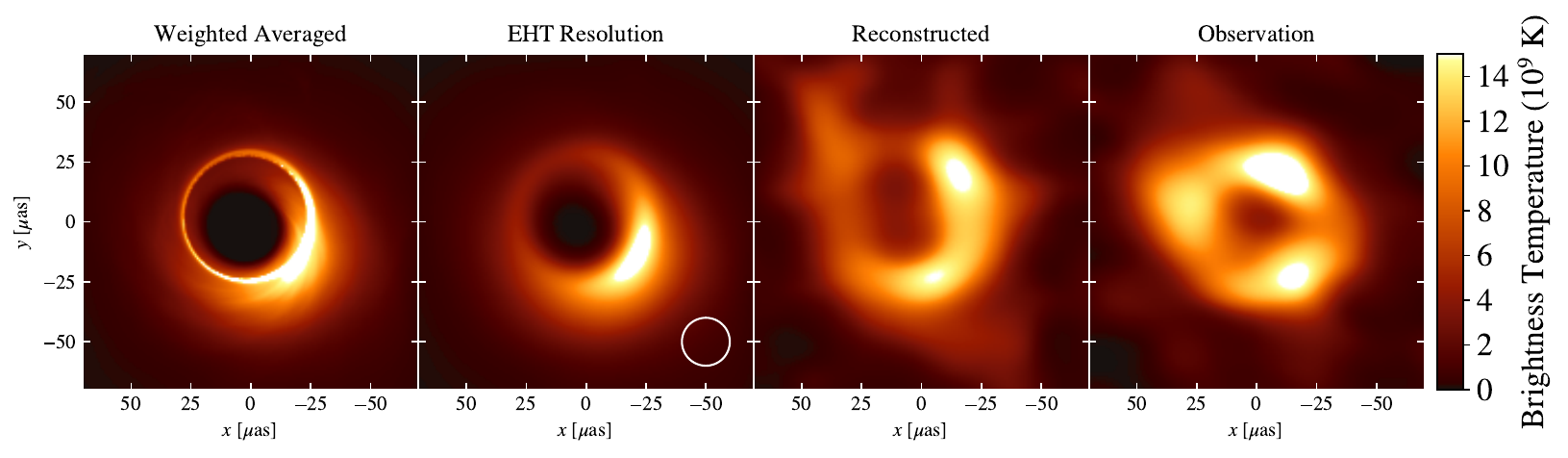}
  \caption{%
    Reconstructions of a fiducial model in the best-bet region,
    compared to the April~7 observation.
    This is the same model shown in
    Figure~\ref{fig:bestbet_imgs_snapshot}.
    The leftmost panel shows an average of the snapshots used in the
    synthetic observation, weighted by the number of baselines at
    each time.
    The second panel shows the averaged image convolved with a 20\uas
    beam, which roughly approximate EHT resolution.
    The third panel shows an average image reconstructed from
    synthetic data using the fiducial model.
    The final image shows the average image reconstructed from April~7
    EHT data.%
  }
  \label{fig:bestbet_imgs}
\end{figure*}

We have made a first comparison of the EHT 2017 \sgra data to a state-of-the-art library of ideal general relativistic magnetohydrodynamics (GRMHD) models.
The models assume that the mass and distance to \sgra are known and that the central object is a black hole described by the Kerr metric.
We use multiple simulation pipelines and find that for a given model configuration, independent simulations are remarkably consistent (Appendix~\ref{app:numerical}).

The model parameters are: whether the horizon magnetic field is strong or weak (MAD or SANE, respectively); the black hole spin $\abh$; and the inclination angle $i$ between the line of sight and the accretion flow orbital angular momentum vector.
The electron distribution function (eDF) also has one or more parameters.
In our ``fiducial'' model set, run with three independent codes, the eDF is determined using the so-called $\Rh$ prescription (Section \ref{sec:models}).
We have also considered exploratory models with alternate eDF prescriptions and alternate initial conditions.

We have selected and applied 11 heterogeneous observational constraints.
Six derive directly from EHT VLBI data, two derive from 86\GHz VLBI observations with the GMVA, one from variability of the 230\GHz light curve, and one each from the 2.2\um flux density and the X-ray luminosity.

Five structural constraints derive from EHT VLBI data.
When combined these constraints reject about 75\% of our fiducial models.
The EHT cut favors $\abh \ge 0$ and avoids edge-on ($i = 90\degree$) models and models with equal ion and electron temperatures ($\Rh = 1$).
We are {\em not} able to constrain the source position angle due to sparse baseline coverage.
The 2017 EHT observations are, nevertheless, quite constraining.  New EHT observations with additional antennas will be even more constraining.

The strongest EHT-derived constraint is \mring width.  The physical interpretation of \mring fits is challenging because fitting is done after the model is observed with limited baseline coverage and limited temporal sampling.  Nevertheless, some interpretation is possible.  For example, there is a trend toward lower width at higher inclination that eliminates many of the edge-on models. This can be understood since many edge-on models have higher peak brightness temperature than face-on models due to Doppler boosting of emission from the approaching side of the disk. The flux density, which must average 2.4Jy, is approximately proportional to the solid angle of the source multiplied by a typical brightness temperature, so when brightness temperature is higher, solid angle must be smaller.  If the source is a ring of fixed radius, then, higher brightness temperature implies a narrower ring.

Four constraints derive from non-EHT data that are contemporaneous or near-contemporaneous.
Combined, the non-EHT constraints reject 94\% of fiducial models.
The non-EHT cut favors strongly magnetized (MAD) models and eliminates most models at $i > 50\degree$ \citep[consistent with interpretations of GRAVITY results][]{2020A&A...643A..56G}, and also eliminates all models with equal ion and electron temperatures.
These results highlight the value of continued multiwavelength monitoring of \sgra.

The non-EHT constraints, like the EHT-derived constraints, exhibit  complicated but interpretable trends across parameter space. A full discussion of fiducial model trends will be explored in later papers, but as an example consider the 86GHz size constraint.  At $\Rh = 1$, the mean 86GHz FWHM of SANE models {\em decreases} as inclination increases.  The origin of this trend is very similar to the origin of the trend in \mring width with inclination.  At $\Rh = 1$ the electrons are hot and the source is optically thin. The peak brightness temperature of edge-on models is higher than that of face-on models due to Doppler boosting of emission from the approaching side of the disk, and thus at fixed flux density the source becomes smaller as inclination increases.  At $\Rh = 160$, by contrast, the trend is reversed and the mean 86GHz FWHM of SANE models {\em increases} as inclination increases.  In $\Rh = 160$ SANE models the electrons are cool in the disk midplane and most emission arises along the walls of the jet (see Figure 4 of \M87PaperV).  The equatorial plane is relatively opaque and Doppler-boosted emission is hidden. In face-on models one is looking down the jet and the source appears relatively small; in edge-on models the jet is extended perpendicular to the line of sight and the source appears larger.  Interestingly, MAD models exhibit only weak trends in 86GHz size with inclination, in part because MAD models are more optically thin than SANE models and also because MAD models are more slowly rotating than SANE models, weakening Doppler boosting.

\sgra is variable, but is not as variable as we expected based on our fiducial models.
We have used two tests to compare the variability of models and data.
One characterizes variability in the 230\GHz light curve (including simultaneous ALMA data) and the other characterizes structural variability expressed through fluctuations in the visibility amplitudes.
The light curve variability is the tightest of all 11 constraints: it rejects 95\% of our fiducial models.
We find that strongly magnetized (MAD) models are more variable than weakly magnetized (SANE) models and, grouped together, both SANE and MAD models are more variable than the data.
The structural variability constraint measures the slope and amplitude of the power spectrum of the VA variability.
Remarkably, we find that the power spectrum slope is consistent for all models, while the power spectrum amplitude is consistent for 43\% of fiducial models.


The higher variability of the MAD models compared to the SANE models is a consequence of the quasi-regular magnetic flux expulsion events that are a defining feature of the MAD models. Magnetic flux through the horizon builds up until it exceeds a threshold, and then escapes over a few dynamical times through the surrounding accretion flow.  As flux is escaping, strongly magnetized low density regions push aside plasma in the region close to the black hole that produces most of the 230GHz emission and this produces large fluctuations in the 230GHz flux density.  The SANE models, by contrast, exhibit relatively steady accretion through the equatorial plane.

The failure of nearly all fiducial models to match the light curve variability is interesting.
It may signal the presence of extended, slowly varying structure that is resolved out by EHT, or it may signal that future models need to incorporate collisionless effects (potentially modeled as viscosity and conductivity) or a more sophisticated treatment of electron thermodynamics including cooling. In addition, different initial geometries and polarities of the magnetic field could lead to more slowly varying structures \citep{2021arXiv211103689N}. If when combined these effects were to reduce \mi{3} by 30\% then many MAD models would be consistent with the data.

None of the fiducial models survive the full gauntlet of 11 constraints.
If we set aside both variability constraints, however, then there 2 fiducial models that pass the remaining 9 constraints in all simulation pipelines (a few more survive in one model set but not the other).
These models in the ``best-bet region'' are strongly magnetized (MAD) and have $\Rh = 160$, positive spin, and low inclination, with ($\abh, i$) = (0.5, 30\degree) and (0.94, 30\degree).
They have accretion rates $\dot{M} = (5.2$--$9.5)\times 10^{-9}\msun\yr^{-1}$, which are consistent with earlier estimates and overlap with accretion rates in wind-fed models, $\sim 10^{-8} \msun\yr^{-1}$ \citep{2020ApJ...896L...6R}.
The $\abh = 0.5$ MAD with $i=30\degree$ model is presented in Figure~\ref{fig:bestbet_imgs_snapshot} as snapshots and in Figure~\ref{fig:bestbet_imgs} as a weighted average, a convolved average, and a reconstructed average image from synthetic data.
One of the reconstructed average images from the 2017 EHT observations is shown in the rightmost panel in Figure~\ref{fig:bestbet_imgs} for comparison.

We produced synthetic SEDs, and therefore bolometric luminosities $L_\mathrm{bol}$, for all fiducial models.
Typically $L_\mathrm{bol}$ is dominated by a synchrotron bump in the submm and for the best-bet region is $(6.8$--$9.8)\times10^{35}\ergps$; the corresponding radiative efficiency $L_\mathrm{bol}/(\dot{M} c^2)$ is $(1.3$--$3.0)\times 10^{-3}$.
The maximum radiative efficiency over the entire fiducial model set is 0.08 (for a MAD, $\abh = 0.94$, $\Rh = 1$ model), which is necessary but not sufficient to justify our neglect of radiative cooling in the GRMHD evolution.

All our fiducial models produce bipolar outflows, and for each we measured the outflow power $P_\mathrm{out}$, defined in Section~\ref{sec:discussions}.
Consistent with earlier work we find that outflow power is higher for strongly magnetized (MAD) models than for comparable weakly magnetized (SANE) models, and increases by more than an order of magnitude from $\abh = 0$ to $|\abh| = 0.94$.
For models in the best-bet region, $P_\mathrm{out} = (1.3$--$4.8) \times 10^{38}\ergps$, corresponding to an outflow efficiency $P_\mathrm{out} /(\dot{M} c^2)$ of 0.25--1.6.
Such large outflow efficiency is only possible if energy is extracted from a spinning black hole via the mechanism proposed by  \cite{1977MNRAS.179..433B}.
It is an open question how these powerful outflows might interact with incoming gas in a self-consistent accretion model that follows plasma over a larger range in radius than our fiducial models.
It is also an open question whether the outflow power could be detected in the dense but crowded galactic center environment.
Notice that this outflow luminosity is comparable to the spindown luminosity of the Crab pulsar.

All fiducial models assume a particular parameterization for the electron distribution function (the $\Rh$ prescription), use a common initial setup (a magnetized torus), and assume the black hole spin vector and torus orbital angular momentum are aligned or anti-aligned.
To partially control for the errors introduced by these assumptions we have included a set of exploratory models.  These include several eDF prescriptions, a wind-fed model that tracks accretion from stellar winds down to the scale of the horizon, and tilted disk models in which the black hole spin and torus angular momentum are misaligned.

Our non-thermal models are remarkably similar to their thermal counterparts.
For the limited set of non-thermal eDF prescriptions we consider here the 230\GHz image structure differs very little for models in which the non-thermal electrons are introduced mainly in the jet.  The 230\GHz variability is not detectably different than corresponding thermal models.\footnote{The non-thermal models are imaged over $5\times 10^3\tg$, so constraints on $\mi{3}$ are weaker than for the fiducial models, which are imaged for 3 times as long.}
The 86\GHz size and flux density, which are the most restrictive non-EHT constraints, are not detectably affected by the addition of non-thermal electrons for most non-thermal models (except $\kappa = 5$ models).
Nonthermal electrons consistently increase the 2.2\um flux density over similar thermal models, however.
Accelerating even a small fraction of the electron population into a non-thermal tail risks overproducing 2.2\um emission.
The 2.2\um (and submm through mid-IR) flux density therefore provides the strongest eDF constraints.
Future EHT analyses would benefit from incorporating submm constraints \citep[e.g.,][]{2019ApJ...881L...2B} and, because model submm SEDs are highly variable, the submm and 2.2\um data should be as close to simultaneous as possible.

The stellar wind-fed models of \citet{2020ApJ...896L...6R} feature the best-motivated treatment of boundary and initial conditions for \sgra models.
They differ from our torus-initialized fiducial models in that they follow plasma from its ejection from stars on known orbits down to the event horizon.
We have imaged these models using an $\Rh$ prescription for the electron temperature, with $\Rh$ adjusted in the otherwise parameter-free models to produce the correct time-averaged 230\GHz flux density.
The two models considered here, both with $\abh = 0$, fail the 86\GHz flux, \mring width, and $\mi{3}$ constraints.
This does {\em not} imply that wind fed models are ruled out; they clearly merit further investigation with longer integrations over a broader range of eDFs and $\abh$.

In general, black hole accretion flows can be tilted in the sense that the orbital angular momentum of the disk and the spin angular momentum of the hole are misaligned.
Tilted disks have not until now been included in EHT analyses because
\emph{i}) it is conceivable that accretion flows align either by consistently oriented long term accretion or by some analog of the Bardeen-Petterson effect \citep{1975ApJ...195L..65B}; and
\emph{ii}) the tilted disk parameter space is larger than the aligned disk parameter space by two dimensions: the tilt angle and the longitude of the observer.
We considered models with tilt $30\degree$ and $60\degree$, observed at a single longitude.
The integrations were too short ($\no{3000}\,\tg$) to provide strong constraints on tilt, but we find that the \mring width test is particularly sensitive to tilt and rejects a progressively larger fraction of the models as tilt increases at the single observing longitude studied here.
Tilted models clearly merit further investigation.

Our fiducial models and variable $\kappa$ non-thermal models have been run with independent GRMHD codes and imaged with independent radiative transfer codes.
The outcomes are largely consistent (see Appendix~\ref{app:numerical} for details).
The code comparisons were valuable and helped identify multiple issues in the independent simulation sets.
The consistency between codes is remarkable given the complexity of the modeling process and the scope for error.
Tracking down the remaining discrepancies (for example, in the 2.2\um flux density) and developing a quantitative error budget is an essential but difficult task for the future.

\begin{acknowledgments}

This paper is dedicated to the memory of John F. Hawley, whose pioneering work on black hole accretion flows made this paper possible.  We are grateful to an anonymous referee whose comments significantly improved this paper.

The Event Horizon Telescope Collaboration thanks the following
organizations and programs: 
National Science Foundation (awards OISE-1743747, AST-1816420, AST-1716536, AST-1440254, AST-1935980);
the Black Hole Initiative, which is funded by grants from the John Templeton Foundation and the Gordon 
and Betty Moore Foundation (although the opinions expressed in this work are those of the author(s) 
and do not necessarily reflect the views of these Foundations);
NASA Hubble Fellowship grant
HST-HF2-51431.001-A awarded by the Space Telescope
Science Institute, which is operated by the Association
of Universities for Research in Astronomy, Inc.,
for NASA, under contract NAS5-26555;
the Academy
of Finland (projects 274477, 284495, 312496, 315721); the Agencia Nacional de Investigación 
y Desarrollo (ANID), Chile via NCN$19\_058$ (TITANs) and Fondecyt 1221421, the Alexander
von Humboldt Stiftung; an Alfred P. Sloan Research Fellowship;
Allegro, the European ALMA Regional Centre node in the Netherlands, the NL astronomy
research network NOVA and the astronomy institutes of the University of Amsterdam, Leiden University and Radboud University;
the Institute for Advanced Study;
the China Scholarship
Council;  Consejo
Nacional de Ciencia y Tecnolog\'{\i}a (CONACYT,
Mexico, projects  U0004-246083, U0004-259839, F0003-272050, M0037-279006, F0003-281692,
104497, 275201, 263356);
the Delaney Family via the Delaney Family John A.
Wheeler Chair at Perimeter Institute; Dirección General
de Asuntos del Personal Académico-—Universidad
Nacional Autónoma de México (DGAPA-—UNAM,
projects IN112417 and IN112820); the European Research Council Synergy
Grant ``BlackHoleCam: Imaging the Event Horizon
of Black Holes" (grant 610058); the Generalitat
Valenciana postdoctoral grant APOSTD/2018/177 and
GenT Program (project CIDEGENT/2018/021); MICINN Research Project PID2019-108995GB-C22;
the European Research Council for advanced grant `JETSET: Launching, propagation and 
emission of relativistic 
jets from binary mergers and across mass scales' (Grant No. 884631); 
the Istituto Nazionale di Fisica
Nucleare (INFN) sezione di Napoli, iniziative specifiche
TEONGRAV; the two Dutch National Supercomputers, Cartesius and Snellius  
(NWO Grant 2021.013);
the International Max Planck Research
School for Astronomy and Astrophysics at the
Universities of Bonn and Cologne; 
DFG research grant ``Jet physics on horizon scales and beyond'' (Grant No. FR 4069/2- 1);
Joint Princeton/Flatiron and Joint Columbia/Flatiron Postdoctoral Fellowships, 
research at the Flatiron Institute is supported by the Simons Foundation; 
the Japanese Government (Monbukagakusho:
MEXT) Scholarship; the Japan Society for
the Promotion of Science (JSPS) Grant-in-Aid for JSPS
Research Fellowship (JP17J08829); the Key Research
Program of Frontier Sciences, Chinese Academy of
Sciences (CAS, grants QYZDJ-SSW-SLH057, QYZDJSSW-SYS008, ZDBS-LY-SLH011); 
the Leverhulme Trust Early Career Research
Fellowship; the Max-Planck-Gesellschaft (MPG);
the Max Planck Partner Group of the MPG and the
CAS; the MEXT/JSPS KAKENHI (grants 18KK0090, JP21H01137,
JP18H03721, JP18K13594, 18K03709, JP19K14761, 18H01245, 25120007); the Malaysian Fundamental Research Grant Scheme (FRGS) FRGS/1/2019/STG02/UM/02/6; the MIT International Science
and Technology Initiatives (MISTI) Funds; 
the Ministry of Science and Technology (MOST) of Taiwan (103-2119-M-001-010-MY2, 105-2112-M-001-025-MY3, 105-2119-M-001-042, 106-2112-M-001-011, 106-2119-M-001-013, 106-2119-M-001-027, 106-2923-M-001-005, 107-2119-M-001-017, 107-2119-M-001-020, 107-2119-M-001-041, 107-2119-M-110-005, 107-2923-M-001-009, 108-2112-M-001-048, 108-2112-M-001-051, 108-2923-M-001-002, 109-2112-M-001-025, 109-2124-M-001-005, 109-2923-M-001-001, 110-2112-M-003-007-MY2, 110-2112-M-001-033, 110-2124-M-001-007, and 110-2923-M-001-001);
the Ministry of Education (MoE) of Taiwan Yushan Young Scholar Program;
the Physics Division, National Center for Theoretical Sciences of Taiwan;
the National Aeronautics and
Space Administration (NASA, Fermi Guest Investigator
grant 80NSSC20K1567, NASA Astrophysics Theory Program grant 80NSSC20K0527, NASA NuSTAR award 
80NSSC20K0645); 
the National
Institute of Natural Sciences (NINS) of Japan; the National
Key Research and Development Program of China
(grant 2016YFA0400704, 2017YFA0402703, 2016YFA0400702); the National
Science Foundation (NSF, grants AST-0096454,
AST-0352953, AST-0521233, AST-0705062, AST-0905844, AST-0922984, AST-1126433, AST-1140030,
DGE-1144085, AST-1207704, AST-1207730, AST-1207752, MRI-1228509, OPP-1248097, AST-1310896,  
AST-1555365, AST-1614868, AST-1615796, AST-1715061, AST-1716327,  AST-2034306); 
the Natural Science Foundation of China (grants 11650110427, 10625314, 11721303, 11725312, 11873028, 11933007, 11991052, 11991053, 12192220, 12192223);
NWO grant number OCENW.KLEIN.113; a 
fellowship of China Postdoctoral Science Foundation (2020M671266); the Natural
Sciences and Engineering Research Council of
Canada (NSERC, including a Discovery Grant and
the NSERC Alexander Graham Bell Canada Graduate
Scholarships-Doctoral Program); the National Youth
Thousand Talents Program of China; the National Research
Foundation of Korea (the Global PhD Fellowship
Grant: grants NRF-2015H1A2A1033752, the Korea Research Fellowship Program:
NRF-2015H1D3A1066561, Basic Research Support Grant 2019R1F1A1059721, 2022R1C1C1005255); the Dutch Organization
for Scientific Research (NWO) VICI award
(grant 639.043.513) and Spinoza Prize SPI 78-409; the YCAA Prize Postdoctoral Fellowship.
LM gratefully acknowledges support from an NSF Astronomy and Astrophysics Postdoctoral Fellowship under award no. AST-1903847. TK is supported by MEXT as "Program for Promoting Researches on the Supercomputer Fugaku" (Toward a unified view of the universe: from large scale structures to planets, JPMXP1020200109) and JICFuS. RPD and IN acknowledge funding by the South African Research Chairs Initiative, through the South African Radio Astronomy Observatory (SARAO, grant ID 77948),  which is a facility of the National Research Foundation (NRF), an agency of the Department of Science and Innovation (DSI) of South Africa.

We thank the Onsala Space Observatory
(OSO) national infrastructure, for the provisioning
of its facilities/observational support (OSO receives
funding through the Swedish Research Council under
grant 2017-00648);  the Perimeter Institute for Theoretical
Physics (research at Perimeter Institute is supported
by the Government of Canada through the Department
of Innovation, Science and Economic Development
and by the Province of Ontario through the
Ministry of Research, Innovation and Science); the Spanish Ministerio de Ciencia e Innovación (grants PGC2018-098915-B-C21, AYA2016-80889-P,
PID2019-108995GB-C21, PID2020-117404GB-C21); 
the University of Pretoria for financial aid in the provision of the new 
Cluster Server nodes and SuperMicro (USA) for a SEEDING GRANT approved towards these 
nodes in 2020;
the State
Agency for Research of the Spanish MCIU through
the ``Center of Excellence Severo Ochoa'' award for
the Instituto de Astrofísica de Andalucía (SEV-2017-
0709); the Toray Science Foundation; the Consejería de Economía, Conocimiento, 
Empresas y Universidad 
of the Junta de Andalucía (grant P18-FR-1769), the Consejo Superior de Investigaciones 
Científicas (grant 2019AEP112);
the M2FINDERS project which has received funding by the European Research Council (ERC) under 
the European Union’s Horizon 2020 Research and Innovation Programme (grant agreement No 101018682);
the US Department
of Energy (USDOE) through the Los Alamos National
Laboratory (operated by Triad National Security,
LLC, for the National Nuclear Security Administration
of the USDOE (Contract 89233218CNA000001);
 the European Union’s Horizon 2020
research and innovation programme under grant agreement
No 730562 RadioNet;
Shanghai Pilot Program for Basic Research, Chinese Academy of Science, 
Shanghai Branch (JCYJ-SHFY-2021-013);
ALMA North America Development
Fund; the Academia Sinica; Chandra DD7-18089X and TM6-
17006X; the GenT Program (Generalitat Valenciana)
Project CIDEGENT/2018/021. This work used the
Extreme Science and Engineering Discovery Environment
(XSEDE), supported by NSF grant ACI-1548562,
and CyVerse, supported by NSF grants DBI-0735191,
DBI-1265383, and DBI-1743442. XSEDE Stampede2 resource
at TACC was allocated through TG-AST170024
and TG-AST080026N. XSEDE JetStream resource at
PTI and TACC was allocated through AST170028.

The simulations were performed in part on the SuperMUC
cluster at the LRZ in Garching, on the
LOEWE cluster in CSC in Frankfurt, and on the
HazelHen cluster at the HLRS in Stuttgart. This
research was enabled in part by support provided
by Compute Ontario (http://computeontario.ca), Calcul
Quebec (http://www.calculquebec.ca) and Compute
Canada (http://www.computecanada.ca). 
CC acknowledges support from the Swedish Research Council (VR).

We thank
the staff at the participating observatories, correlation
centers, and institutions for their enthusiastic support.
This paper makes use of the following ALMA data:
ADS/JAO.ALMA\#2016.1.01154.V. ALMA is a partnership
of the European Southern Observatory (ESO;
Europe, representing its member states), NSF, and
National Institutes of Natural Sciences of Japan, together
with National Research Council (Canada), Ministry
of Science and Technology (MOST; Taiwan),
Academia Sinica Institute of Astronomy and Astrophysics
(ASIAA; Taiwan), and Korea Astronomy and
Space Science Institute (KASI; Republic of Korea), in
cooperation with the Republic of Chile. The Joint
ALMA Observatory is operated by ESO, Associated
Universities, Inc. (AUI)/NRAO, and the National Astronomical
Observatory of Japan (NAOJ). The NRAO
is a facility of the NSF operated under cooperative agreement
by AUI.
This research used resources of the Oak Ridge Leadership Computing Facility at the Oak Ridge National
Laboratory, which is supported by the Office of Science of the U.S. Department of Energy under Contract
No. DE-AC05-00OR22725. We also thank the Center for Computational Astrophysics, National Astronomical Observatory of Japan.

Support for this work was also provided by the NASA Hubble Fellowship 
grant HST-HF2-51431.001-A awarded 
by the Space Telescope Science Institute, which is operated by the Association of Universities for 
Research in Astronomy, Inc., for NASA, under contract NAS5-26555.
HO and GM were supported by Virtual Institute of Accretion (VIA) postdoctoral fellowships from the Netherlands Research School for Astronomy (NOVA).
APEX is a collaboration between the
Max-Planck-Institut f{\"u}r Radioastronomie (Germany),
ESO, and the Onsala Space Observatory (Sweden). The
SMA is a joint project between the SAO and ASIAA
and is funded by the Smithsonian Institution and the
Academia Sinica. The JCMT is operated by the East
Asian Observatory on behalf of the NAOJ, ASIAA, and
KASI, as well as the Ministry of Finance of China, Chinese
Academy of Sciences, and the National Key Research and Development
Program (No. 2017YFA0402700) of China
and Natural Science Foundation of China grant 11873028.
Additional
funding support for the JCMT is provided by the Science
and Technologies Facility Council (UK) and participating
universities in the UK and Canada. 
Simulations were performed in part on the SuperMUC cluster at the LRZ in Garching, 
on the 
LOEWE cluster in CSC in Frankfurt, on the HazelHen cluster at the HLRS in Stuttgart, 
and on the Pi2.0 and Siyuan Mark-I at Shanghai Jiao Tong University.
The computer resources of the Finnish IT Center for Science (CSC) and the Finnish Computing 
Competence Infrastructure (FCCI) project are acknowledged.
JO was supported by Basic Science Research Program through the National Research
Foundation of Korea(NRF) funded by the Ministry of Education(NRF-2021R1A6A3A01086420;
2022R1C1C1005255).
We thank Martin Shepherd for the addition of extra features in the Difmap software 
that were used for the CLEAN imaging results presented in this paper.
The computing cluster of Shanghai VLBI correlator supported by the Special Fund 
for Astronomy from the Ministry of Finance in China is acknowledged.
This work was supported by the Brain Pool Program through the National Research
Foundation 
of Korea (NRF) funded by the Ministry of Science and ICT (019H1D3A1A01102564).
This research is part of the Frontera computing project at the Texas Advanced 
Computing Center through the Frontera Large-Scale Community Partnerships allocation
AST20023. Frontera is made possible by National Science Foundation award OAC-1818253.
This research was carried out using resources provided by the Open Science Grid, 
which is supported by the National Science Foundation and the U.S. Department of 
Energy Office of Science.

The LMT is a project operated by the Instituto Nacional
de Astrófisica, Óptica, y Electrónica (Mexico) and the
University of Massachusetts at Amherst (USA). The
IRAM 30-m telescope on Pico Veleta, Spain is operated
by IRAM and supported by CNRS (Centre National de
la Recherche Scientifique, France), MPG (Max-Planck-Gesellschaft, Germany) 
and IGN (Instituto Geográfico
Nacional, Spain). The SMT is operated by the Arizona
Radio Observatory, a part of the Steward Observatory
of the University of Arizona, with financial support of
operations from the State of Arizona and financial support
for instrumentation development from the NSF.
Support for SPT participation in the EHT is provided by the National Science Foundation through award OPP-1852617 
to the University of Chicago. Partial support is also 
provided by the Kavli Institute of Cosmological Physics at the University of Chicago. The SPT hydrogen maser was 
provided on loan from the GLT, courtesy of ASIAA.
Support for this work was provided by NASA through the NASA Hubble Fellowship grant
\#HST--HF2--51494.001 awarded by the Space Telescope Science Institute, which is operated 
by the Association of Universities for Research in Astronomy, Inc., for NASA, 
under contract NAS5--26555.
Jongho Park acknowledges financial support through the EACOA Fellowship awarded by the East Asia Core
Observatories Association, which consists of the Academia Sinica Institute of Astronomy and
Astrophysics, the National Astronomical Observatory of Japan, Center for Astronomical Mega-Science,
Chinese Academy of Sciences, and the Korea Astronomy and Space Science Institute.

The EHTC has
received generous donations of FPGA chips from Xilinx
Inc., under the Xilinx University Program. The EHTC
has benefited from technology shared under open-source
license by the Collaboration for Astronomy Signal Processing
and Electronics Research (CASPER). The EHT
project is grateful to T4Science and Microsemi for their
assistance with Hydrogen Masers. This research has
made use of NASA’s Astrophysics Data System. We
gratefully acknowledge the support provided by the extended
staff of the ALMA, both from the inception of
the ALMA Phasing Project through the observational
campaigns of 2017 and 2018. We would like to thank
A. Deller and W. Brisken for EHT-specific support with
the use of DiFX. We acknowledge the significance that
Maunakea, where the SMA and JCMT EHT stations
are located, has for the indigenous Hawaiian people.


\end{acknowledgments}

\facility{EHT, the Global mm VLBI Array, Atacama Large Millimeter Array, Chandra X-ray Observatory;
Frontera supercomputer (Texas Advanced Computing Center), Puma Supercomputer (Arizona), Open Science Grid, CyVerse}

\software{\bhac, \bhoss, \difmap \citep{1997ASPC..125...77S}, \dmc, \ehtim \citep{2019zndo...2614016C}, \foci,
  \grmonty, \hallmark, \hamr, \ipole, \kharma, \koral, \mockservation,
  \smili, \themis, \vida, VisIt, numpy \citep{vanderwalt2011}, scipy \citep{oliphant2007}, matplotlib \citep{hunter2007}}

\appendix

\section{Numerical Methods}\label{app:numerical}

\subsection{Consistency of Radiative Transfer Simulations}\label{app:radtrans}

Two studies have been undertaken within the EHT Collaboration to evaluate the consistency of radiative transfer codes.

The first, \citet{2020ApJ...897..148G}, evaluated the consistency between general relativistic ray-traced radiative transfer (GRRT) codes when tracing geodesics and when integrating the unpolarized radiative transfer equation.  \citeauthor{2020ApJ...897..148G} compares \bhoss and \ipole, which are the two transfer codes used in this paper, and  also compares to {\tt grtrans}, {\tt raptor}, {\tt odyssey}, {\tt gray2} and {\tt raikou}.  Code consistency was found to be excellent, with sub-percent level variations between codes when run with standard numerical parameters, i.e. without accuracy parameters tuned for consistency.

The second, \citet{Prather_et_al_2022}, evaluates code performance when imaging GRMHD simulation output and when integrating the equations of polarized radiative transfer.  \citeauthor{Prather_et_al_2022} includes \ipole, {\tt grtrans}, {\tt odyssey}, and {\tt raptor}.  Code consistency was also found to be excellent.

Uncertainty in the radiative transfer calculation is therefore unlikely to contribute significantly to the model error budget.

\subsection{GRMHD Simulations Consistency and Convergence}\label{app:resolution_study}

As evident in Table~\ref{tab:GRMHDmodels} the thermal models have been calculated for an identical parameter space from two different codes, namely \kharma and \bhac for the GRMHD simulations and \ipole and \bhoss codes for the GRRT calculations.
This allows us to perform an in depth comparison between the different numerical methods used in this work in addition to the EHTC code comparison projects \citep{2019ApJS..243...26P,2020ApJ...897..148G}.

\begin{figure*}
  \centering
  \includegraphics[width=0.8\textwidth]{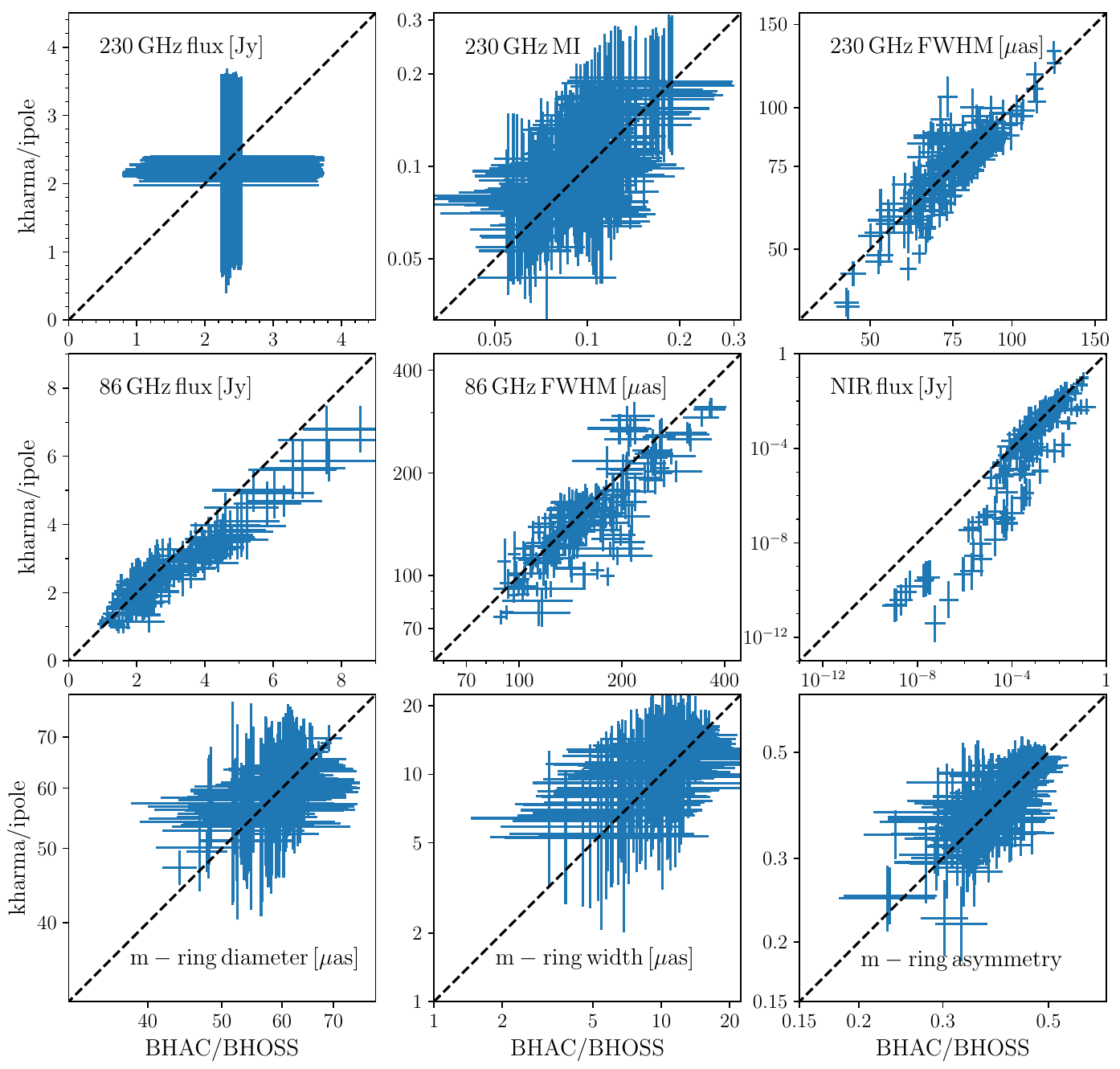}
  \caption{Correlation between \bhac and \kharma models for 9 model constraints.
The horizontal axis is the constraint value from \bhac/\bhoss, and the vertical axis shows the constraint value from \kharma/\ipole.
Each point corresponds to a single model, with the width of the distribution shown by the error bars.
See text for details.}
  \label{fig:modelcorrelation}
\end{figure*}

In Figure \ref{fig:modelcorrelation} we show the correlation between the thermal \kharma and \bhac models for constraints where we have predictions from both models.
The top row shows from left to right the 230\,GHz flux density, \mi{3}, and the 230\GHz image size obtained from image moments.
Since we normalize the 230\GHz images to an average flux of 2.4\,Jy within a time window of 5000\,M (28.5 h for \sgra), the scatter around this value is small.
The deviation from an ideal correlation reflects the precision and number of GRMHD snapshots included during normalization procedure.

The correlation in $\mi{3}$ spreads over $\Delta \mi{3}=0.75$, which serves as a measure of intra-code (e.g., MAD vs. SANE accretion) and inter-code (\bhac vs. \kharma) differences.
Despite these differences the models show a strong correlation throughout the investigated models and parameter space.

We also find a strong correlation between models and codes for the image size computed from image moments, i.e. second moments analysis.

The middle row presents the correlation plots for the 86\GHz flux density (left), the 86\,GHz image size using second moments (middle), and the NIR flux (right).
The 86\GHz flux and 86\GHz image size exhibit a shift toward larger values for the \bhac models.
This difference can be explained by the larger field of view used for the \bhac models at 86\GHz during the radiative transfer calculations.
Thus, more extended structure and therefore a larger total flux is included in the \bhac models.
This affects mainly models with large inclinations $i\geq70^\degree$ and jet dominated emission models ($\Rh \geq 40$).

The NIR fluxes show a tight correlation over four orders of magnitude and systematically larger flux for the \bhac models for low NIR fluxes ($\log_{10}({\rm NIR}/{\rm Jy}) < -7$).
These fluxes are far below the NIR constraints of $\sim 1\,\mathrm{mJy}$, and therefore they do not affect the passing or failing of the models.
In the thermal models the NIR flux is generated from the tail of the electron distribution function and is thus very sensitive to the electron temperature.
Small differences in the distribution and value of the electron temperature between the two codes explain the observed de-correlation at very low NIR flux.

The correlation between models for the m-ring parameters is presented in the third row of Figure~\ref{fig:modelcorrelation}.
The correlation of the diameter of the m-ring is plotted in the left panel.
The spread covers nearly the same extent as the 230\GHz image size (top row, right panel) however the scatter in the correlation is larger.
The same is true for the width of the m-ring (middle panel in the last row of Figure~\ref{fig:modelcorrelation}).
Compared to the diameter and width of the m-ring, the asymmetry of the m-ring is less correlated (right panel).
Notice that horizontal and vertical limits in the asymmetries occur because the parameter hits the boundary of the allowed range, which is $0.5$.

The smaller correlation of the m-ring parameters as compared to the other parameters presented in Figure~\ref{fig:modelcorrelation} is a consequence of the noisy nature of the m-ring fits.
Still, the distributions are quite symmetric under reflection across the diagonal, so the models are at least not biased with respect to each other.
Notice also that these plots do not capture all the information that is contained in the distribution of m-ring parameters, just the central value.

We are somewhat surprised by the strength of the correlations seen in Figure~\ref{fig:modelcorrelation}.
The range of each constraint is substantially larger than the width of the correlation, so the variations between models are real, detectable, and reproducible with independent codes.
The question of the origin of the systematic offsets between models for some constraints (for example, in the NIR) is interesting but beyond the scope of this paper.

\section{Variability of GRMHD models}\label{app:variability}

Nearly all models fail to recover the variability of \sgra in 230\GHz flux density as measured by \mi{3}.
In this Appendix we discuss and dismiss four possible causes for this variability excess.

\subsection{Effect of resolution}

For the 50\% change in resolution considered in the comparison shown in the preceding Appendix (between \kharma and \bhac simulations) we find no
evidence for systematic changes in \mi{3} with resolution.  This is not a large
range in resolution, however, and much higher resolution simulations \citep{2022ApJ...924L..32R, 2020ApJ...900..100R, 2021arXiv211103689N}
show the emergence of qualitatively
new structures (plasmoids) in current sheets that could affect 230GHz variability.
A deeper study of the resolution dependence of variability is clearly warranted
but is beyond the scope of this paper.

\subsection{Simulations Duration}\label{app:narayan}

\begin{figure}
  \centering
  \includegraphics[width=\columnwidth]{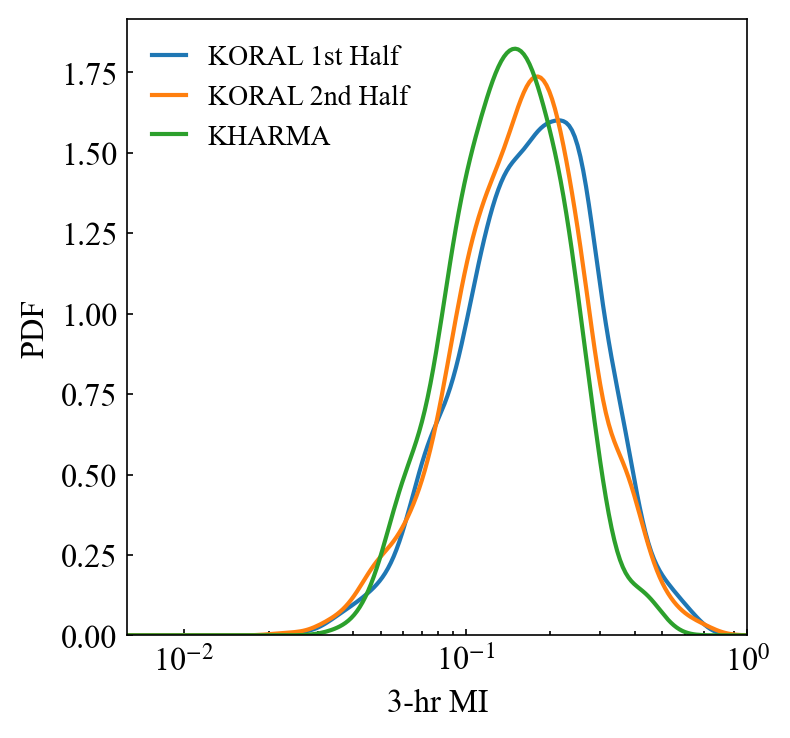}
  \caption{Distribution of \mi{3} from \koral models, divided between the first and second half of the simulation, and from the fiducial \kharma models. We choose the models that are at similar points in the parameter space, limiting the comparison to spin -0.9, -0.5, 0.0 and 0.9 over all inclinations for the \koral models and MAD, \Rh 40, spin -0.94, -0.5, 0.0, and 0.94 for the \kharma models.}
  \label{fig:koral_MI}
\end{figure}

The fiducial models are evolved for $\sim \no{30000}\,\tg$.
Figure ~\ref{fig:koral_MI} compares $\mi{3}$ distributions from a fiducial \kharma simulation to a \koral model with similar parameters that was evolved and imaged for approximately three times longer.
The $\mi{3}$ distributions have similar mean and standard deviation regardless of time interval chosen for comparison.

\subsection{Effect of \texorpdfstring{$\Rl$}{Rlow}}

The $\Rh$ prescription (Equation \ref{eq:rhigh_prescription}) has three free parameters: $\Rh$, $\Rl$ and $\beta_\mathrm{crit}$.
In the main text the $\Rh$ parameter is varied while $\Rl$ and $\beta_\mathrm{crit}$ are set to unity.

The $\Rl$ parameter determines the electron temperature in regions of low $\beta$, i.e. in and near the funnel.
Increasing $\Rl$ mimics rapid electron cooling.
We are particularly interested in the effect of increasing $\Rl$ on \mi{3}.

\begin{figure*}
  \centering
  \includegraphics[width=0.95\textwidth]{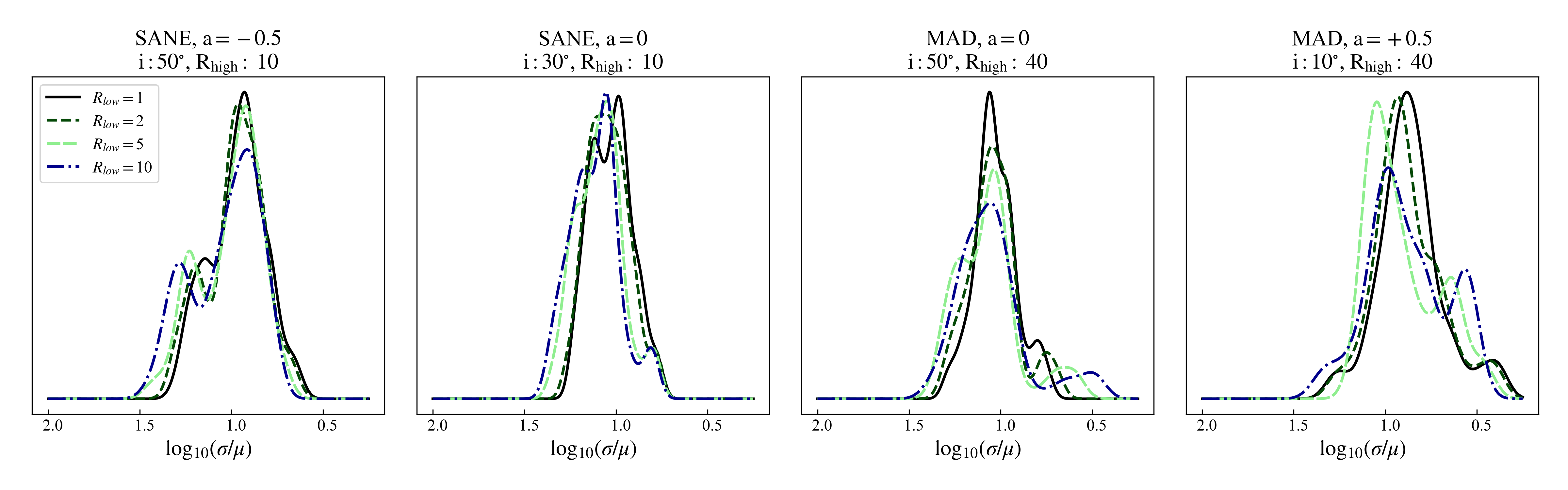}
  \caption{Modulation index computed over 3 hour intervals $\mi{3}$ for a subset of the thermal models (\kharma datasets).
For this analysis, we considered the 25,000--30,000$\tg$ time interval.}
  \label{fig:mi_rlow}
\end{figure*}

Figure~\ref{fig:mi_rlow} shows the \mi{3} distribution for a set of four \kharma models with four values of $\Rl$ (1,2,5, and 10).
Evidently the \mi{3} distribution does not exhibit a clear trend with $\Rh$, and is still inconsistent with the observed distribution even at $\Rl = 10$.

\subsection{Effect of self-consistent electron heating}

\citealt{2015MNRAS.454.1848R} provide a formulation to model electron thermodynamics during the fluid evolution.
Numerical dissipation at the grid scale sources entropy generation and is used to heat the electrons based on a microphysical, sub-grid heating prescription.
Local fluid and electromagnetic variables are used to compute the electron entropy which, along with the ideal gas equation of state, can be converted into a temperature $\Theta_{e}$.
This approach allows computing the electron temperature at each timestep of the simulation rather than post-processing, as is done in the $\Rh$ and Critical-$\beta$ prescriptions.

We consider three sub-grid heating models that prescribe the partition of dissipated energy into electrons and ions.
\cite{2010MNRAS.409L.104H} computed the ratio of ion-to-electron heating due to dissipation of Alfv\'enic turbulent cascade, while \cite{10.1093/mnras/stx2530} and \cite{Rowan_2017} considered magnetic reconnection as the source of energy dissipation at sub-grid scales.
These studies provide approximate fitting formulae for the ion-to-electron heating rate $Q_{i}/Q_{e}$ based on local ion-to-electron temperature ratio $T_{i}/T_{e}$ and local magnetic field strength---parameterized by $\sigma$ or $\beta$.

We use a subset of the simulations analyzed in \citealt{2020MNRAS.494.4168D}.
These include MAD and SANE accretion flows at spins, $\abh = 0,+1/2,+15/16$.
We compute the 3 hour modulation index $\mi{3}$, over the time interval 5,000--10,000$GM/c^{3}$.
The average $\mi{3}$ values are comparable to similar $\Rh$ models, with SANE reconnection models exhibiting a slightly reduced variability as compared to the corresponding turbulent heating models.
However, the $\mi{3}$ distribution is still inconsistent with the historical data.

\subsection{Effect of fluid adiabatic index}

We expect the ions and electrons in hot accretion flows to be thermally decoupled and the resulting plasma to be two-temperature \citep{1976ApJ...204..187S, Quataert_1998, 10.1093/mnras/stw3116, Ryan_2018, Chael2018}.
The electrons are relativistic and can be modeled as a fluid with an adiabatic index $\Gamma_{e}=4/3$, while the ions are nonrelativistic with adiabatic index $\Gamma_{i}=5/3$.

The adiabatic index of the fluid assumes a value between $\Gamma_{e}$ and $\Gamma_{i}$ dictated by the thermodynamics of the ions and electrons (cf. Figure 4 in \citealt{10.1093/mnras/stw3116}).
If the electrons and ions have equal temperature then in the relativistic electron/nonrelativistic ion regime the fluid adiabatic index is 13/9.

Our simulations are not fully consistent in their treatment of the adiabatic index.
All use a fixed $\Gamma_\mathrm{ad}$, but some set $\Gamma_\mathrm{ad} = 4/3$ while other use $13/9$ or $5/3$.

Two-temperature simulations can self-consistently evolve adiabatic indices of electrons and ions and compute the net fluid adiabatic index with contributions from both species \citep{10.1093/mnras/stw3116}.
These two-temperature simulations often show variation of the adiabatic index with polar angle, with the fluid energy dominated by hot electrons near the poles ($\Gamma = 4/3$) and by cooler ions and  electrons in the midplane ($\Gamma=5/3$).

We evaluate the effect of $\Gamma_\mathrm{ad}$ on light curve variability by comparing $\mi{3}$ for thermal, GRMHD simulations with varying  $\Gamma_\mathrm{ad}$.
This includes MAD models with $\Gamma_\mathrm{ad}=13/9$ (see Section~\ref{app:narayan} and  \citealt{2022MNRAS.511.3795N}) and SANE models with $\Gamma_\mathrm{ad}=5/3$.
The models exhibit light curve variability similar to the fiducial models and all have \mi{3} distributions that are inconsistent with the historical data.

\section{Pass/Fail Plots}\label{app:tables}

The full set of constraint results for the fiducial models is presented below in graphical form.

\begin{figure*}
  \centering
  \includegraphics[width=0.8\textwidth]{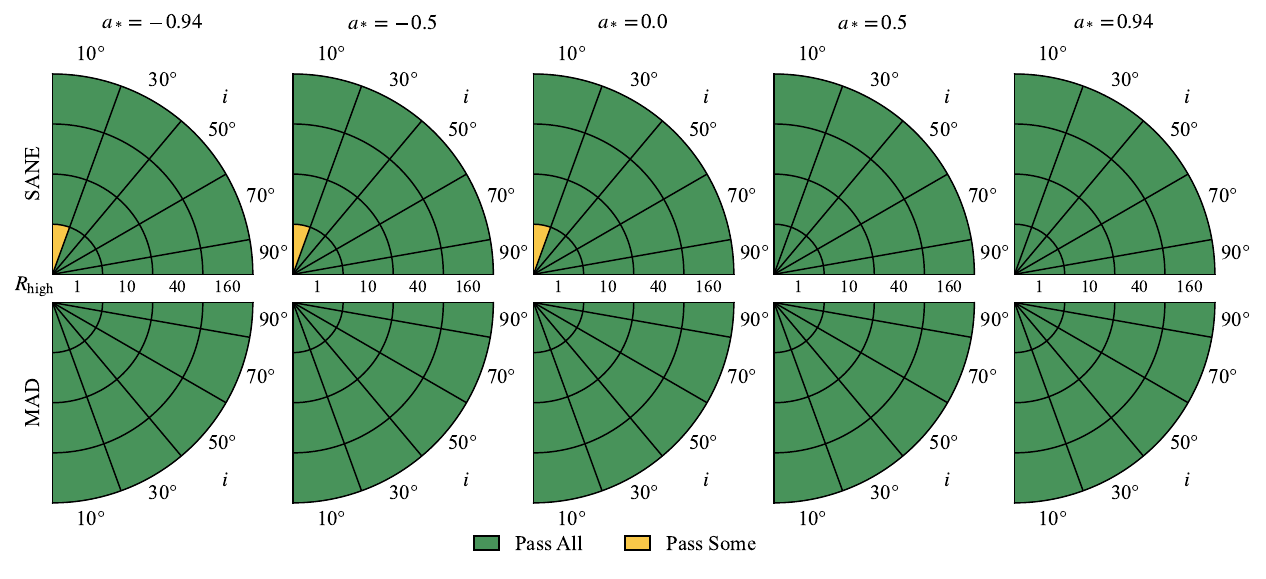}
  \caption{2nd Moment Constraint.  Green indicates that the \kharma, \bhac, and \hamr models pass, yellow that one or two of the fiducial models fail, and red that all three fail.}
  \label{fig:230GHz_size_pizza}
\end{figure*}

\begin{figure*}
  \centering
  \includegraphics[width=0.8\textwidth]{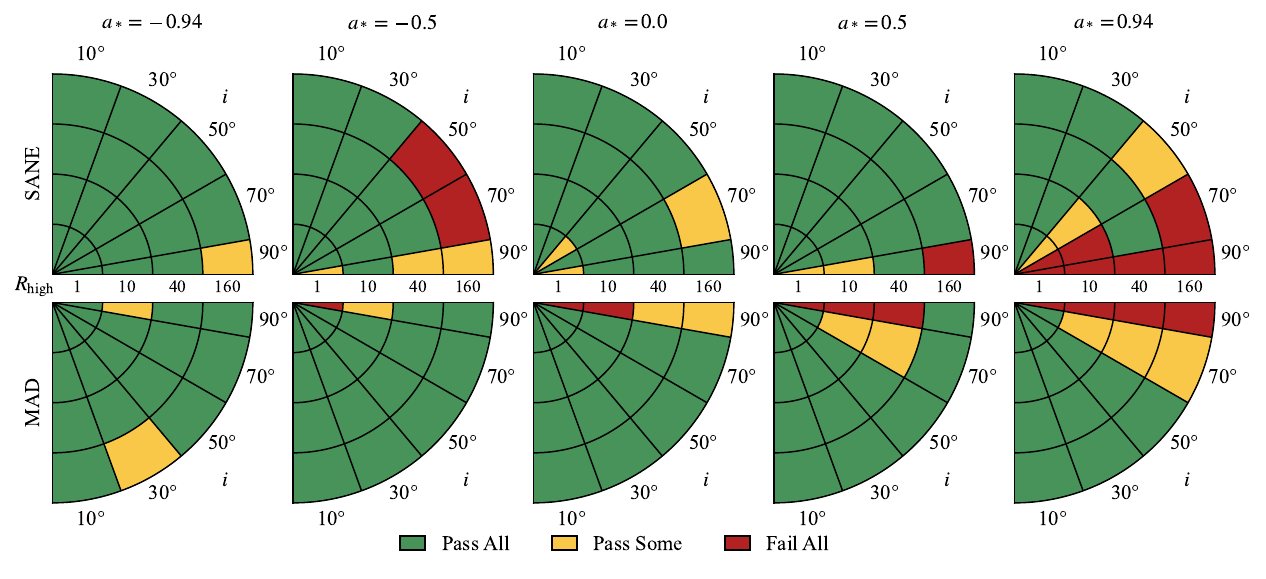}
  \caption{Null Location Constraint.  Green indicates that the \kharma, \bhac, and \hamr models pass, yellow that one or two of the fiducial models fail, and red that all three fail.}
  \label{fig:null_pizza}
\end{figure*}

\begin{figure*}
  \centering
  \includegraphics[width=0.8\textwidth]{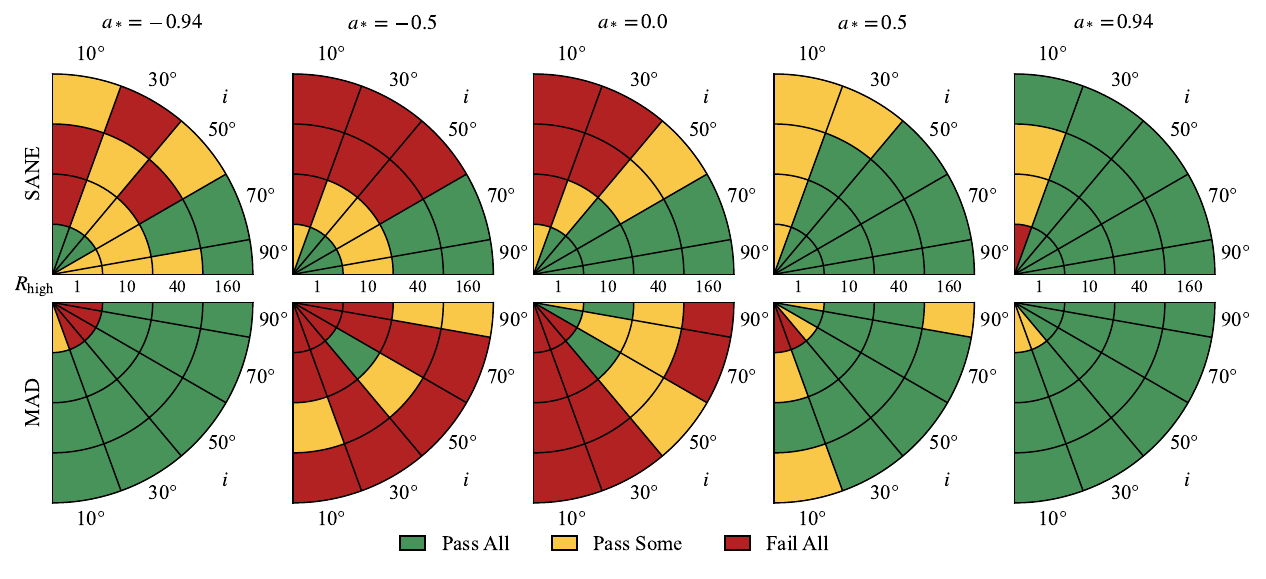}
  \caption{M-ring Diameter Constraints.  Green indicates that the \kharma, \bhac, and \hamr models pass, yellow that one or two of the fiducial models fail, and red that all three fail.}
  \label{fig:mring_diam_pizza}
\end{figure*}

\begin{figure*}
  \centering
  \includegraphics[width=0.8\textwidth]{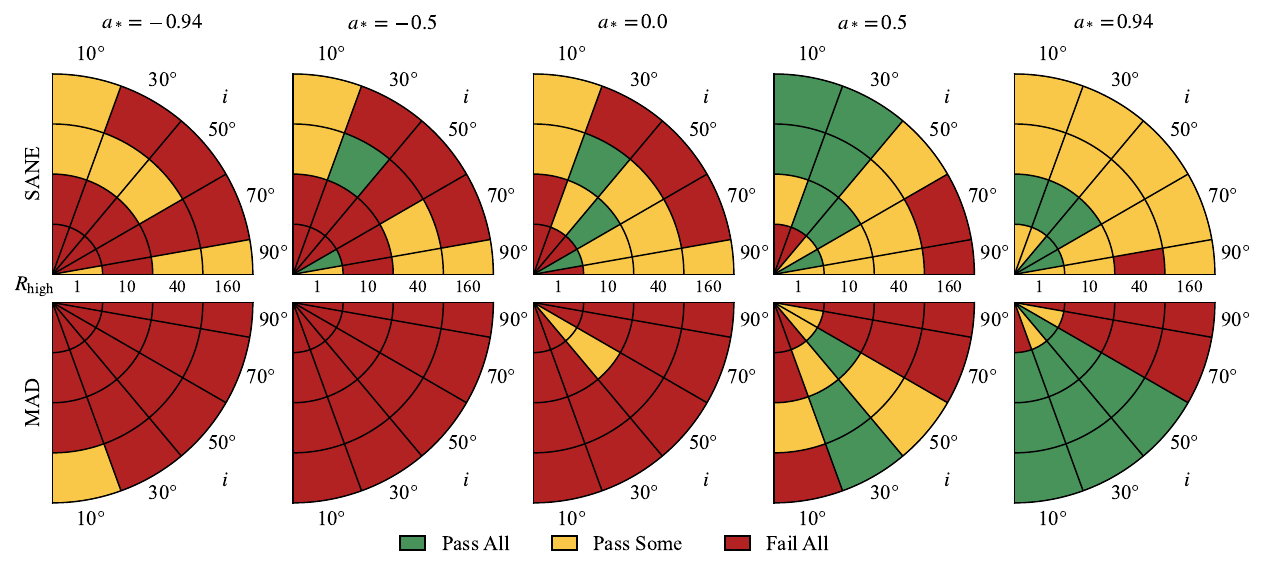}
  \caption{M-ring Width Constraints.  Green indicates that the \kharma, \bhac, and \hamr models pass, yellow that one or two of the fiducial models fail, and red that all three fail.}
  \label{fig:mring_width_pizza}
\end{figure*}

\begin{figure*}
  \centering
  \includegraphics[width=0.8\textwidth]{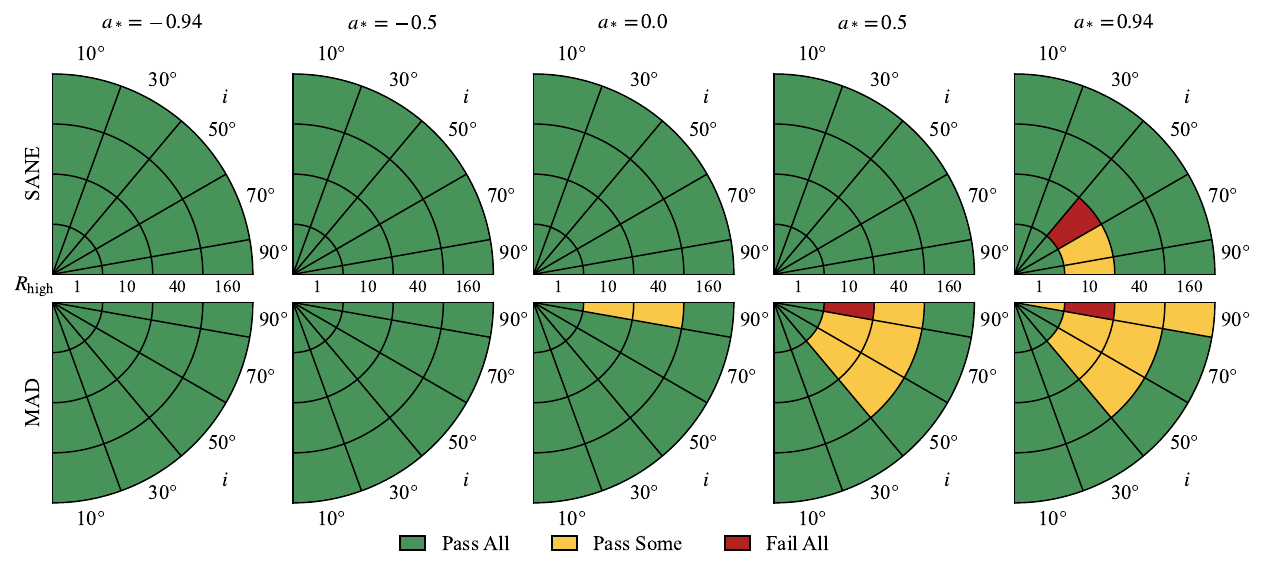}
  \caption{M-ring Asymmetry Constraints.  Green indicates that the \kharma, \bhac, and \hamr models pass, yellow that one or two of the fiducial models fail, and red that all three fail.}
  \label{fig:mring_asymm_pizza}
\end{figure*}

\begin{figure*}
  \centering
  \includegraphics[width=0.8\textwidth]{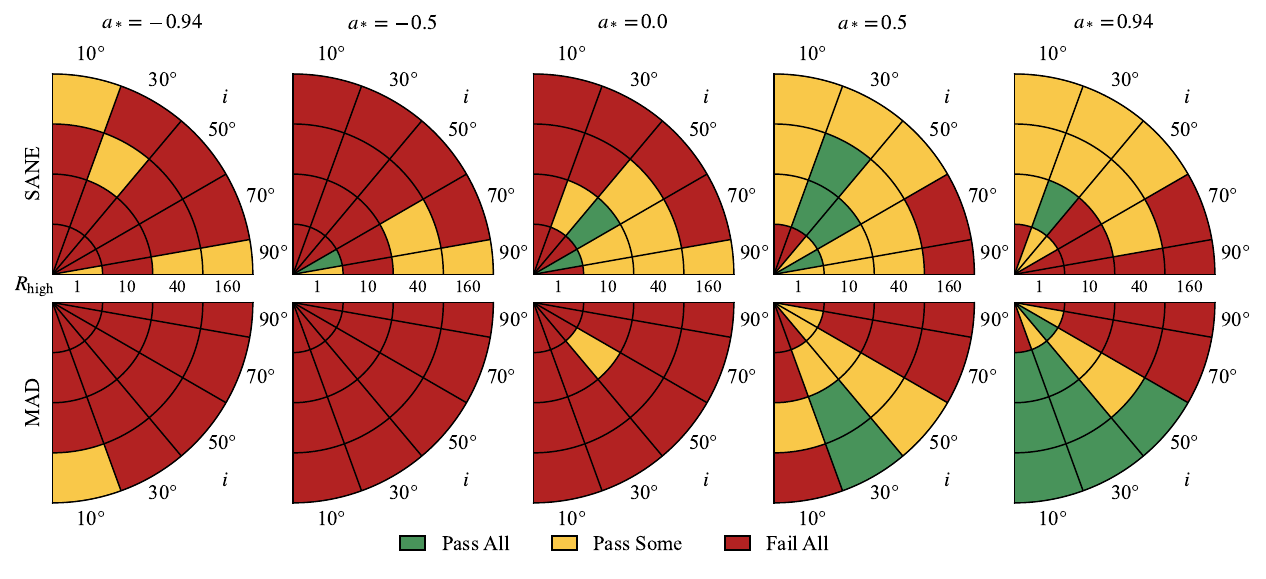}
  \caption{Combined EHT Constraints.  Green indicates that the \kharma, \bhac, and \hamr models pass, yellow that one or two of the fiducial models fail, and red that all three fail.}
  \label{fig:eht_comb_pizza}
\end{figure*}

﻿We start with the EHT constraints.
﻿Figure~\ref{fig:230GHz_size_pizza} shows the 230\GHz 2nd moment constraint and
Figure~\ref{fig:null_pizza} shows the null location constraint.
Figures~\ref{fig:mring_diam_pizza}--\ref{fig:mring_asymm_pizza} show m-ring diameter, width, and asymmetry constraints, respectively.
Figure~\ref{fig:eht_comb_pizza} combines all the EHT constraints listed above.

\begin{figure*}
  \centering
  \includegraphics[width=0.8\textwidth]{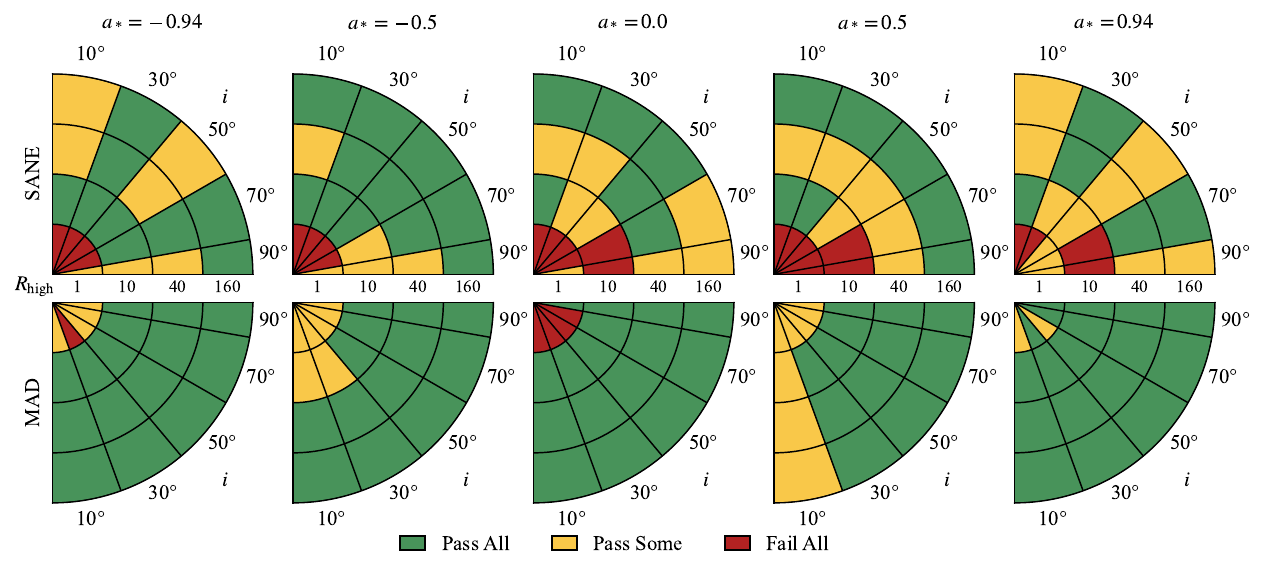}
  \caption{86GHz Flux Constraints.  Green indicates that the \kharma, \bhac, and \hamr models pass, yellow that one or two of the fiducial models fail, and red that all three fail.}
  \label{fig:86GHz_flux_pizza}
\end{figure*}

\begin{figure*}
  \centering
  \includegraphics[width=0.8\textwidth]{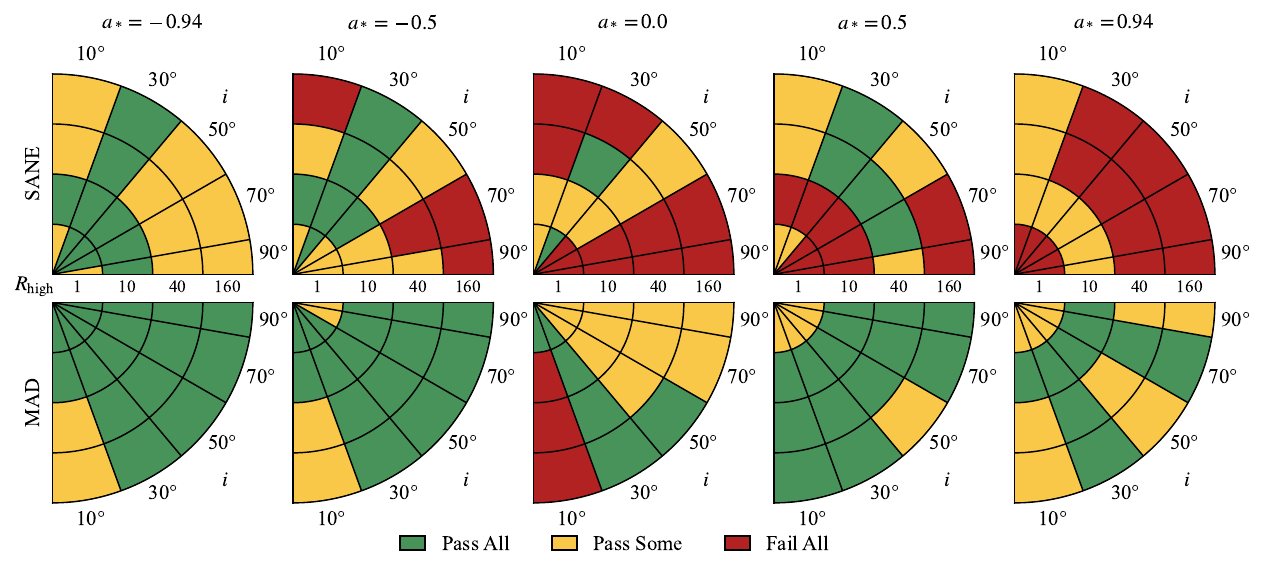}
  \caption{86GHz Size Constraints.  Green indicates that the \kharma, \bhac, and \hamr models pass, yellow that one or two of the fiducial models fail, and red that all three fail.}
  \label{fig:86GHz_size_pizza}
\end{figure*}

\begin{figure*}
  \centering
  \includegraphics[width=0.8\textwidth]{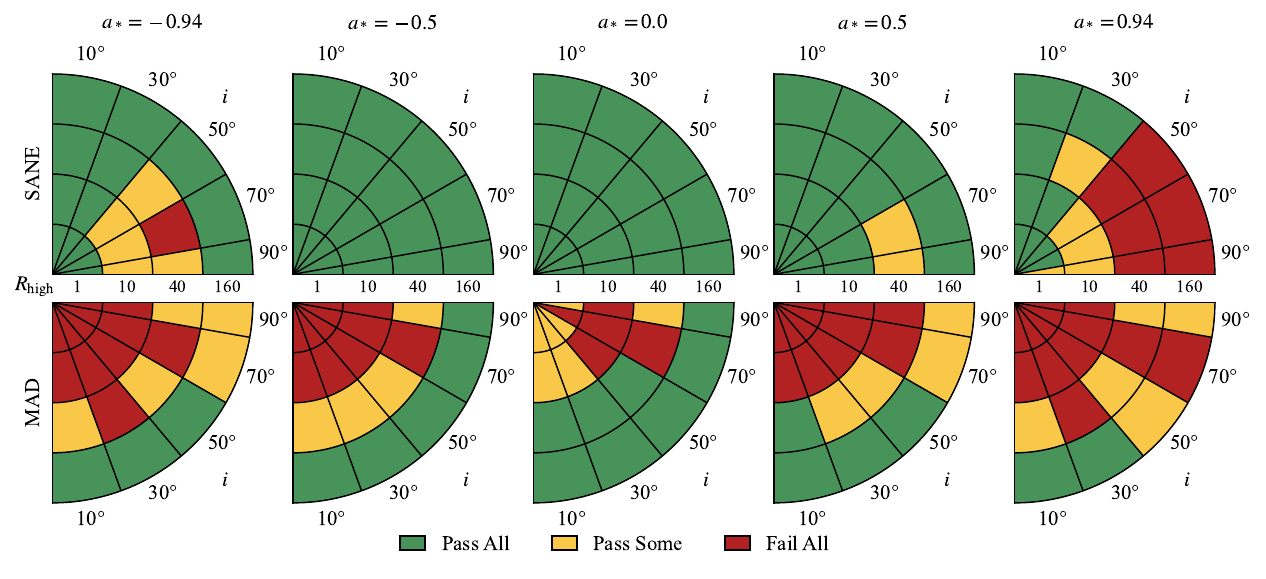}
  \caption{$2.2\mu$m Flux Constraints.  Green indicates that the \kharma, \bhac, and \hamr models pass, yellow that one or two of the fiducial models fail, and red that all three fail.}
  \label{fig:2um_flux_pizza}
\end{figure*}

\begin{figure*}
  \centering
  \includegraphics[width=0.8\textwidth]{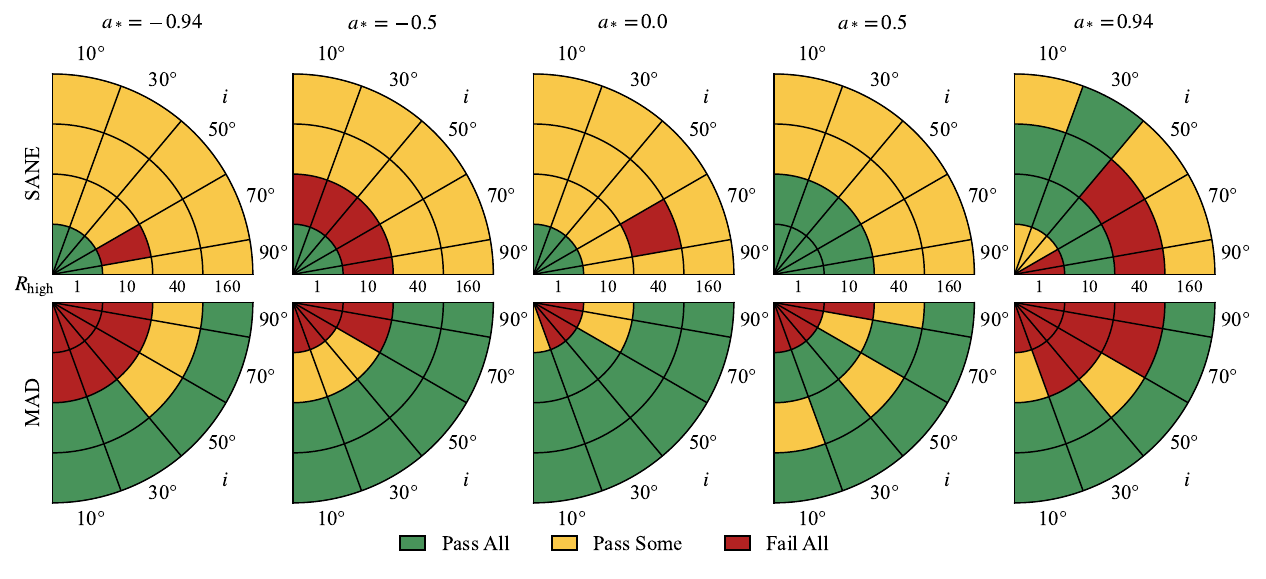}
  \caption{X-Ray Luminosity Constraints.  Green indicates that the \kharma, \bhac, and \hamr models pass, yellow that one or two of the fiducial models fail, and red that all three fail.}
  \label{fig:xray_pizza}
\end{figure*}

\begin{figure*}
  \centering
  \includegraphics[width=0.8\textwidth]{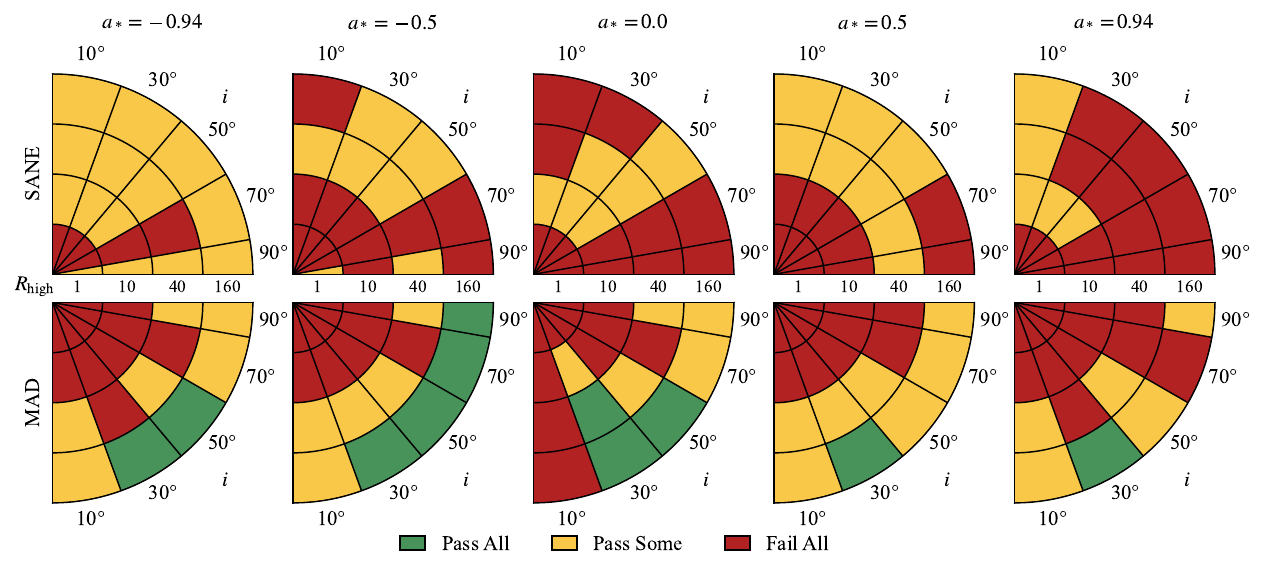}
  \caption{Combined non-EHT Constraints.  Green indicates that the \kharma, \bhac, and \hamr models pass, yellow that one or two of the fiducial models fail, and red that all three fail.}
  \label{fig:noneht_pizza}
\end{figure*}

﻿We then ﻿show the non-EHT constriants.
Figures~\ref{fig:86GHz_flux_pizza} and \ref{fig:86GHz_size_pizza} are the 86\GHz flux and size constraints, respectively.
Figures~\ref{fig:2um_flux_pizza} and \ref{fig:xray_pizza} show the 2.2\um and x-ray constraints.
Figure~\ref{fig:noneht_pizza} combines all the non-EHT constraints.

\begin{figure*}
  \centering
  \includegraphics[width=0.8\textwidth]{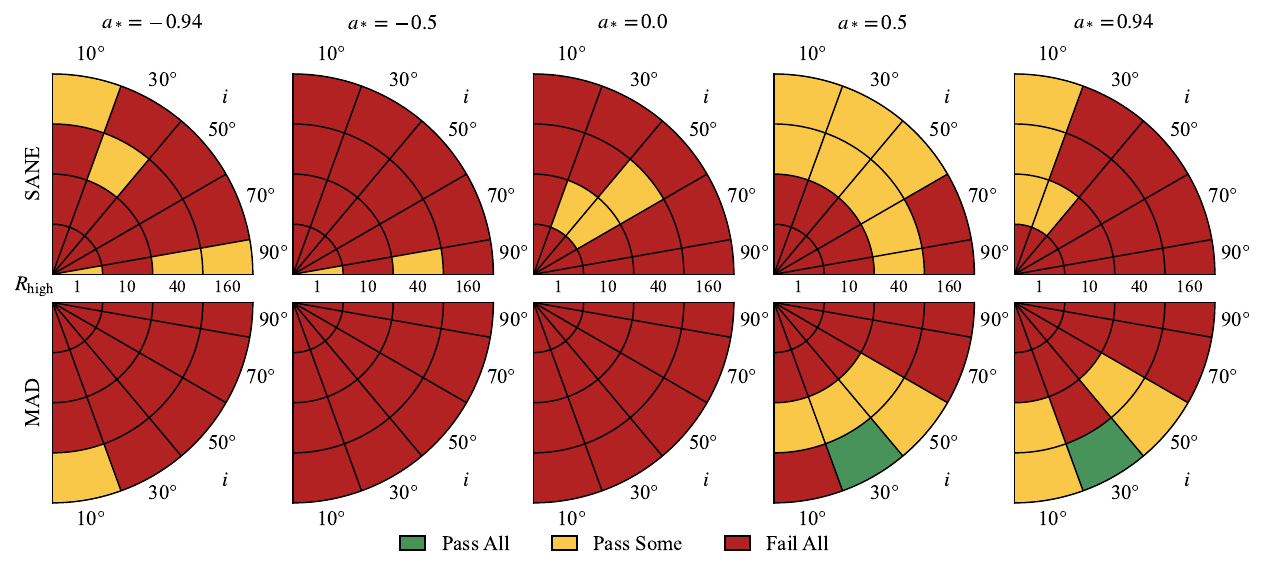}
  \caption{Combined constraints without structural or flux variability.  Green indicates that the \kharma, \bhac, and \hamr models pass, yellow that one or two of the fiducial models fail, and red that all three fail.}
  \label{fig:all_pizza_app}
\end{figure*}

\begin{figure*}
  \centering
  \includegraphics[width=0.8\textwidth]{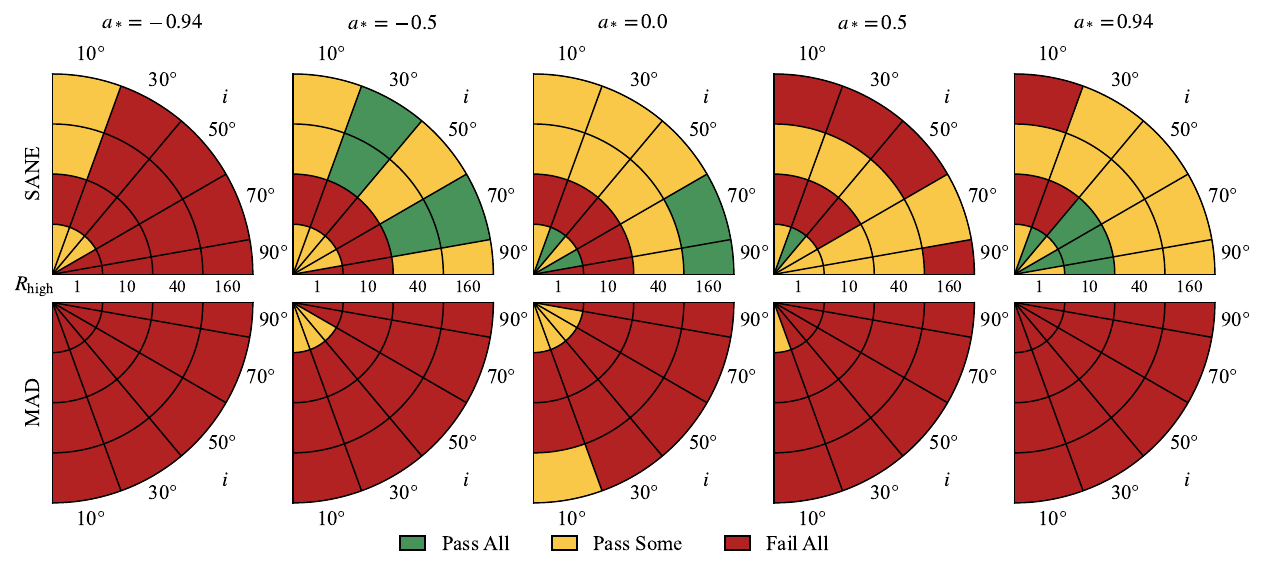}
  \caption{$M_3$ Constraints.  Green indicates that the \kharma, \bhac, and \hamr models pass, yellow that one or two of the fiducial models fail, and red that all three fail.}
  \label{fig:m3_pizza}
\end{figure*}

\begin{figure*}
  \centering
  \includegraphics[width=0.8\textwidth]{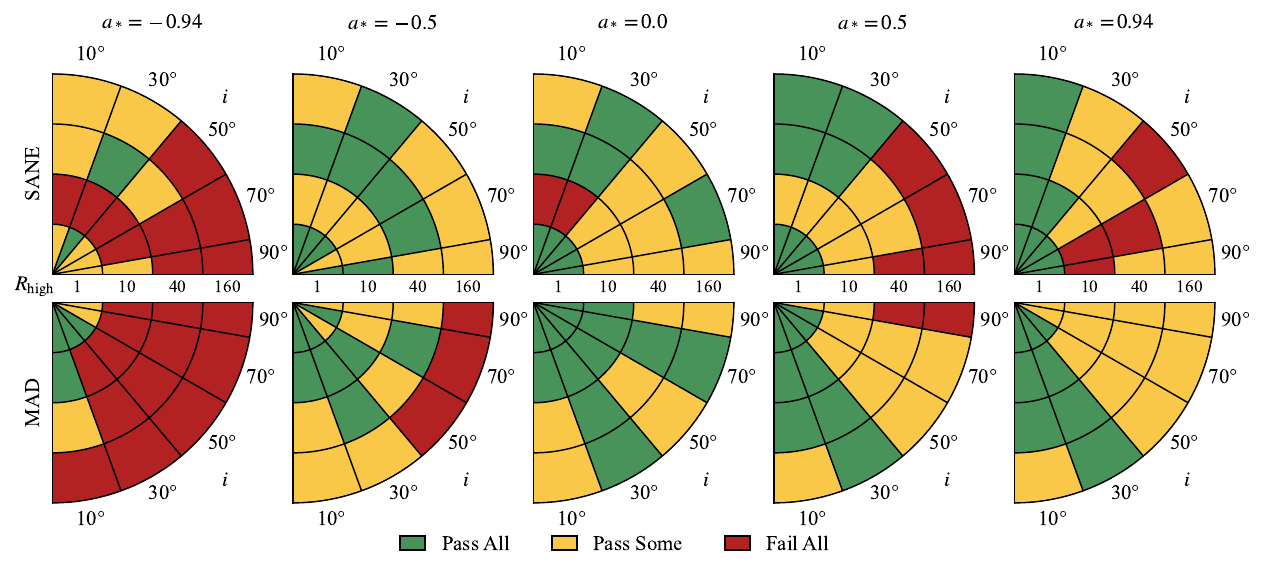}
  \caption{EHT Structural Variability Constraints.  Green indicates that the \kharma, \bhac, and \hamr models pass, yellow that one or two of the fiducial models fail, and red that all three fail.}
  \label{fig:ehtvar_pizza}
\end{figure*}

Figure~\ref{fig:all_pizza_app} shows the ﻿combined constraints without structural or flux variability.
Finally, Figures~\ref{fig:m3_pizza} and \ref{fig:ehtvar_pizza} show the \mi constraints and structural variability constraints, respectively.
These results are summarized in Section~\ref{sec:summarythermal}.

In each plot the green models pass the constraint or constraints for \kharma, \bhac, and \hamr version of the model, while the red models fail the constraint or constraints for \kharma, \bhac, and \hamr.
The yellow models have different results for the different codes.
Notice that \hamr models are available only for $i = 10, 50, 90$deg; at $30, 70$deg only the \kharma and \bhac models are compared.


\clearpage

\allauthors

\nocite{M87PaperI}
\nocite{M87PaperII}
\nocite{M87PaperIII}
\nocite{M87PaperIV}
\nocite{M87PaperV}
\nocite{M87PaperVI}
\nocite{M87PaperVII}
\nocite{M87PaperVIII}
\nocite{PaperI}
\nocite{PaperII}
\nocite{PaperIII}
\nocite{PaperIV}
\nocite{PaperV}
\nocite{PaperVI}

\bibliographystyle{yahapj}
\bibliography{main,refs,EHTCPapers}

\end{document}